\documentclass[a4paper,11pt]{article}
\pdfoutput=1

\usepackage{jinstpub}
\usepackage{caption}
\usepackage{subcaption}
\usepackage{ptdr-definitions}
\usepackage{orcidlink}

\newcommand{\dedx}{\ensuremath{dE\!/\!dx}\xspace}
\newcommand{\KE}{\ensuremath{E_K}\xspace}
\newcommand{\KEtrue}{\ensuremath{E_{K}^{\rm{true}}}\xspace}
\newcommand{\KErange}{\ensuremath{E_{K}^{\rm{range}}}\xspace}
\newcommand{\KEfull}{\ensuremath{E_{K}^{\rm{full}}}\xspace}
\newcommand{\KEtle}{\ensuremath{E_{K}^{\rm{TLE}}}\xspace}
\newcommand{\dqdx}{\ensuremath{dQ/\!dx}\xspace}

\title{The track-length extension fitting algorithm for energy measurement of interacting particles in liquid argon TPCs and its performance with ProtoDUNE-SP data}

\collaboration{The DUNE Collaboration}
\affiliation[0]{Abilene Christian University, Abilene, TX 79601, USA}
\affiliation[1]{University of Albany, SUNY, Albany, NY 12222, USA}
\affiliation[2]{Institute of Nuclear Physics at Almaty, Almaty 050032, Kazakhstan
}
\affiliation[3]{University of Amsterdam, NL-1098 XG Amsterdam, The Netherlands}
\affiliation[4]{Antalya Bilim University, 07190 D\"o{\c s}emealtı/Antalya, Turkey}
\affiliation[5]{University of Antananarivo, Antananarivo 101, Madagascar}
\affiliation[6]{University of Antioquia, Medell\'in, Colombia}
\affiliation[7]{Universidad Antonio Nari\~no, Bogot\'a, Colombia}
\affiliation[8]{Argonne National Laboratory, Argonne, IL 60439, USA}
\affiliation[9]{University of Arizona, Tucson, AZ 85721, USA}
\affiliation[10]{Universidad Nacional de Asunci\'on, San Lorenzo, Paraguay}
\affiliation[11]{University of Athens, Zografou GR 157 84, Greece}
\affiliation[12]{Universidad del Atl\'antico, Barranquilla, Atl\'antico, Colombia}
\affiliation[13]{Augustana University, Sioux Falls, SD 57197, USA}
\affiliation[14]{University of Bern, CH-3012 Bern, Switzerland}
\affiliation[15]{Beykent University, Istanbul, Turkey}
\affiliation[16]{University of Birmingham, Birmingham B15 2TT, United Kingdom}
\affiliation[17]{Universit\`a di Bologna, 40127 Bologna, Italy}
\affiliation[18]{Boston University, Boston, MA 02215, USA}
\affiliation[19]{University of Bristol, Bristol BS8 1TL, United Kingdom}
\affiliation[20]{Brookhaven National Laboratory, Upton, NY 11973, USA}
\affiliation[21]{University of Bucharest, Bucharest, Romania}
\affiliation[22]{University of California Berkeley, Berkeley, CA 94720, USA}
\affiliation[23]{University of California Davis, Davis, CA 95616, USA}
\affiliation[24]{University of California Irvine, Irvine, CA 92697, USA}
\affiliation[25]{University of California Los Angeles, Los Angeles, CA 90095, USA}
\affiliation[26]{University of California Riverside, Riverside CA 92521, USA}
\affiliation[27]{University of California Santa Barbara, Santa Barbara, CA 93106, USA}
\affiliation[28]{California Institute of Technology, Pasadena, CA 91125, USA}
\affiliation[29]{University of Cambridge, Cambridge CB3 0HE, United Kingdom}
\affiliation[30]{Universidade Estadual de Campinas, Campinas - SP, 13083-970, Brazil}
\affiliation[31]{Universit\`a di Catania, 2 - 95131 Catania, Italy}
\affiliation[32]{Universidad Cat\'olica del Norte, Antofagasta, Chile}
\affiliation[33]{Centro Brasileiro de Pesquisas F\'isicas, Rio de Janeiro, RJ 22290-180, Brazil}
\affiliation[34]{IRFU, CEA, Universit\'e Paris-Saclay, F-91191 Gif-sur-Yvette, France}
\affiliation[35]{CERN, The European Organization for Nuclear Research, 1211 Meyrin, Switzerland}
\affiliation[36]{Institute of Particle and Nuclear Physics of the Faculty of Mathematics and Physics of the Charles University, 180 00 Prague 8, Czech Republic }
\affiliation[37]{University of Chicago, Chicago, IL 60637, USA}
\affiliation[38]{Chung-Ang University, Seoul 06974, South Korea}
\affiliation[39]{CIEMAT, Centro de Investigaciones Energ\'eticas, Medioambientales y Tecnol\'ogicas, E-28040 Madrid, Spain}
\affiliation[40]{University of Cincinnati, Cincinnati, OH 45221, USA}
\affiliation[41]{Centro de Investigaci\'on y de Estudios Avanzados del Instituto Polit\'ecnico Nacional (Cinvestav), Mexico City, Mexico}
\affiliation[42]{Universidad de Colima, Colima, Mexico}
\affiliation[43]{University of Colorado Boulder, Boulder, CO 80309, USA}
\affiliation[44]{Colorado State University, Fort Collins, CO 80523, USA}
\affiliation[45]{Columbia University, New York, NY 10027, USA}
\affiliation[46]{Comisi\'on Nacional de Investigaci\'on y Desarrollo Aeroespacial, Lima, Peru}
\affiliation[47]{Centro de Tecnologia da Informacao Renato Archer, Amarais - Campinas, SP - CEP 13069-901}
\affiliation[48]{Central University of South Bihar, Gaya, 824236, India
}
\affiliation[49]{Institute of Physics, Czech Academy of Sciences, 182 00 Prague 8, Czech Republic}
\affiliation[50]{Czech Technical University, 115 19 Prague 1, Czech Republic}
\affiliation[51]{Laboratoire d'Annecy de Physique des Particules, Universit\'e Savoie Mont Blanc, CNRS, LAPP-IN2P3, 74000 Annecy, France}
\affiliation[52]{Daresbury Laboratory, Cheshire WA4 4AD, United Kingdom}
\affiliation[53]{Dordt University, Sioux Center, IA 51250, USA}
\affiliation[54]{Drexel University, Philadelphia, PA 19104, USA}
\affiliation[55]{Duke University, Durham, NC 27708, USA}
\affiliation[56]{Durham University, Durham DH1 3LE, United Kingdom}
\affiliation[57]{University of Edinburgh, Edinburgh EH8 9YL, United Kingdom}
\affiliation[58]{Universidad EIA, Envigado, Antioquia, Colombia}
\affiliation[59]{E\"otv\"os Lor\'and University, 1053 Budapest, Hungary}
\affiliation[60]{Erciyes University, Kayseri, Turkey}
\affiliation[61]{Faculdade de Ci\^encias da Universidade de Lisboa - FCUL, 1749-016 Lisboa, Portugal}
\affiliation[62]{Universidade Federal de Alfenas, Po{\c c}os de Caldas - MG, 37715-400, Brazil}
\affiliation[63]{Universidade Federal de Goias, Goiania, GO 74690-900, Brazil}
\affiliation[64]{Universidade Federal do ABC, Santo Andr\'e - SP, 09210-580, Brazil}
\affiliation[65]{Universidade Federal do Rio de Janeiro, Rio de Janeiro - RJ, 21941-901, Brazil}
\affiliation[66]{Fermi National Accelerator Laboratory, Batavia, IL 60510, USA}
\affiliation[67]{University of Ferrara, Ferrara, Italy}
\affiliation[68]{University of Florida, Gainesville, FL 32611-8440, USA}
\affiliation[69]{Florida State University, Tallahassee, FL, 32306 USA}
\affiliation[70]{Fluminense Federal University, 9 Icara\'i Niter\'oi - RJ, 24220-900, Brazil }
\affiliation[71]{Universit\`a degli Studi di Genova, Genova, Italy}
\affiliation[72]{Georgian Technical University, Tbilisi, Georgia}
\affiliation[73]{University of Granada \& CAFPE, 18002 Granada, Spain}
\affiliation[74]{Gran Sasso Science Institute, L'Aquila, Italy}
\affiliation[75]{Laboratori Nazionali del Gran Sasso, L'Aquila AQ, Italy}
\affiliation[76]{University Grenoble Alpes, CNRS, Grenoble INP, LPSC-IN2P3, 38000 Grenoble, France}
\affiliation[77]{Universidad de Guanajuato, Guanajuato, C.P. 37000, Mexico}
\affiliation[78]{Harish-Chandra Research Institute, Jhunsi, Allahabad 211 019, India}
\affiliation[79]{University of Hawaii, Honolulu, HI 96822, USA}
\affiliation[80]{Hong Kong University of Science and Technology, Kowloon, Hong Kong, China}
\affiliation[81]{University of Houston, Houston, TX 77204, USA}
\affiliation[82]{University of  Hyderabad, Gachibowli, Hyderabad - 500 046, India}
\affiliation[83]{Idaho State University, Pocatello, ID 83209, USA}
\affiliation[84]{Instituto de F\'isica Corpuscular, CSIC and Universitat de Val\`encia, 46980 Paterna, Valencia, Spain}
\affiliation[85]{Instituto Galego de F\'isica de Altas Enerx\'ias, University of Santiago de Compostela, Santiago de Compostela, 15782, Spain}
\affiliation[86]{Indian Institute of Technology Kanpur, Uttar Pradesh 208016, India}
\affiliation[87]{Illinois Institute of Technology, Chicago, IL 60616, USA}
\affiliation[88]{Imperial College of Science, Technology and Medicine, London SW7 2BZ, United Kingdom}
\affiliation[89]{Indian Institute of Technology Guwahati, Guwahati, 781 039, India}
\affiliation[90]{Indian Institute of Technology Hyderabad, Hyderabad, 502285, India}
\affiliation[91]{Indiana University, Bloomington, IN 47405, USA}
\affiliation[92]{Istituto Nazionale di Fisica Nucleare Sezione di Bologna, 40127 Bologna BO, Italy}
\affiliation[93]{Istituto Nazionale di Fisica Nucleare Sezione di Catania, I-95123 Catania, Italy}
\affiliation[94]{Istituto Nazionale di Fisica Nucleare Sezione di Ferrara, I-44122 Ferrara, Italy}
\affiliation[95]{Istituto Nazionale di Fisica Nucleare Laboratori Nazionali di Frascati, Frascati, Roma, Italy}
\affiliation[96]{Istituto Nazionale di Fisica Nucleare Sezione di Genova, 16146 Genova GE, Italy}
\affiliation[97]{Istituto Nazionale di Fisica Nucleare Sezione di Lecce, 73100 - Lecce, Italy}
\affiliation[98]{Istituto Nazionale di Fisica Nucleare Sezione di Milano Bicocca, 3 - I-20126 Milano, Italy}
\affiliation[99]{Istituto Nazionale di Fisica Nucleare Sezione di Milano, 20133 Milano, Italy}
\affiliation[100]{Istituto Nazionale di Fisica Nucleare Sezione di Napoli, I-80126 Napoli, Italy}
\affiliation[101]{Istituto Nazionale di Fisica Nucleare Sezione di Padova, 35131 Padova, Italy}
\affiliation[102]{Istituto Nazionale di Fisica Nucleare Sezione di Pavia,  I-27100 Pavia, Italy}
\affiliation[103]{Istituto Nazionale di Fisica Nucleare Laboratori Nazionali di Pisa, Pisa PI, Italy}
\affiliation[104]{Istituto Nazionale di Fisica Nucleare Sezione di Roma, 00185 Roma RM, Italy}
\affiliation[105]{Istituto Nazionale di Fisica Nucleare Roma Tor Vergata , 00133 Roma RM, Italy}
\affiliation[106]{Istituto Nazionale di Fisica Nucleare Laboratori Nazionali del Sud, 95123 Catania, Italy}
\affiliation[107]{Universidad Nacional de Ingenier\'ia, Lima 25, Per\'u}
\affiliation[108]{University of Insubria, Via Ravasi, 2, 21100 Varese VA, Italy}
\affiliation[109]{University of Iowa, Iowa City, IA 52242, USA}
\affiliation[110]{Iowa State University, Ames, Iowa 50011, USA}
\affiliation[111]{Institut de Physique des 2 Infinis de Lyon, 69622 Villeurbanne, France}
\affiliation[112]{Institute for Research in Fundamental Sciences, Tehran, Iran}
\affiliation[113]{Instituto Superior T\'ecnico - IST, Universidade de Lisboa, 1049-001 Lisboa, Portugal}
\affiliation[114]{Instituto Tecnol\'ogico de Aeron\'autica, Sao Jose dos Campos, Brazil}
\affiliation[115]{Iwate University, Morioka, Iwate 020-8551, Japan}
\affiliation[116]{Jackson State University, Jackson, MS 39217, USA}
\affiliation[117]{Jawaharlal Nehru University, New Delhi 110067, India}
\affiliation[118]{Jeonbuk National University, Jeonrabuk-do 54896, South Korea}
\affiliation[119]{Jyv\"askyl\"a University, FI-40014 Jyv\"askyl\"a, Finland}
\affiliation[120]{Kansas State University, Manhattan, KS 66506, USA}
\affiliation[121]{Kavli Institute for the Physics and Mathematics of the Universe, Kashiwa, Chiba 277-8583, Japan}
\affiliation[122]{High Energy Accelerator Research Organization (KEK), Ibaraki, 305-0801, Japan}
\affiliation[123]{Korea Institute of Science and Technology Information, Daejeon, 34141, South Korea}
\affiliation[124]{National Institute of Technology, Kure College, Hiroshima, 737-8506, Japan}
\affiliation[125]{Taras Shevchenko National University of Kyiv, 01601 Kyiv, Ukraine}
\affiliation[126]{Lancaster University, Lancaster LA1 4YB, United Kingdom}
\affiliation[127]{Lawrence Berkeley National Laboratory, Berkeley, CA 94720, USA}
\affiliation[128]{Laborat\'orio de Instrumenta{\c c}\~ao e F\'isica Experimental de Part\'iculas, 1649-003 Lisboa and 3004-516 Coimbra, Portugal}
\affiliation[129]{University of Liverpool, L69 7ZE, Liverpool, United Kingdom}
\affiliation[130]{Los Alamos National Laboratory, Los Alamos, NM 87545, USA}
\affiliation[131]{Louisiana State University, Baton Rouge, LA 70803, USA}
\affiliation[132]{Laboratoire de Physique des Deux Infinis Bordeaux - IN2P3, F-33175 Gradignan, Bordeaux, France, }
\affiliation[133]{University of Lucknow, Uttar Pradesh 226007, India}
\affiliation[134]{Madrid Autonoma University and IFT UAM/CSIC, 28049 Madrid, Spain}
\affiliation[135]{Johannes Gutenberg-Universit\"at Mainz, 55122 Mainz, Germany}
\affiliation[136]{University of Manchester, Manchester M13 9PL, United Kingdom}
\affiliation[137]{Massachusetts Institute of Technology, Cambridge, MA 02139, USA}
\affiliation[138]{University of Medell\'in, Medell\'in, 050026 Colombia }
\affiliation[139]{University of Michigan, Ann Arbor, MI 48109, USA}
\affiliation[140]{Michigan State University, East Lansing, MI 48824, USA}
\affiliation[141]{Universit\`a di Milano Bicocca , 20126 Milano, Italy}
\affiliation[142]{Universit\`a degli Studi di Milano, I-20133 Milano, Italy}
\affiliation[143]{University of Minnesota Duluth, Duluth, MN 55812, USA}
\affiliation[144]{University of Minnesota Twin Cities, Minneapolis, MN 55455, USA}
\affiliation[145]{University of Mississippi, University, MS 38677 USA}
\affiliation[146]{Universit\`a degli Studi di Napoli Federico II , 80138 Napoli NA, Italy}
\affiliation[147]{Nikhef National Institute of Subatomic Physics, 1098 XG Amsterdam, Netherlands}
\affiliation[148]{National Institute of Science Education and Research (NISER), Odisha 752050, India}
\affiliation[149]{University of North Dakota, Grand Forks, ND 58202-8357, USA}
\affiliation[150]{Northern Illinois University, DeKalb, IL 60115, USA}
\affiliation[151]{Northwestern University, Evanston, Il 60208, USA}
\affiliation[152]{University of Notre Dame, Notre Dame, IN 46556, USA}
\affiliation[153]{University of Novi Sad, 21102 Novi Sad, Serbia}
\affiliation[154]{Occidental College, Los Angeles, CA  90041}
\affiliation[155]{Ohio State University, Columbus, OH 43210, USA}
\affiliation[156]{Oregon State University, Corvallis, OR 97331, USA}
\affiliation[157]{University of Oxford, Oxford, OX1 3RH, United Kingdom}
\affiliation[158]{Pacific Northwest National Laboratory, Richland, WA 99352, USA}
\affiliation[159]{Universt\`a degli Studi di Padova, I-35131 Padova, Italy}
\affiliation[160]{Panjab University, Chandigarh, 160014, India}
\affiliation[161]{Universit\'e Paris-Saclay, CNRS/IN2P3, IJCLab, 91405 Orsay, France}
\affiliation[162]{Universit\'e Paris Cit\'e, CNRS, Astroparticule et Cosmologie, Paris, France}
\affiliation[163]{University of Parma,  43121 Parma PR, Italy}
\affiliation[164]{Universit\`a degli Studi di Pavia, 27100 Pavia PV, Italy}
\affiliation[165]{University of Pennsylvania, Philadelphia, PA 19104, USA}
\affiliation[166]{Pennsylvania State University, University Park, PA 16802, USA}
\affiliation[167]{Physical Research Laboratory, Ahmedabad 380 009, India}
\affiliation[168]{Universit\`a di Pisa, I-56127 Pisa, Italy}
\affiliation[169]{University of Pittsburgh, Pittsburgh, PA 15260, USA}
\affiliation[170]{Pontificia Universidad Cat\'olica del Per\'u, Lima, Per\'u}
\affiliation[171]{University of Puerto Rico, Mayaguez 00681, Puerto Rico, USA}
\affiliation[172]{Punjab Agricultural University, Ludhiana 141004, India}
\affiliation[173]{Queen Mary University of London, London E1 4NS, United Kingdom
}
\affiliation[174]{Radboud University, NL-6525 AJ Nijmegen, Netherlands}
\affiliation[175]{Rice University, Houston, TX 77005}
\affiliation[176]{University of Rochester, Rochester, NY 14627, USA}
\affiliation[177]{Royal Holloway College London, London, TW20 0EX, United Kingdom}
\affiliation[178]{Rutgers University, Piscataway, NJ, 08854, USA}
\affiliation[179]{STFC Rutherford Appleton Laboratory, Didcot OX11 0QX, United Kingdom}
\affiliation[180]{Universit\`a del Salento, 73100 Lecce, Italy}
\affiliation[181]{Universidad del Magdalena, Santa Marta - Colombia}
\affiliation[182]{Sapienza University of Rome, 00185 Roma RM, Italy}
\affiliation[183]{Universidad Sergio Arboleda, 11022 Bogot\'a, Colombia}
\affiliation[184]{University of Sheffield, Sheffield S3 7RH, United Kingdom}
\affiliation[185]{SLAC National Accelerator Laboratory, Menlo Park, CA 94025, USA}
\affiliation[186]{University of South Carolina, Columbia, SC 29208, USA}
\affiliation[187]{South Dakota School of Mines and Technology, Rapid City, SD 57701, USA}
\affiliation[188]{South Dakota State University, Brookings, SD 57007, USA}
\affiliation[189]{Southern Methodist University, Dallas, TX 75275, USA}
\affiliation[190]{Stony Brook University, SUNY, Stony Brook, NY 11794, USA}
\affiliation[191]{Sanford Underground Research Facility, Lead, SD, 57754, USA}
\affiliation[192]{University of Sussex, Brighton, BN1 9RH, United Kingdom}
\affiliation[193]{Syracuse University, Syracuse, NY 13244, USA}
\affiliation[194]{Universidade Tecnol\'ogica Federal do Paran\'a, Curitiba, Brazil}
\affiliation[195]{Tel Aviv University, Tel Aviv-Yafo, Israel}
\affiliation[196]{Texas A\&M University, College Station, Texas 77840}
\affiliation[197]{Texas A\&M University - Corpus Christi, Corpus Christi, TX 78412, USA}
\affiliation[198]{University of Texas at Arlington, Arlington, TX 76019, USA}
\affiliation[199]{University of Texas at Austin, Austin, TX 78712, USA}
\affiliation[200]{University of Toronto, Toronto, Ontario M5S 1A1, Canada}
\affiliation[201]{Tufts University, Medford, MA 02155, USA}
\affiliation[202]{Universidade Federal de S\~ao Paulo, 09913-030, S\~ao Paulo, Brazil}
\affiliation[203]{Ulsan National Institute of Science and Technology, Ulsan 689-798, South Korea}
\affiliation[204]{University College London, London, WC1E 6BT, United Kingdom}
\affiliation[205]{Universidad Nacional Mayor de San Marcos, Lima, Peru}
\affiliation[206]{Valley City State University, Valley City, ND 58072, USA}
\affiliation[207]{University of Vigo, E- 36310 Vigo Spain}
\affiliation[208]{Virginia Tech, Blacksburg, VA 24060, USA}
\affiliation[209]{University of Warsaw, 02-093 Warsaw, Poland}
\affiliation[210]{University of Warwick, Coventry CV4 7AL, United Kingdom}
\affiliation[211]{Wellesley College, Wellesley, MA 02481, USA}
\affiliation[212]{Wichita State University, Wichita, KS 67260, USA}
\affiliation[213]{William and Mary, Williamsburg, VA 23187, USA}
\affiliation[214]{University of Wisconsin Madison, Madison, WI 53706, USA}
\affiliation[215]{Yale University, New Haven, CT 06520, USA}
\affiliation[216]{Yerevan Institute for Theoretical Physics and Modeling, Yerevan 0036, Armenia}
\affiliation[217]{York University, Toronto M3J 1P3, Canada}
\author[35]{A.~Abed Abud,}
\author[157]{B.~Abi,}
\author[66]{R.~Acciarri,}
\author[12]{M.~A.~Acero,}
\author[194]{M.~R.~Adames,}
\author[72]{G.~Adamov,}
\author[66]{M.~Adamowski,}
\author[20]{D.~Adams,}
\author[19]{M.~Adinolfi,}
\author[30]{C.~Adriano,}
% ...

\emailAdd{sungbino@fnal.gov}

\abstract{This paper introduces a novel track-length extension fitting algorithm for measuring the kinetic energies of inelastically interacting particles in liquid argon time projection chambers (LArTPCs). The algorithm finds the most probable offset in track length for a track-like object by comparing the measured ionization density as a function of position with a theoretical prediction of the energy loss as a function of the energy, including models of electron recombination and detector response. The algorithm can be used to measure the energies of particles that interact before they stop, such as charged pions that are absorbed by argon nuclei. The algorithm's energy measurement resolutions and fractional biases are presented as functions of particle kinetic energy and number of track hits using samples of stopping secondary charged pions in data collected by the ProtoDUNE-SP detector, and also in a detailed simulation. Additional studies describe the impact of the $dE\!/\!dx$ model on energy measurement performance. The method described in this paper to characterize the energy measurement performance can be repeated in any LArTPC experiment using stopping secondary charged pions.}

\keywords{Noble liquid detectors (scintillation, ionization, single-phase), Time projection chambers, Large detector systems for particle and astroparticle physics}

\begin{document}
\maketitle
\flushbottom

\section{Introduction}
\label{sec:intro}
The observation of neutrino oscillations~\cite{PhysRevLett.81.1562,PhysRevLett.89.011301} has opened a new window for the field of particle physics. The fact that neutrinos oscillate in flavor implies not only that some neutrino mass eigenstates have nonzero masses, but it also provides the possibility to contribute to our understanding of the broken baryon number asymmetry of the universe~\cite{Fukugita:1986hr}.

The liquid argon time projection chamber (LArTPC)~\cite{Rubbia:1977zz} is a leading detector technology which has been proposed to answer the important questions described above~\cite{DUNE:2022aul}.
In LArTPCs, the argon serves simultaneously as the target for neutrino interactions and as the
sensitive detector material. Neutrinos interact primarily with argon nuclei, but they may also scatter off the electrons in the argon atoms. Particles produced by these interactions travel through the liquid argon, ionizing argon atoms along their paths. In an applied electric field, some of the electrons liberated from the argon atoms will drift towards collection wires or pixels, while other electrons will recombine with the positively-charged argon ions. Scintillation light with a wavelength of 128~nm is produced abundantly along the paths of ionizing tracks in liquid argon. The liquid itself is transparent to this light. By collecting the drifting electrons and the scintillation photons, LArTPCs can record neutrino interactions with 10\% level energy resolution~\cite{Friedland:2018vry}, mm-scale position resolution based on collection wire or pixel spacing, and ns order timing resolution from photon detection systems.

Neutrino beams for experiments using LArTPCs typically have wide energy spectra over several GeV with average energies of the order of a GeV. Neutrino oscillation probabilities depend on $L/E$, where $L$ is the distance from the production point to the detector, and $E$ is the neutrino energy. The distance $L$ is well known, but the energy $E$ must be measured for each neutrino scattering event, both in the near detectors and the far detectors. The neutrino energy must be estimated from the observed secondary particles produced in the interaction of the neutrino and an argon atom. These particles consist of charged leptons, protons, charged pions, photons, neutrons, kaons, and other shorter-lived particles that decay into the particles listed. Because the beams have broad energy spectra, a number of different scattering process categories are possible: quasielastic scattering, resonant scattering and deep inelastic scattering~\cite{Formaggio:2012cpf}.

The dense detection material in a LArTPC introduces a challenge by causing secondary interactions of hadrons originating from the neutrino scatter with other argon atoms in the detector. Because the lengths of the tracks of interacting particles are shorter than those of stopping particles with the same energies, range-based energy estimations are insufficient for interacting particles. Furthermore, interactions between hadrons and nuclei typically produce particles for which the energy is difficult or impossible to determine, such as neutrons and heavy nuclear fragments. Calorimetric energy measurements therefore only collect a fraction of the total energy of the incident particle, and that fraction varies from interaction to interaction in a random way which is difficult to model. Usually, calorimetric methods must be calibrated in situ with particles of known species and energies.

In this paper, we introduce a new method for energy measurement which can be utilized in a LArTPC.
It is based on a fit for the best track-length extension using calorimetric and position information of hits on a reconstructed track.
For an inelastically interacting particle, the method calculates the total track length the particle would have traveled if it had not interacted inelastically.
Therefore, the method addresses the challenges introduced in the above paragraph.
Data and simulation samples of the ProtoDUNE-SP detector~\cite{Abed_Abud_2022,Abi_2020} are used to evaluate the performance.
The method shows good performance for energy measurement on short tracks produced by interacting charged pions.

This paper is organized as follows.
In Section~\ref{sec:methodology}, we introduce the track-length extension fitting (TLEFit) algorithm.
Section~\ref{sec:ProtoDUNE} introduces the ProtoDUNE-SP detector and beam instrumentation,
and Section~\ref{sec:sample_reco} describes event reconstruction and the detector simulation.
In Section~\ref{sec:Performance}, studies for the performance of the TLEFit algorithm are presented.
Then, the paper concludes with a summary in Section~\ref{sec:summary}.

\section{The track-length extension fitting algorithm}
\label{sec:methodology}

Before introducing the track-length extension fitting (TLEFit) algorithm, it is important to discuss the basic operational principle of a LArTPC. A particle's activity inside liquid argon liberates electrons from argon atoms. A portion of these electrons recombines with argon ions, emitting photons. In general, a LArTPC experiment consists of a cathode plane and one or more anode planes to apply an electric field. The electrons drift toward the anode planes in this field, inducing electric signals in collection wires or pixels around the anode plane. Photon detection systems also collect the photons. The signals in the collection wires or pixels are recorded as functions of time within a time window. Since photons propagate much faster than drifting electrons, the trigger system that defines the starting point of the time window relies on photon detection systems.

For LArTPC experiments with collection wires, 2D space points are reconstructed by identifying peaks in the waveform of the electric signal. The combination of multiple collection wire planes with different orientations provides the capability to reconstruct 3D space points. In LArTPC experiments with readout pixels, a peak in the waveform represents a 3D space point. The deposited energy for a 3D space point is measured using the area around the peak in the waveform. Physical objects corresponding to traces of particles that traveled inside the LArTPC are produced by clustering the 3D space points with their positions and deposited energies (or hits hereafter). These physical objects are generally categorized as shower-like objects or track-like objects.

Particles such as muons, charged pions and protons ionize argon atoms in long, narrow trails. The drifted electrons are used to reconstruct track trajectories in the detector.
The average energy loss per unit length (\dedx) is well described by the Bethe--Bloch formula~\cite{Workman:2022ynf}, 
\begin{equation}
  { \Bigl \langle \frac{dE}{dx} \Bigl \rangle = \frac{\rho K Z}{A {\beta}^{2}} \left[ \frac{1}{2} \ln\left( \frac{2 {m}_{e}c^2 {\beta}^{2}{\gamma}^{2} {W}_{\rm{max}}}{{I}^{2}} \right)  - {\beta}^{2} - \frac{\delta}{2} \right]},
  \label{eq:Bethe-Blcoh}
\end{equation}
where ${m}_{e}=0.511~\MeV\!/c^2$ is the electron mass and $K = 0.307075~\MeV\, {\rm{cm}}^{2}\!/{\rm{mol}}$.
The parameters $\beta=v/c$ and $\gamma=\left(1-{\beta}^{2}\right)^{-1/2}$ are for the particle traveling inside the material.
For liquid argon, the density $\rho = 1.39~{\rm{g}}/{\rm{cm}}^{3}$, the atomic number $Z = 18$, the atomic mass $A= 39.948~{\rm{g/mol}}$,
and the average excitation energy $I = 188.0\times{10}^{-6}$~MeV\!.
The maximum transferable energy ${W\!}_{\rm{max}}$ is defined as
\begin{equation}
  {W\!}_{\rm{max}} = \frac{2 {m}_{e}c^2 {\beta}^{2}{\gamma}^{2}}{1 + \frac{2\gamma {m}_{e}}{M} +  \frac{{m}_{e}^{2}}{{M}^{2}}},
  \label{eq:Wmax}
\end{equation}
where $M$ is the mass of the traveling particle.
The last term of Eq~\ref{eq:Bethe-Blcoh} is the density correction,
\begin{equation}
  \delta = \left\{
    \begin{array}{@{}ll@{}}
      2\ln(\beta \gamma), & \text{if}\  {\log}_{10}(\beta \gamma) > {y}_{1}\\
      0, & \text{if}\ {\log}_{10}(\beta \gamma) < {y}_{0}\\
      2\ln(\beta \gamma) - C + a{[{y}_{1} -  {\log}_{10}(\beta \gamma)  ]}^{k}, & \text{otherwise,}
    \end{array}\right.
\end{equation}
where ${y}_{0} = 0.2$, ${y}_{1} = 3.0$, $C= 5.2146$, $a=0.19559$, and $k = 3.0$.
For kinetic energies above tens of \MeV, which is the threshold for particles to be reconstructed as tracks, nuclear effects in energy deposition is negligible.

While the energy lost by an ionizing particle in a thin slice of material is distributed according to the Landau distribution~\cite{Workman:2022ynf}, a particle traveling a long distance (more than a few $\rm{cm}$) through a material will experience an energy loss with an approximately Gaussian distribution due to the large number of ionization interactions and the central limit theorem.
As a result, the continuous slowing down approximation (CSDA) based on Eq~\ref{eq:Bethe-Blcoh} for the
average energy loss can be used to relate a particle's initial energy to its range in the material~\cite{Workman:2022ynf}. If a particle only undergoes multiple scattering and ionization energy loss before it stops, its initial energy can be estimated from its range with percent-level accuracy~\cite{MicroBooNE:2017tkp}.

If a particle interacts inelastically, however, the range cannot be used to estimate its energy.
Its track is cut off due to the interaction, so the range would result in an energy estimate lower than the true value.
Therefore, neutrino experiments with LArTPCs have been using the visible energy~\cite{LArIAT:2019kzd,MicroBooNE:2021sfa}, which is defined to be the sum of energy deposits left by the track object and all of its daughters (and their daughters) created in interactions with the liquid argon.
Unfortunately, it works well only when all daughters have enough energy to be reconstructed and the contribution from undetected neutrons is small.

\begin{figure*}[htbp]
\begin{center}
  \includegraphics[width=0.9\textwidth]{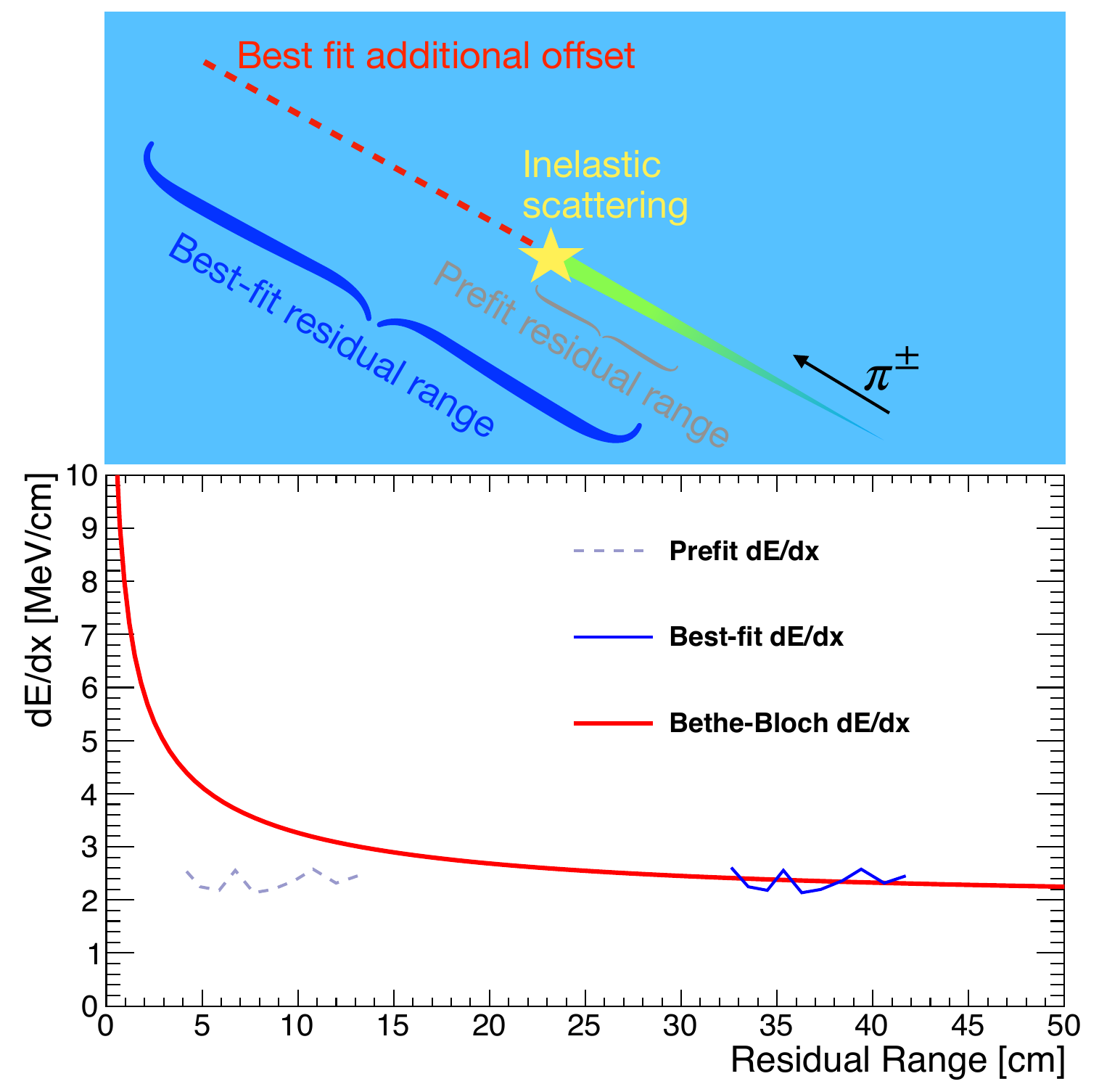}
  \caption{This figure illustrates the TLEFit method.
    In the bottom figure, the red solid line shows the mean energy loss $\left\langle\dedx\right\rangle$ as a function of distance from the stopping point (residual range) using Eq.~\ref{eq:Bethe-Blcoh} for a charged pion in LAr.
    An example of measured \dedx for an interacting charged pion is shown with a dashed line.
    The solid blue line shows the same measured \dedx curve with an additional offset on range with best match with the red solid line.
  }
  \label{fig:Figure_001}
\end{center}
\end{figure*}

The TLEFit algorithm introduced in this paper is a new energy measurement method which overcomes the weaknesses of the range-based estimation and visible energy-based estimation. Figure~\ref{fig:Figure_001} shows an example with hits of a reconstructed charged pion track. The abscissa is the residual range, which is defined for a hit on a charged particle's trajectory as the length from the trajectory's end point to the hit, following along the trajectory. The ordinate is \dedx in MeV\!/cm units. The average \dedx distribution of charged pions as a function of residual range is shown with the red solid line based on the Bethe-Bloch formula in liquid argon (LAr).
A \dedx distribution of an inelastically interacting charged pion is shown with a prefit \dedx label.
The measured \dedx values are obtained from the measured ionization charge density corrected for electron drift lifetime and recombination effects, as well as the charge yield expected per $\mathrm{MeV}$ of energy loss in LAr.
There is a gap between the prefit \dedx distribution and the vertical axis because first and last several hits are not included in the fit. They could be hits with incomplete \dedx or only noise. This is discussed in more detail in Section~\ref{sec:technical_detail}.
To measure the energy of a charged pion that has interacted inelastically with the argon, we look for an offset in the residual range which makes the residual range and \dedx distribution agree best with the known physics for passage of charged particle through liquid argon.
That offset is added to the original track length to estimate the best track-length.
This is the track length that the particle with the prefit \dedx distribution would have traveled until it stopped if it did not interact inelastically.
Finally, energy at the starting point of the track is measured using the CSDA and the best track-length.
The post fit \dedx distribution with the best offset is also shown.

Based on the residual range vs. \dedx curve derived from the Bethe-Bloch formula, shown in figure~\ref{fig:Figure_001}, it could be easily expected that energy measurement performance will highly depend on particle's initial kinetic energy (\KE).
Short residual range region that corresponds to low \KE of charged pions exhibits stiff slopes, commonly referred to as the Bragg peak region. Since \dedx varies rapidly as a function of residual range, it is easier to find the best offset in this region compared to longer residual range region, known as the MIP (minimum-ionizing particle) region, that has much smaller slopes.
For an inelastically interacting charged pion track for which all hits have MIP-like energy deposit around 2 $\mathrm{MeV/cm}$, the TLEFit algorithm cannot work well.
This is shown in Section~\ref{sec:Performance}.

One of the most important parts of the TLEFit is the procedure to calculate the best offset.
This paper describes and compares two methods. One method is based on minimizing a ${\chi}^{2}$ function assuming Gaussian distributions of inputs, and the second method is based on maximizing a likelihood function.

\subsection{Gaussian approximation}
\label{sec:gaus_approx}
The simplest way to score the agreement between the Bethe--Bloch curve and the measured \dedx distribution is by calculating the ${\chi}^{2}$ between them.
Many LArTPC experiments have employed ${\chi}^{2}$-based particle identification scores~\cite{Anderson:2012vc}.
Therefore, we describe this method before introducing a likelihood-based method.

For a given position offset ${L}^{\prime}$, we define
\begin{equation}
  {\chi}^{2}({L}^{\prime}) = \displaystyle\sum_{i = 1}^{{N}_{\rm{hits}}} { \left[ {\frac{dE}{dx}}({\rm{Measured}};~i) - \frac{dE}{dx}({\textrm{Bethe--Bloch}};~{\rm{range}}_{i} + {L}^{\prime}) \right]}^{2}.
  \label{eq:Gaussain_chi2}
\end{equation}
The \dedx term noted with the Bethe--Bloch on the right-hand side in Eq.~\ref{eq:Gaussain_chi2} is the expected \dedx from the CSDA using the track length given by the sum of hit's residual range and an additional offset, ${L}^{\prime}$.
By looking for the offset which gives the minimum ${\chi}^{2}$ value, the energy of the interacted particle can be measured.
We should note that Eq.~\ref{eq:Gaussain_chi2} is based on an assumption that measured \dedx follows Gaussian distributions with mean values given by the Bethe-Bloch formula in Eq.~\ref{eq:Bethe-Blcoh}.

\subsection{Maximum-likelihood}
The measured \dedx does not always follow a Gaussian distribution. 
This approximation is valid when there are sufficient interactions
so that outlying energy loss such as high-energy $\delta$-ray emission contribute a small fraction of the total energy loss.

\begin{figure*}[htbp]
\begin{center}
  \includegraphics[width=0.45\textwidth]{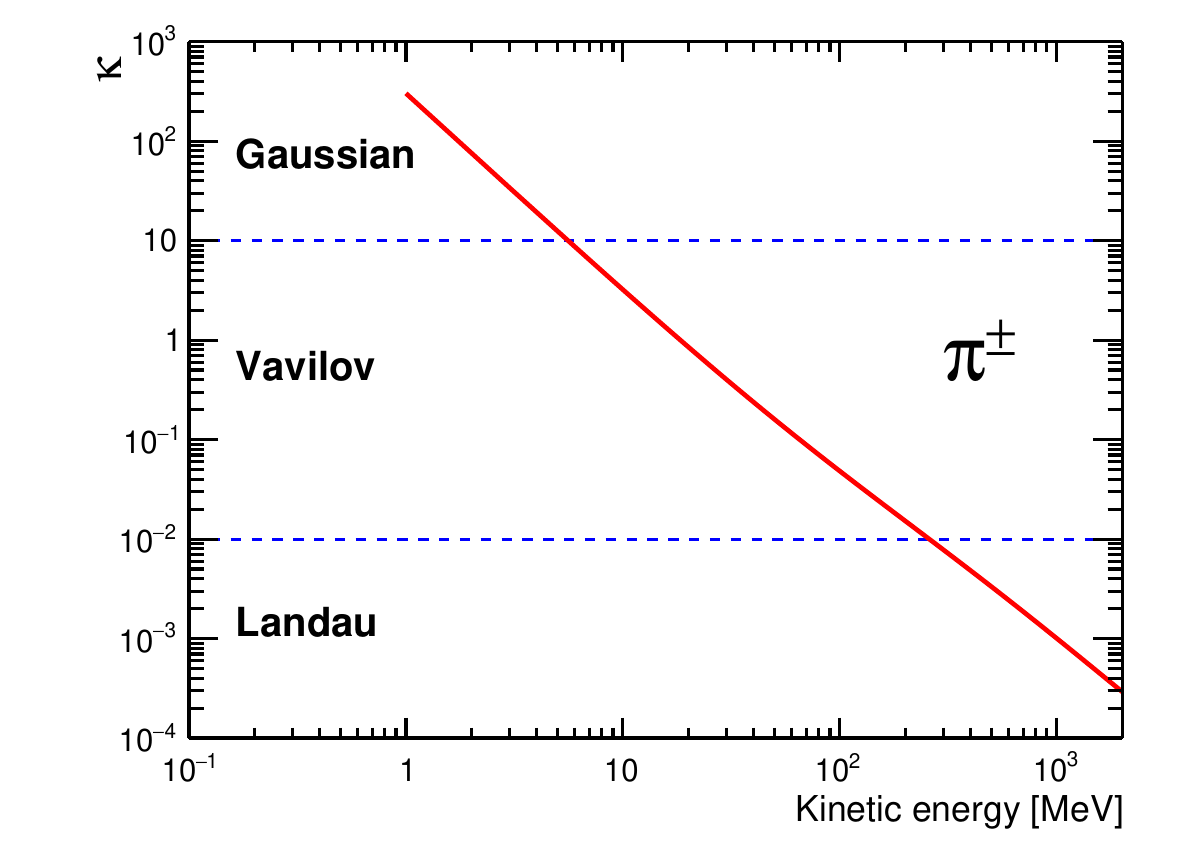}
  \includegraphics[width=0.45\textwidth]{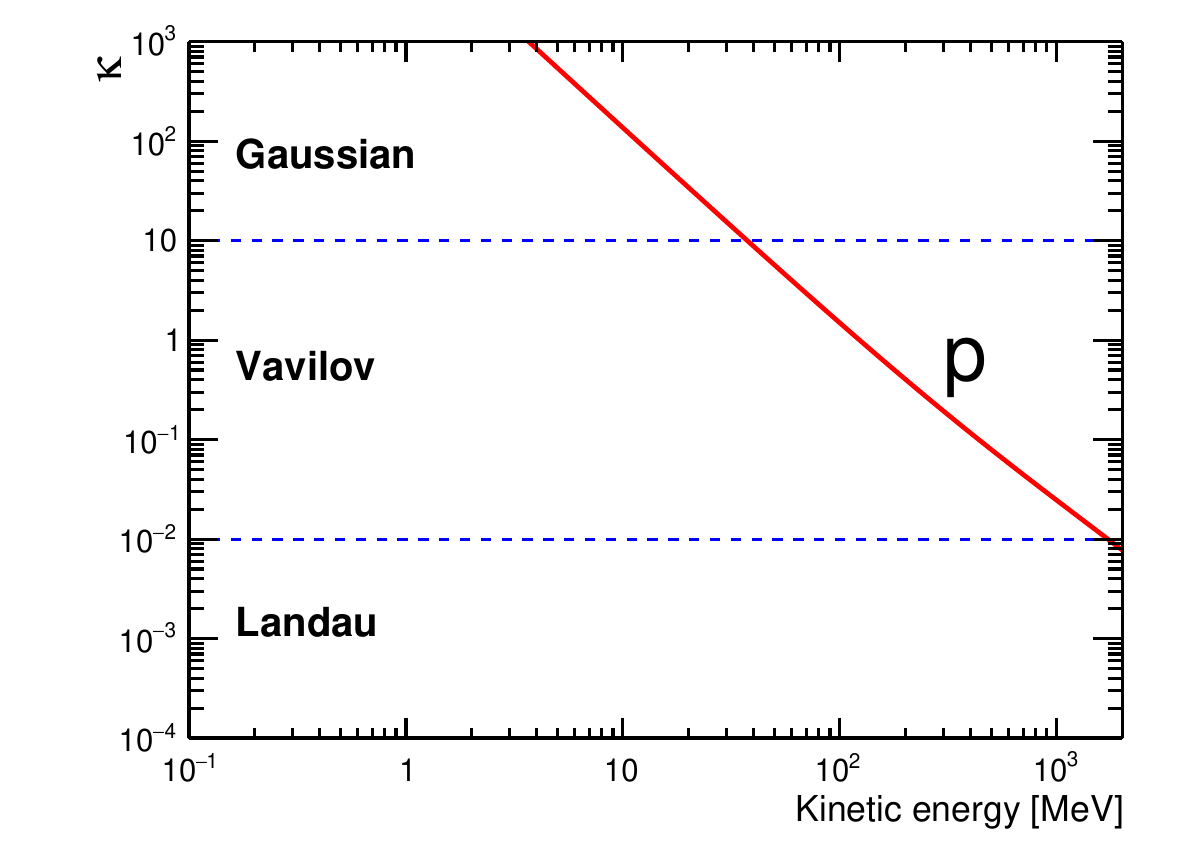}
  \caption{Three regions are defined using the $\kappa$ value. If $\kappa > 10$, the Gaussian PDF is used. If $\kappa < 0.01$, the Landau PDF is used.
    Otherwise, the Vavilov PDF is used. Plots are drawn for charged pions (left) and protons (right) in LAr using typical pitch (0.65 $\rm{cm}$) of the ProtoDUNE-SP.
  }
  \label{fig:Figure_002}
\end{center}
\end{figure*}

It is well known that the probability density function (PDF) of \dedx for ionizing particle is described by the Vavilov function~\cite{Vavilov:1957zz}.
In this paper, the significance parameter, $\kappa$, is used to define regions for the Vavilov function and its approximated versions, the Landau and the Gaussian functions.
The parameter $\kappa$ is defined as
\begin{equation}
  \kappa = \frac{\xi}{{W}_{\rm{max}}},
  \label{eq:kappa}
\end{equation}
where
\begin{equation}
  \xi = \frac{\rho K Z }{2A{\beta}^{2}} \delta x,
  \label{eq:xi}
\end{equation}
with $\delta x$ denoting the distance that the particle traveled in cm. The other variables are defined in the same way as in Eq.~\ref{eq:Bethe-Blcoh}. The $\kappa$ ranges that define PDFs for \dedx are selected as
\begin{equation}
  \mathrm{PDF~for~\dedx} : \left\{
    \begin{array}{@{}ll@{}}
      \mathrm{Gaussian} & , \text{if}\  \kappa > 10\\
      \mathrm{Vavilov} & , \text{if}\  0.01 <\kappa < 10\\
      \mathrm{Landau} & , \text{if}\  \kappa < 0.01\\
    \end{array}\right.
\end{equation}
following the Section PHYS332 of Geant~\cite{Brun:1994aa}. The $\kappa$ distribution as a function of \KE is shown in figure~\ref{fig:Figure_002} for charged pion and proton in LAr. Note that $\kappa$ is a function of $\delta x$ and the typical pitch of hits in ProtoDUNE-SP, 0.65 cm, is used in the figure.

\begin{figure*}[htbp]
\begin{center}
  \begin{subfigure}[b]{0.48\textwidth}
    \includegraphics[width=\textwidth]{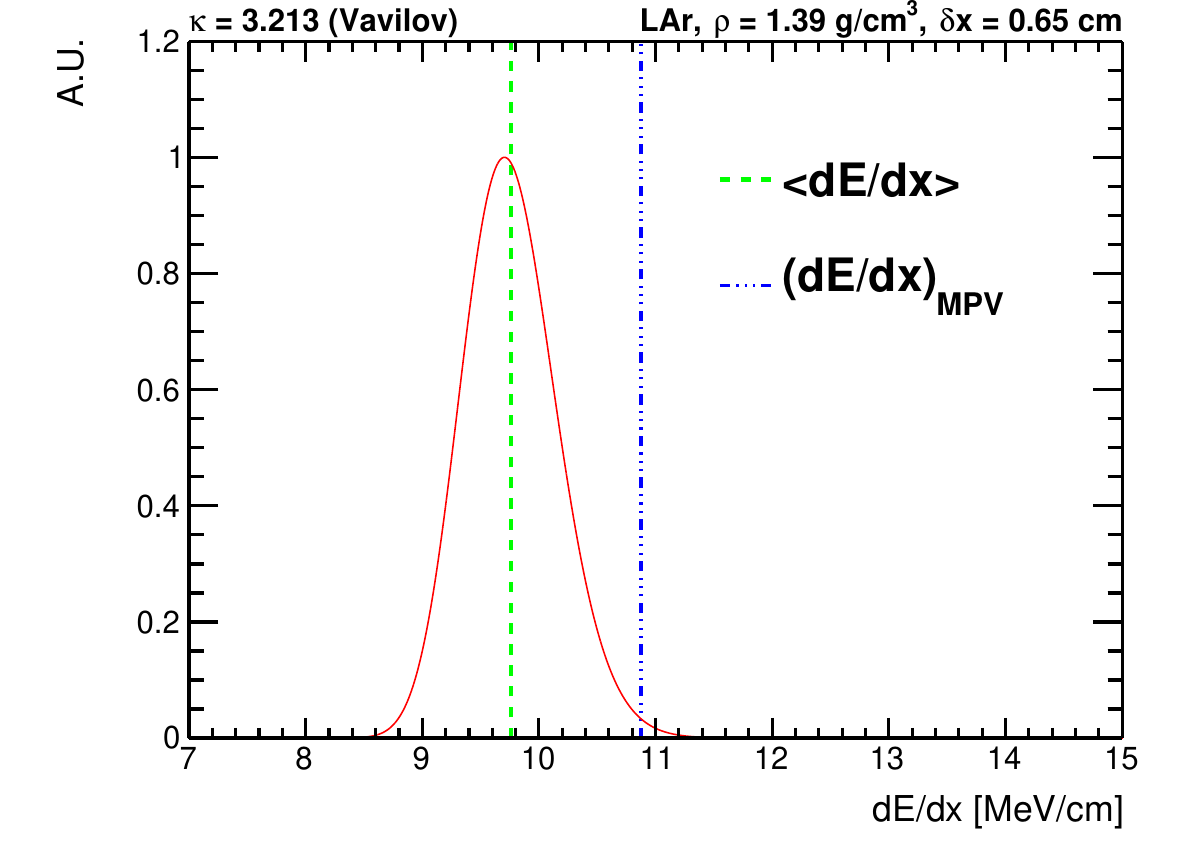}
    \caption{}
    \label{fig:Figure_003_a}
  \end{subfigure}
  \begin{subfigure}[b]{0.48\textwidth}
    \includegraphics[width=\textwidth]{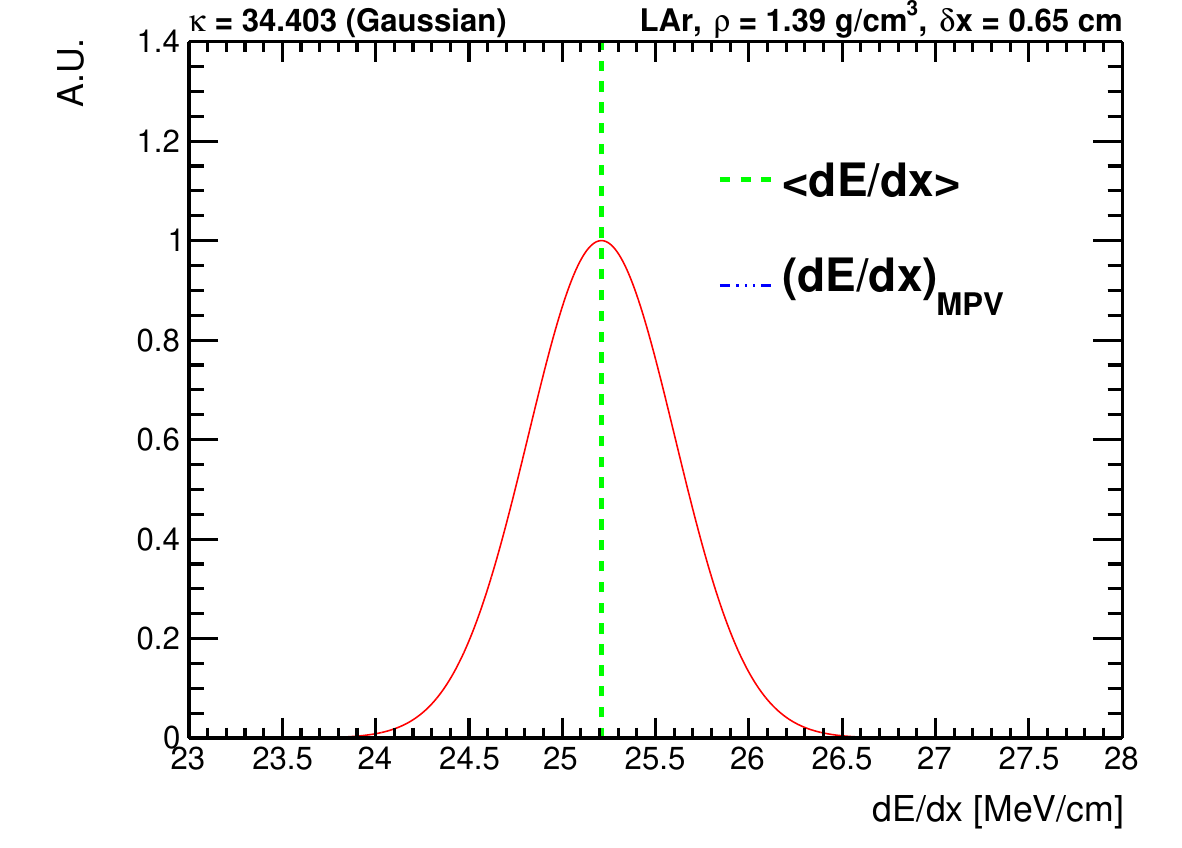}
    \caption{}
    \label{fig:Figure_003_b}
  \end{subfigure}
  \begin{subfigure}[b]{0.48\textwidth}
    \includegraphics[width=\textwidth]{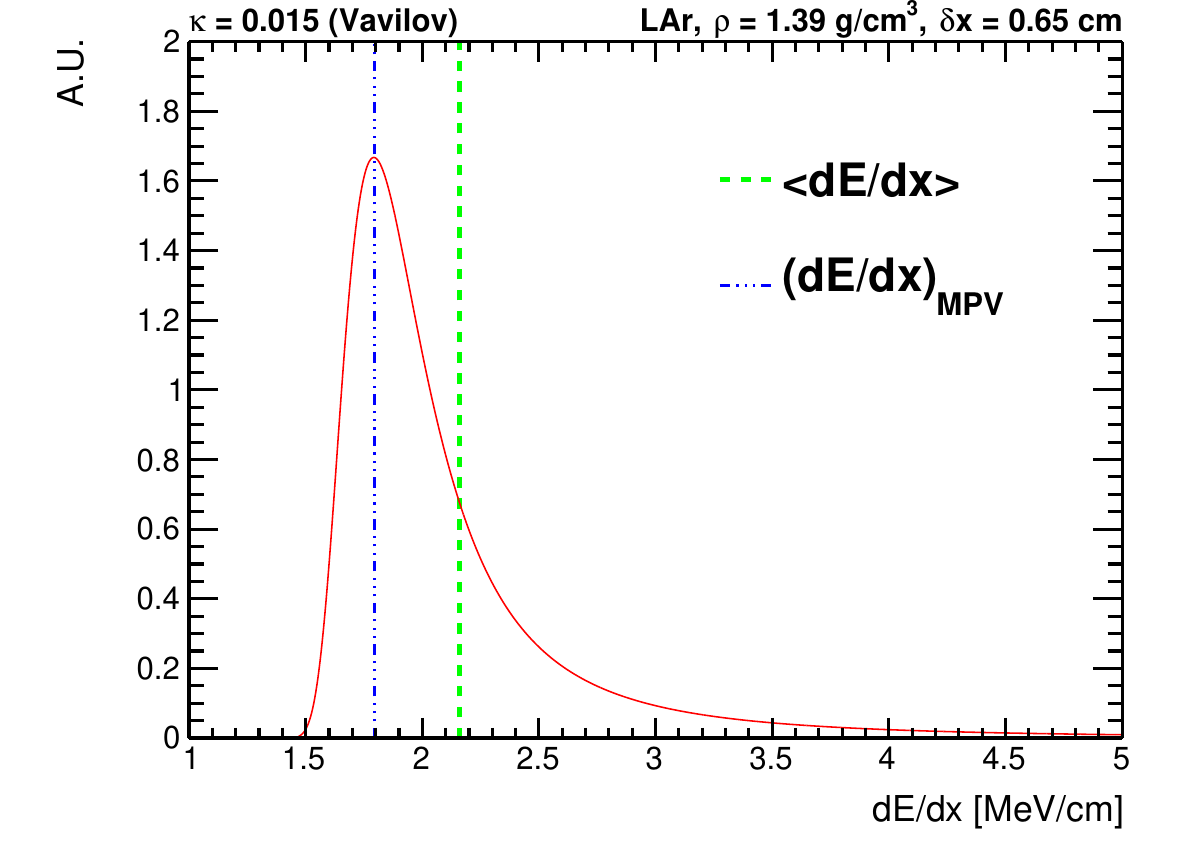}
    \caption{}
    \label{fig:Figure_003_c}
  \end{subfigure}
  \begin{subfigure}[b]{0.48\textwidth}
    \includegraphics[width=\textwidth]{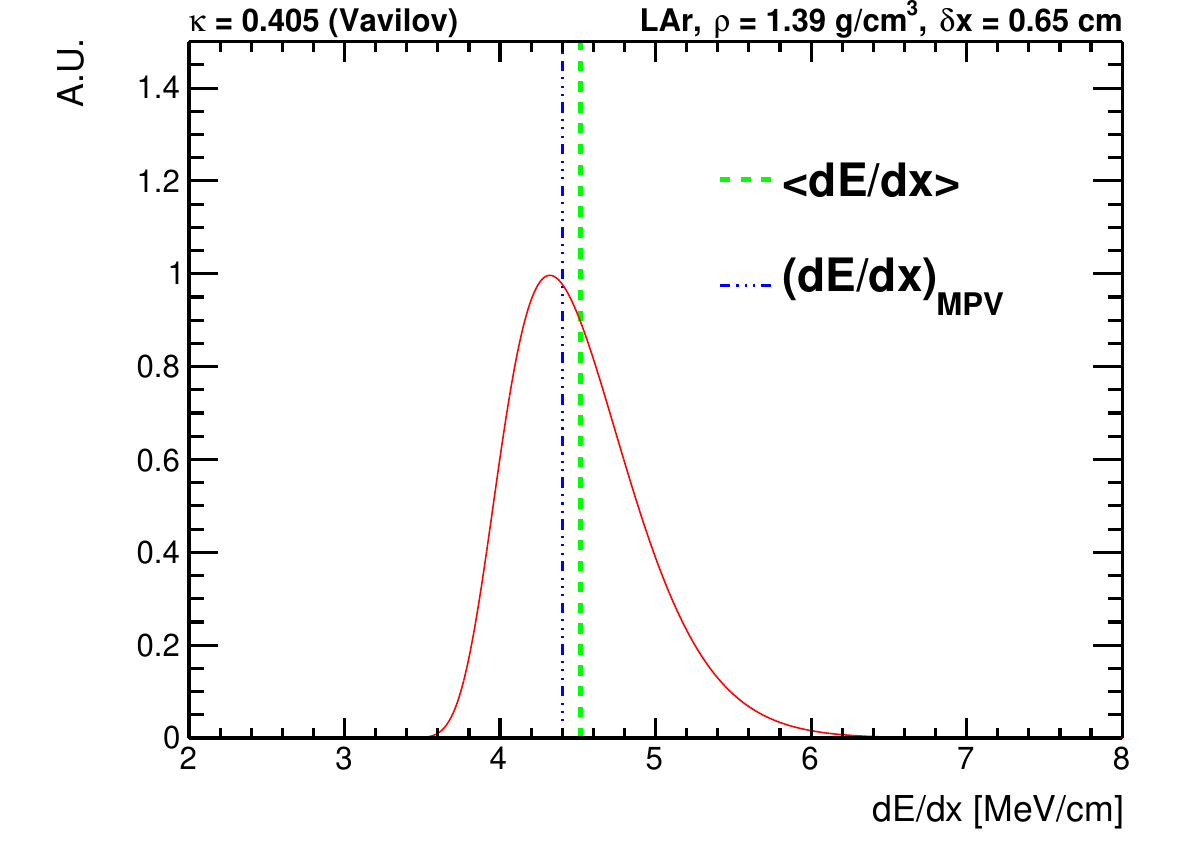}
    \caption{}
    \label{fig:Figure_003_d}
  \end{subfigure}
  \begin{subfigure}[b]{0.48\textwidth}
    \includegraphics[width=\textwidth]{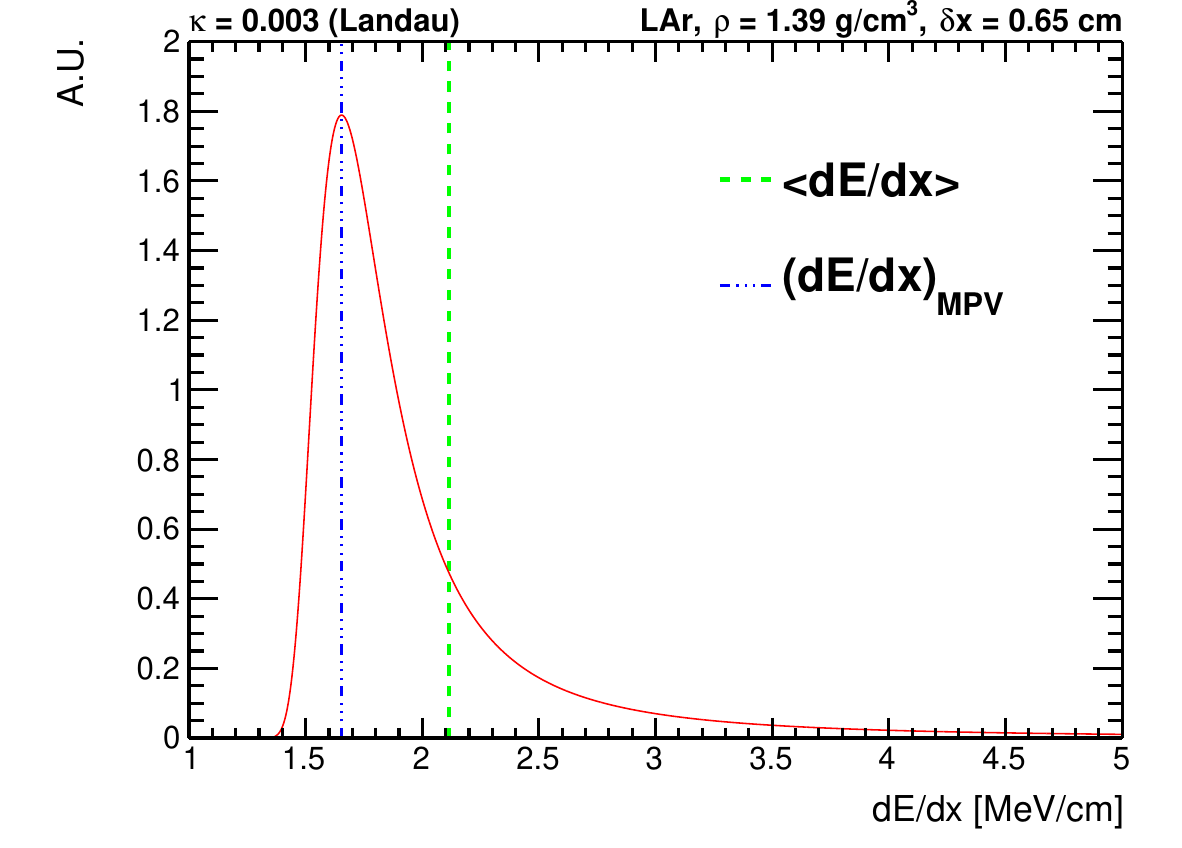}
    \caption{}
    \label{fig:Figure_003_e}
  \end{subfigure}
  \begin{subfigure}[b]{0.48\textwidth}
    \includegraphics[width=\textwidth]{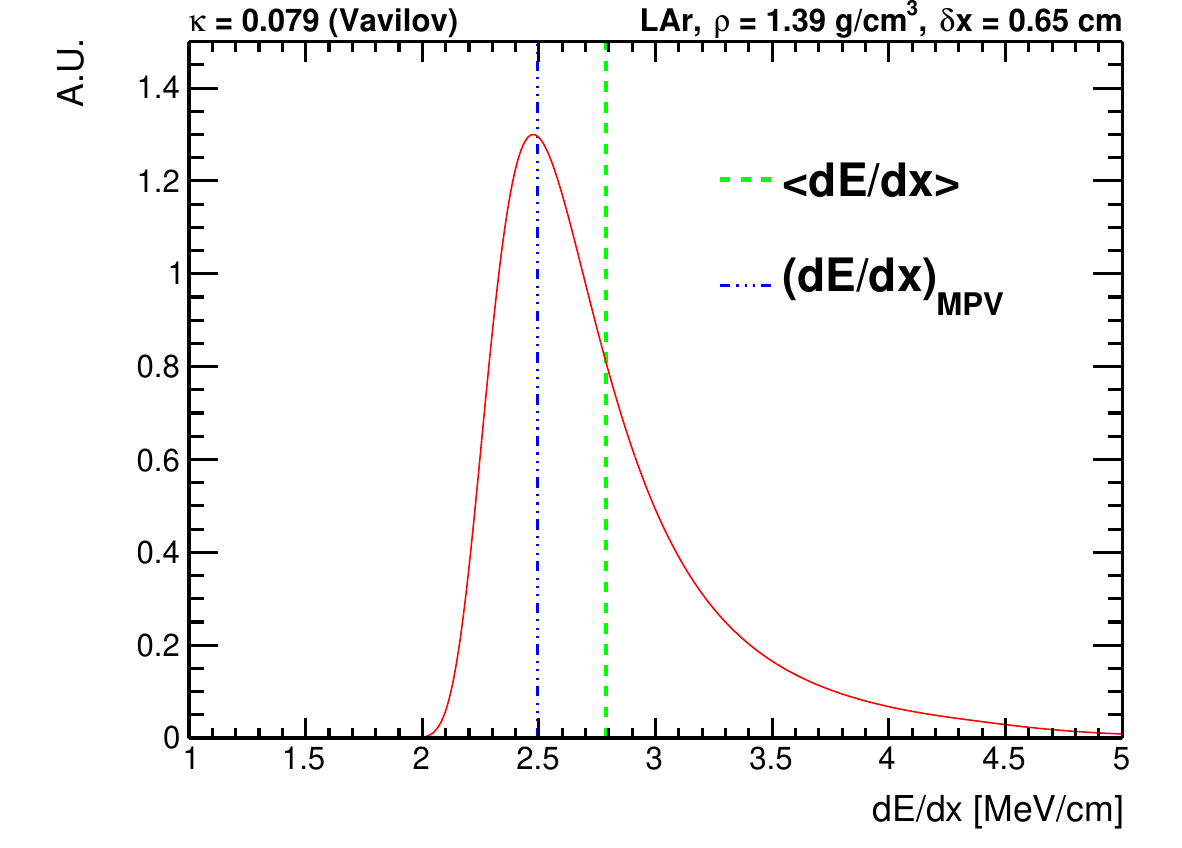}
    \caption{}
    \label{fig:Figure_003_f}
  \end{subfigure}
  \caption{The \dedx PDFs for charged pions (left) and protons (right) in LAr. For charged pions, PDFs are shown with \KE of 10 \MeV (top), 200 \MeV (center), and 500 \MeV (bottom). For protons, PDFs are shown with \KE of 20 \MeV (top), 200 \MeV (center), and 500 \MeV (bottom). Green lines show mean \dedx values from Eq.~\ref{eq:Bethe-Blcoh} and blue lines show the most probable values (MPV) from the Landau-Vavilov-Bichsel formula. The $\kappa$ values and used functions for PDFs are noted on top-left corners of plots.
  }
  \label{fig:Figure_003}
\end{center}
\end{figure*}

The PDFs with several \KE values for charged pions and protons are shown in figure~\ref{fig:Figure_003}. Mean \dedx value from the Bethe-Bloch formula and the most probable values (MPVs) of \dedx from the Landau-Vavilov-Bichsel formula~\cite{Bichsel:1988if} are presented together in dashed green lines and dashed blue lines, respectively.
We can see that the Gaussian approximation introduced in Section~\ref{sec:gaus_approx} is not a good choice for particles with $\kappa$ much smaller than unity. In this case, the Bethe-Bloch formula cannot describe the peak of the PDF, and the shape of the PDF is also significantly different from the Gaussian.
On the other hand, with $\kappa$ values bigger than unity, the Bethe-Bloch formula shows good agreement with the peak position, and the PDF shapes become similar with the Gaussian.

Using the PDFs $P\left(\frac{dE}{dx} \big| \KE\right)$ where $P$ stands for the probability, a joint likelihood (${\mathcal{L}}$) is calculated by multiplying the \dedx probabilities for each hit.
For the TLEFit, the expected \KE of particle at a hit is calculated using the CSDA and the length given by sum of hit's residual range and an additional offset.
As a result, the total likelihood for a given offset, ${L}^{\prime}$, is written as
\begin{equation}
  -2\ln{\mathcal{L}}({L}^{\prime}) = -2\displaystyle\sum_{i = 1}^{{N}_{\rm{hits}}} \ln{P}\left( \frac{dE}{dx}({\rm{Measured}};~i)~\bigg|~\KE^{\rm{CSDA}}({\rm{range}}_{i} + {L}^{\prime}) \right).
  \label{eq:tot_likelihood}
\end{equation}
By looking for the best offset which gives the minimum $-2\ln{\mathcal{L}}$ value, energy of the interacted particle can be estimated through CSDA.

\section{The ProtoDUNE-SP detector}
\label{sec:ProtoDUNE}
\begin{figure*}[htbp]
\begin{center}
  \includegraphics[width=0.6\textwidth]{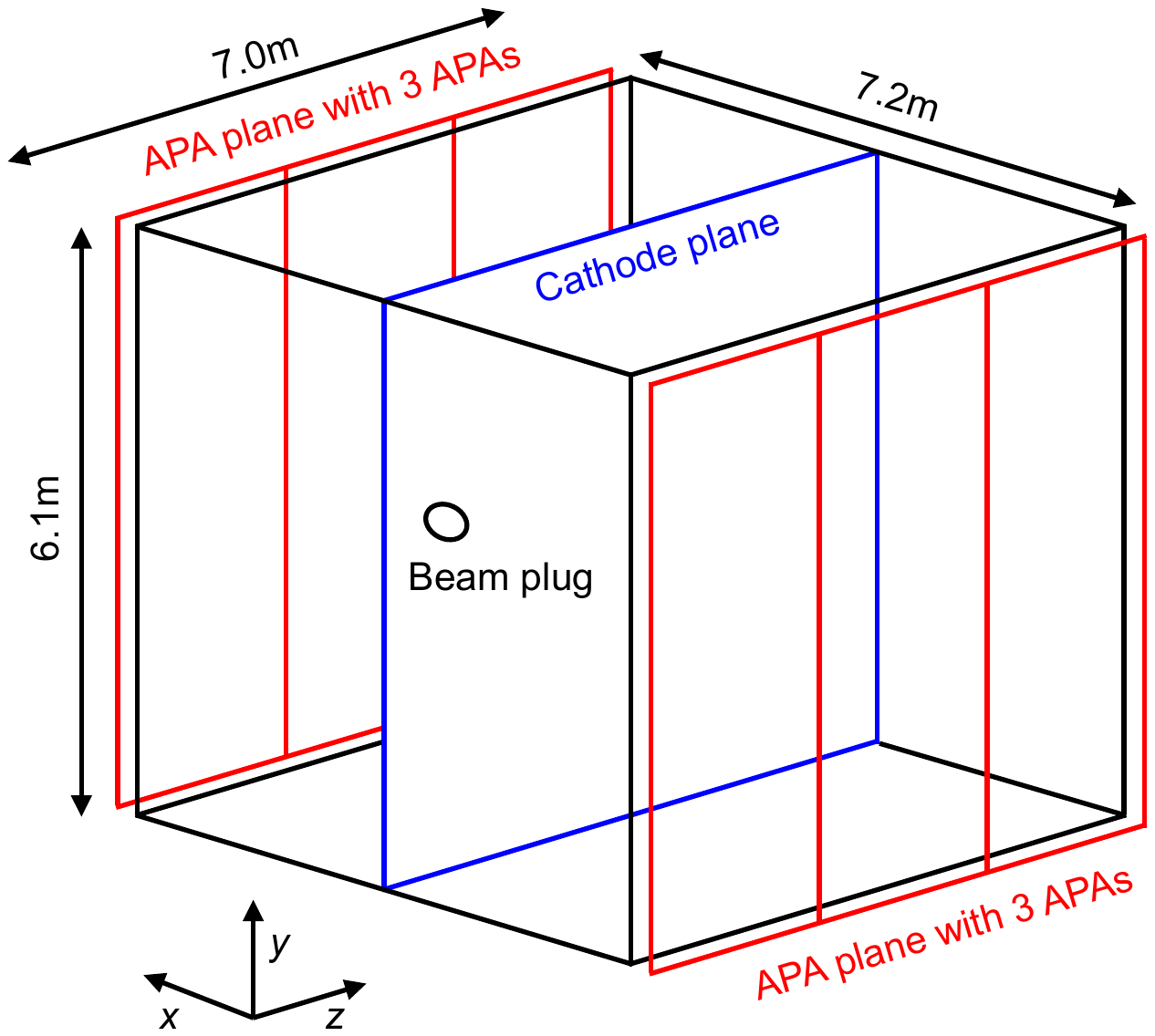}
  \includegraphics[width=0.3\textwidth]{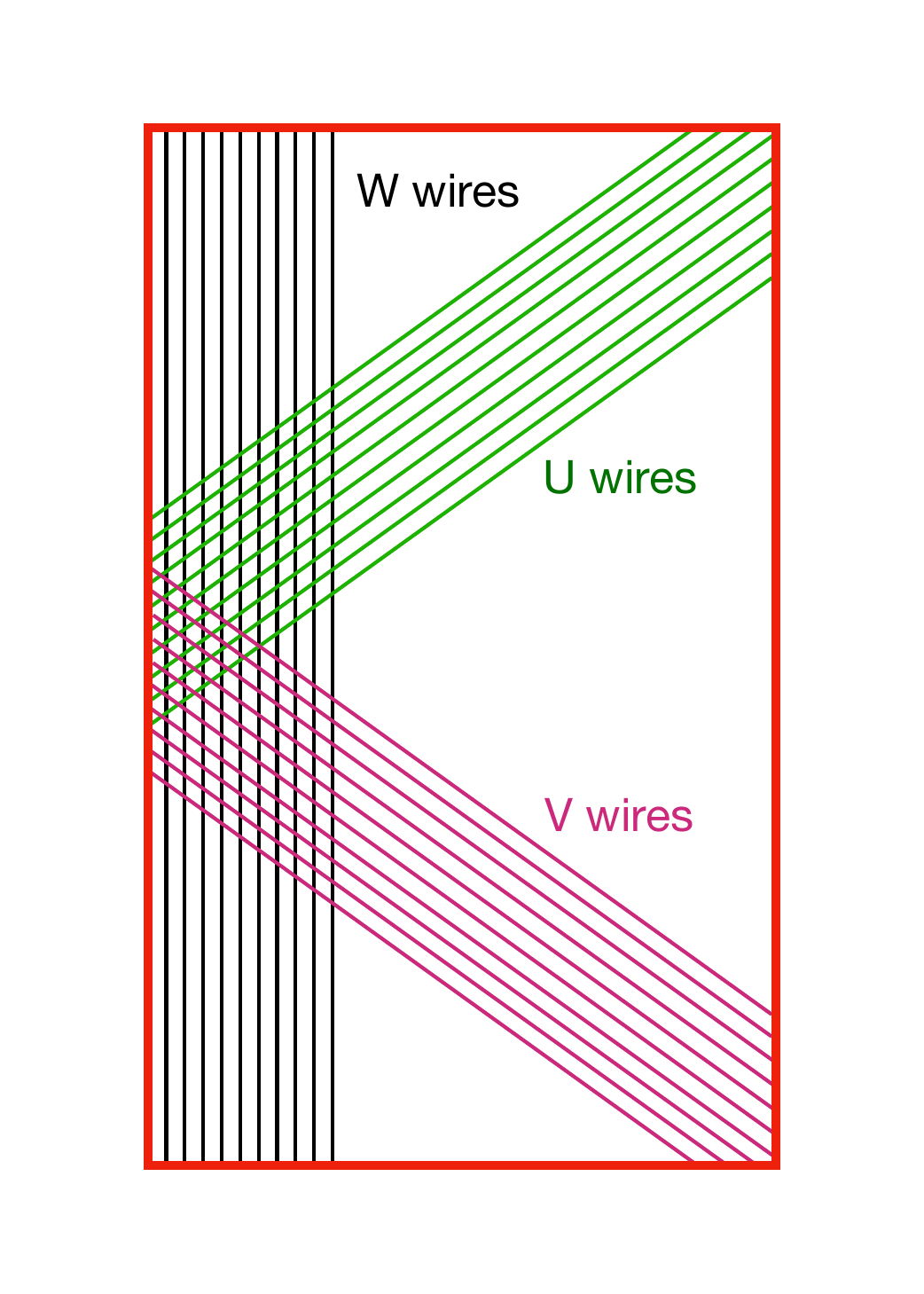}
  \caption{
    A simple drawing of the ProtoDUNE-SP detector (left) and an illustration of the three wire planes on an APA (right) are shown. A black box in the left figure represents the active volume, divided into two parts by the cathode at the center. The six APAs are arranged into two anode planes, each consisting of three side-by-side APAs. The test beam enters through the beam plug, close to the right side of the cathode. The right-handed coordinate system is shown in addition to the dimensions of the active volume. For wire planes, only ten wires for each plane are shown for clarity.
  }
  \label{fig:Figure_004}
\end{center}
\end{figure*}

The ProtoDUNE single-phase (ProtoDUNE-SP) apparatus at CERN~\cite{DUNE:2017pqt,Abed_Abud_2022} is designed as a test bed and full-scale prototype for the elements of the first far detector module of the Deep Underground Neutrino Experiment (DUNE)~\cite{DUNE:2020txw}.
It is a single-phase LArTPC with an active volume of $7.2\times6.1\times7.0$~m$^3$. The coordinate system is right-handed.
The $x$ axis is horizontal and is parallel with the nominal electric field directions and is perpendicular to the wire planes.
The $y$ axis is vertical (positive pointing up), and the $z$ axis is horizontal and points approximately along the beam direction. A simplified schematic of the detector is shown in figure~\ref{fig:Figure_004}.

The time projection chamber (TPC) is divided into two parts by the cathode plane at the center $(x = 0~{\rm{cm}})$.
The anode planes are placed at two sides opposite to the cathode plane.
The nominal electric field strength is 500~V/cm. The maximum drift length in ProtoDUNE-SP is 3572~mm. This results in a maximum drift time of 2.25~ms.

The ProtoDUNE-SP TPC was built to test the segmented design of the DUNE horizontal drift far detector module, with full-size components.
As a result, ProtoDUNE-SP contains six anode plane assemblies (APAs) arranged into the two anode planes, each consisting of three side-by-side APAs.
The single cathode plane is composed of 18 cathode plane assembly (CPA) modules.
The photon detector system comprises 60 optical modules, ten of which are installed in each APA.
It is used to collect scintillation light produced by passing particles and test different photon collection technologies proposed for the DUNE far detector modules.

Each APA has four wire planes. For reconstruction of particle tracks and showers, three planes ($U$, $V$ and $X$) are used. 
The $X$ plane has wires in parallel to the $y$ axis with 4.79~mm pitch.
The $U$ and $V$ planes have wires with 4.67~mm pitch which are oriented ${\pm35.7}^{\circ}$ with respect to the $y$ axis. The fourth plane is an uninstrumented grid plane on the drift side of the $U$ plane. The spacing between the planes is 4.75~mm.

The construction and installation of the detector was finished in early July~2018.
After commissioning, it started its operation from September~2018.
For two months, October and November~2018, the ProtoDUNE-SP TPC was exposed to a tagged and momentum-analyzed particle beam with momentum settings ranging from 0.3~GeV/$c$ to 7~GeV/$c$~\cite{PhysRevAccelBeams.20.111001,PhysRevAccelBeams.22.061003}
and collected more than four million events with incident beam particle inside the TPC.
The standard deviation of the beam momentum distribution was measured to be approximately 7\% by fitting Gaussian distributions to beam spectrometer data. The test beam entered the detector at mid-height and about 30~cm away from the cathode plane, on the negative $x$ side with angles ${11}^{\circ}$ down from the horizontal and ${10}^{\circ}$ to the right of the $z$ direction. At this angle, the spacing between three-dimensional reconstructed hits is approximately 0.65~cm.
A beam plug was installed on the low-$z$, negative-$x$ side of the TPC to minimize beam energy loss due to upstream materials.
The cosmic ray tagger (CRT) with $6.8~{\rm{m}} \times 6.8~{\rm{m}}$ scintillation panels at both upstream and downstream of the ProtoDUNE-SP cryostat is used to tag cosmic and beam-halo muons.
The detector continued to operate through July~2020, collecting data to test and validate the technologies for the future DUNE far detector modules, demonstrate operational stability, and explore operational parameters.

Additional information for design, construction, and operation of ProtoDUNE-SP can be found in~\cite{Abed_Abud_2022}.
First results for the detector calibration and response are published in~\cite{Abi_2020}.

\section{Beam instrumentation}
The upstream beam spectrometer measures the momenta of incident particles using a magnetic field and three planes of scintillating fibers. To remove ambiguity coming from multiple hits in a single plane, there should be exactly one particle reconstructed in the beam spectrometer for an event. The data sample collected with a nominal beam momentum of 1~GeV/$c$ is used in this study. To veto events with activities inside the TPC coming from beam halo, it is further required that both upstream and downstream CRTs should not have hits.

The beam instrumentation provides good particle identification performance between $\rm{{\mu}^{+}/{\pi}^{+}}$ and protons for 1~GeV/$c$ mode using the time-of-flight ($\Delta \rm{{t}_{F}}$) information as described in~\cite{PhysRevAccelBeams.22.061003}.
In this paper, $\Delta \rm{{t}_{F}} < 110~\rm{ns}$ requirement is used to study secondary charged pions coming from interactions between beam charged pions and argon atoms.

\section{Simulation and reconstruction}
\label{sec:sample_reco}

In this paper, data events with incident beam particles inside ProtoDUNE-SP are used to study the performance of the TLEFit method. 

\subsection{Simulation}
To generate Monte-Carlo (MC) simulated samples, the incident beam is modeled using the Geant4~\cite{AGOSTINELLI2003250} based package, G4beamline~\cite{g4beamline}.
Detailed information for the beam line and its simulation is given in~\cite{PhysRevAccelBeams.20.111001,PhysRevAccelBeams.22.061003}.

Cosmic rays are simulated using CORSIKA v7.4~\cite{Heck:1998vt}. To completely cover the 3 ms detector readout window, cosmic rays are generated over a 6 ms time range with the center on the trigger time.

For the TPC simulation, Geant4 v4.10.3 with the QGSP\_BERT physics list~\cite{AGOSTINELLI2003250} is used to describe particle propagation and interactions. The detector response is modeled with LArSoft~\cite{Church:2013hea} using the WireCell toolkit~\cite{MicroBooNE:2021ojx}.

\subsection{Reconstruction of events in the TPC}
The data processing procedure which reconstructs physical objects using waveforms from the anode wire planes is performed as described in~\cite{Abi_2020}. It is summarized briefly below.

For each collection wire, the TPC readout electronics produces a digitized waveform using analog-to-digital converters (ADCs) that sample the current at 2~MHz.
After noise removal, each waveform is deconvolved to reproduce the ionization charge distribution as a function of the drift time. Then, the hit finding algorithm fits the peaks in the deconvolved wire waveforms with Gaussian shapes. A hit represents a charge deposition on a single wire at a given time.

To reconstruct interactions inside the TPC, pattern recognition is performed with the Pandora software package~\cite{DUNE:2022wlc}. The first step starts with two-dimensional clustering of hits in each detector readout plane. For 3D reconstruction, the 2D clusters are matched between different layers using timing and shape information. The angles between the wires in the readout planes provide the necessary information to solve for the third coordinate. If there is ambiguity, the original two-dimensional clustering is changed using information from all three views until consistent matches between two-dimensional clusters is made. Finally, 3D hits are constructed and particle interaction hierarchies are created.

The Pandora software package also executes various algorithms to reconstruct an overall picture in the ProtoDUNE-SP TPC. First, all clusters are analyzed by an cosmic-ray hypothesis algorithm to identify and to remove clear cosmic-ray candidates. After removing energy deposits coming from these clear cosmic-ray candidates, a 3D slicing algorithm divides the detector into spatial regions with two hypotheses, cosmic ray and test beam, where a slice contains all of the hits from a single parent particle interaction. Then, a boosted decision tree (BDT) algorithm selects slices that contain clusters that originate from the test beam.

\section{Energy measurement performance}
\label{sec:Performance}

The performance of the TLEFit method is tested using data and MC simulation samples of the ProtoDUNE-SP experiment. For this purpose, stopping charged pions are selected from the secondary particles resulting from the interaction between the charged pion beam and argon atoms.
So, we introduce the selection for stopping charged pions first.
It is validated that the CSDA can describe well the true \KE for selected stopping charged pion candidates.
As a result, \KE from the CSDA is used as reference to compare performance of the TLEFit method between data and MC simulation samples.
Tunable parameters of the TLEFit algorithm are also introduced.
Then, \KE measurement performance of the TLEFit algorithm is presented.
The fitting is performed using subsets of stopping charged pions' hits to mimic inelastically interacting charged pions.
Therefore, algorithm performance is presented as functions of \KE and number of hits that are used for the fitting.
For the MC simulation sample, the energy measurement performance is also presented using true \KE as reference with the same selection that is used for comparison between data and MC simulation samples.
Finally, a systematic study related with impact of \dedx modeling is performed to understand energy measurement scale difference between data and MC simulation samples.

\subsection{Stopping charged pions as a validation sample}
\label{subsubsec:selection}

To characterize the energy measurement performance, a reference value of the true energy is required. For MC samples, the true energies are known for each particle at each step along their trajectories. But for data, there is no such perfect reference. Fortunately, the CSDA can provide a good reference for the \KE values of stopping charged pions. It is therefore important to filter out inelastically interacting charged pions from reconstructed charged pion candidates.

\begin{figure*}[htbp]
\begin{center}
  \includegraphics[width=0.45\textwidth]{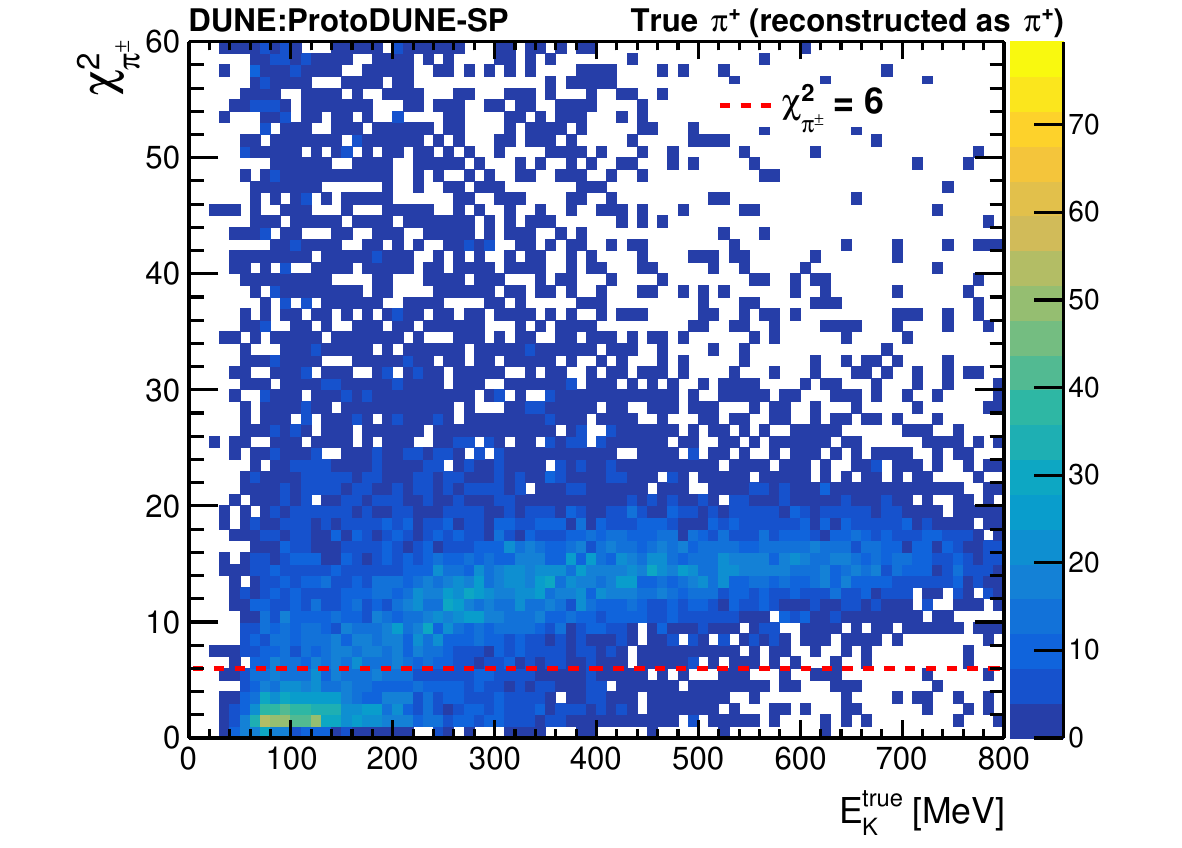}
  \includegraphics[width=0.45\textwidth]{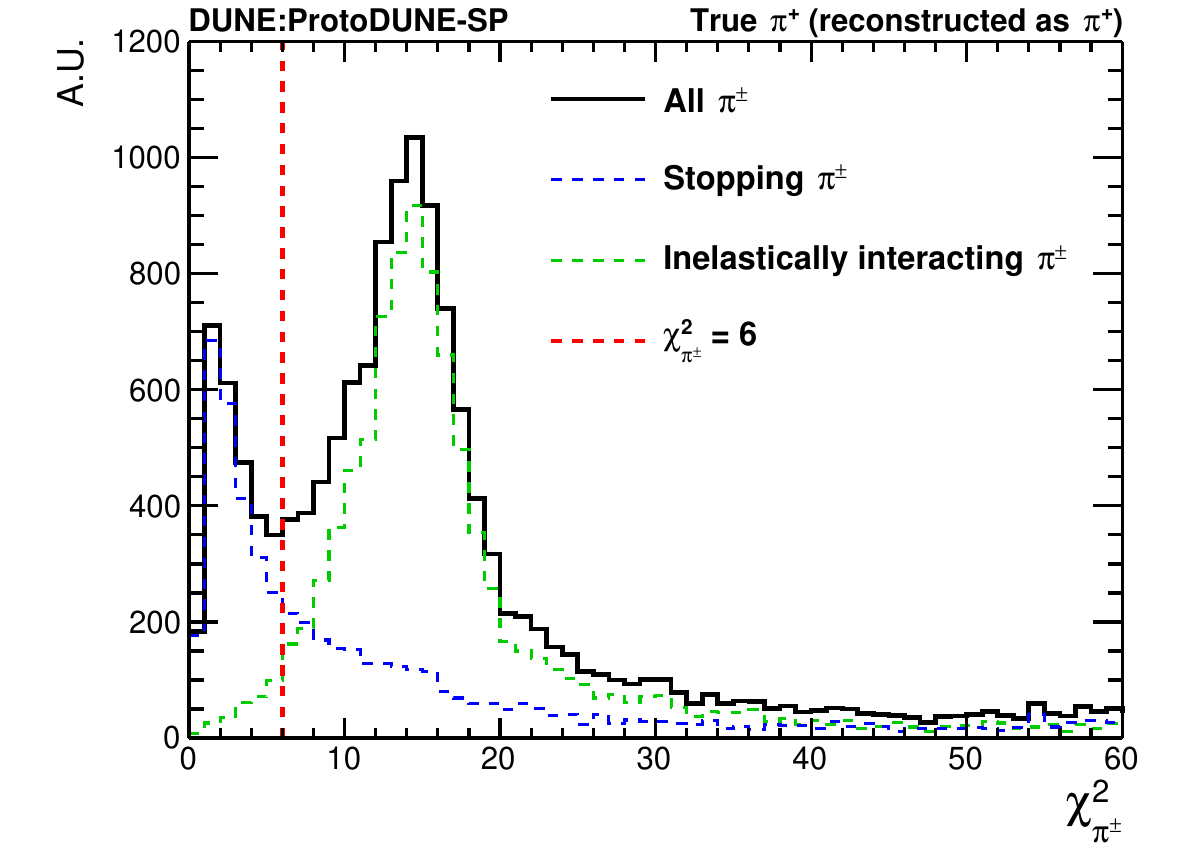}
  \caption{The ${\chi}^{2}_{{\pi}^{\pm}}$ distributions of charged pions in a MC sample are shown. Left plot shows a two dimensional distribution as a function of truth \KE. Right plot shows one dimensional distribution of ${\chi}^{2}_{{\pi}^{\pm}}$. Reconstructed charged pions which are matched with truth-level charged pions are used in this plot.
  }
  \label{fig:Figure_005}
\end{center}
\end{figure*}

\begin{figure*}[htbp]
\begin{center}
  \includegraphics[width=0.45\textwidth]{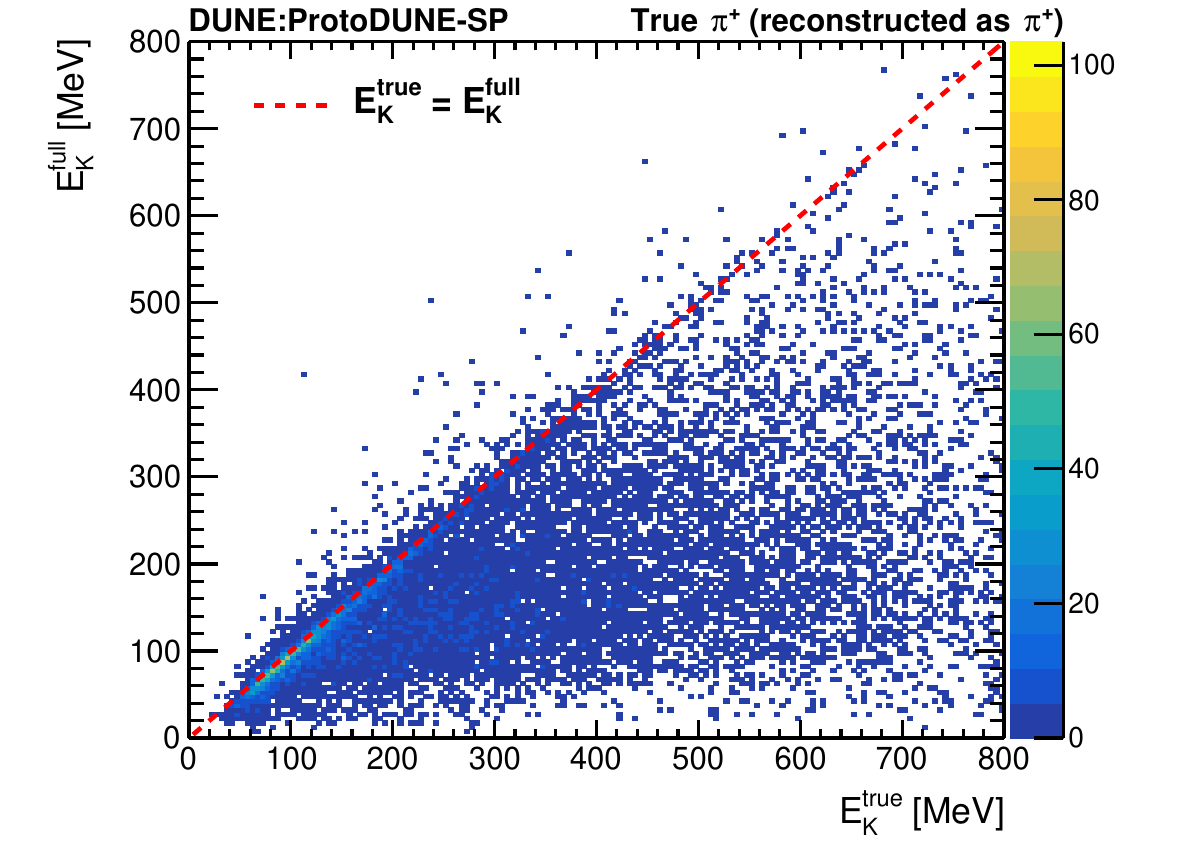}
  \includegraphics[width=0.45\textwidth]{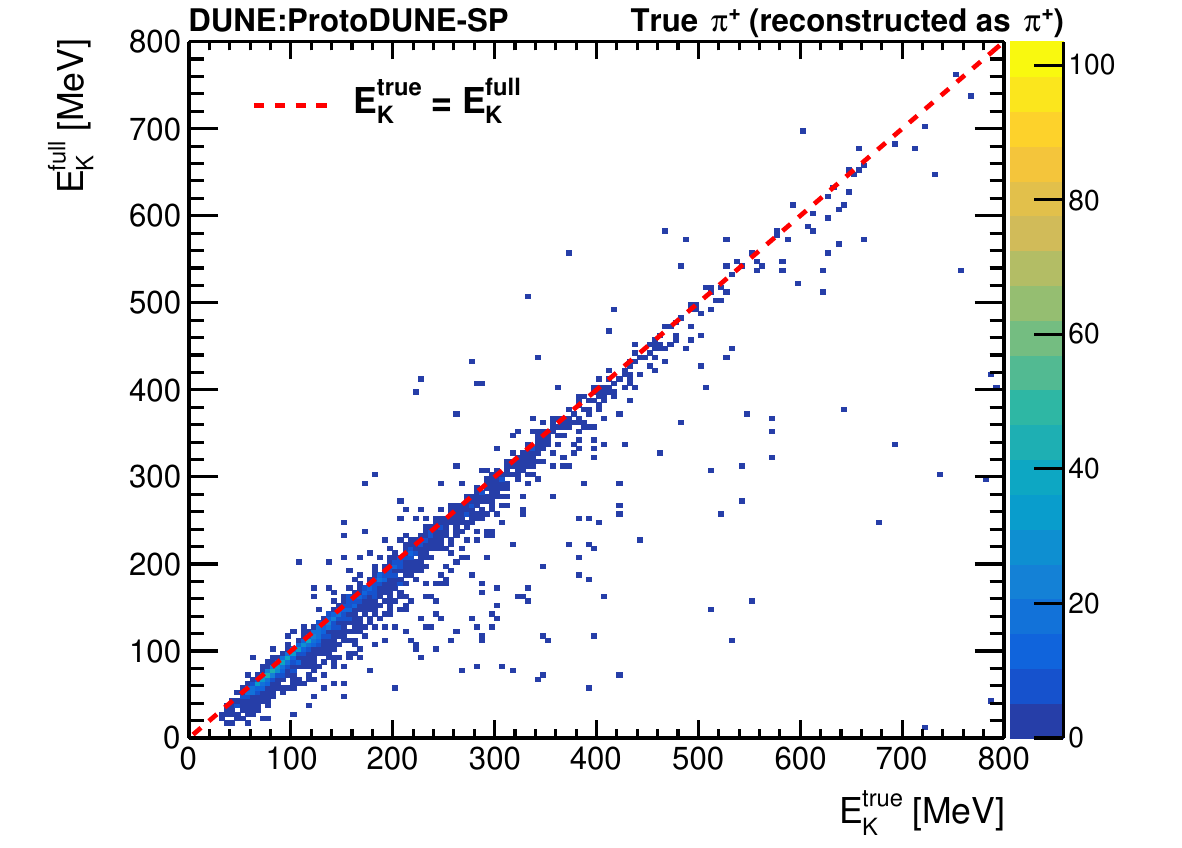}
  \caption{Plots show the relationship between truth-level and range-based kinetic energies for charged pions before (left) and after (right) applying the ${\chi}^{2}_{{\pi}^{\pm}}$\!< 6 cut, respectively. The same charged pion selection used in figure~\ref{fig:Figure_005} is used.
  }
  \label{fig:Figure_006}
\end{center}
\end{figure*}

In this study, the discrimination is provided by ${\chi}^{2}_{{\pi}^{\pm}}$, which is defined to be
\begin{equation}
  {\chi}^{2}_{{\pi}^{\pm}} = \displaystyle \frac{1}{N_{\rm{Hits}}^{\rm{range} < 26 \rm{cm}}}\sum_{i = 0}^{{\rm{range}}_{i} < 26 {\rm{cm}}} {\left( \frac{dE}{dx}({\rm{Measured}}; i) -  \frac{dE}{dx}({\textrm{Bethe--Bloch}}, {{\pi}^{\pm}}; {\rm{range}}_{i})   \right)}^{2} / ~{\sigma}^{2}.
  \label{eq:chi2_pion}
\end{equation}
Here, $\frac{dE}{dx}({\textrm{Bethe--Bloch}}; {\rm{range}}_{i})$ is the expected \dedx for a given residual range coming from the CSDA using the Bethe-Bloch formula in Eq.~\ref{eq:Bethe-Blcoh}, and $\sigma$ contains uncertainties of $\frac{dE}{dx}$ for both from the CSDA and the TPC's energy measurement resolution. Since Eq.~\ref{eq:chi2_pion} uses hits with residual range less than 26~\rm{cm} to make the ${\chi}^{2}_{{\pi}^{\pm}}$ not to be dominated by MIP hits, stopping charge pions with visible Bragg peaks have smaller $\chi^2$ values compared to inelastically interacting charged pions. To produce a normalized ${\chi}^{2}$, number of hits included into the ${\chi}^{2}_{{\pi}^{\pm}}$, $N_{\rm{Hits}}^{\rm{range} < 26 \rm{cm}}$, is divided. Figure~\ref{fig:Figure_005} shows the ${\chi}^{2}_{{\pi}^{\pm}}$ distribution as a function of true \KE in a MC sample of charged pions. A peak with ${\chi}^{2}_{{\pi}^{\pm}}$ value about from 2 to 4 is coming from stopping charged pions. The other peak that has ${\chi}^{2}_{{\pi}^{\pm}}$ value around 12 to 16 is coming from inelastically interacting charged pions that left tracks with length almost 26 cm or longer where all hits are MIP-like. That is the reason why the ${\chi}^{2}_{{\pi}^{\pm}}$ distribution of the band is approximately constant as a function of true \KE in the two dimensional distribution. Very high ${\chi}^{2}_{{\pi}^{\pm}}$ values greater than 20 are coming from inelastically interacted charged pions that left short tracks. In this study, ${\chi}^{2}_{{\pi}^{\pm}}$ value for the secondary charged pions is required to be smaller than 6.0 to select the stopping charged pions.
Figure~\ref{fig:Figure_006} shows that the CSDA with the full track length (\KEfull) estimates the true \KE (\KEtrue) well after applying the ${\chi}^{2}_{{\pi}^{\pm}} < 6$ cut.

\subsection{Algorithm parameters}
\label{sec:technical_detail}

Before the algorithm introduced in Sec.~\ref{sec:methodology} can be applied to real data, several issues which affect the energy measurement performance must be addressed. We describe parameters of the algorithm that can be tuned in order to optimize the performance.

The first parameter is the minimum number of usable hits to include in the fit.
In the case where two \dedx PDFs overlap significantly between two different values of \KE, including only a small number of hits in the fit can lead to a huge uncertainty in the energy measurement.
For example, the \dedx PDFs of charged pions with 200 \MeV and 500 \MeV of \KE values have a large overlap as shown in figure~\ref{fig:Figure_003}. Therefore, fitting a small number of hits which have random \dedx values given by PDFs will perform poorly. On the other hand, requiring many hits excludes low-energy particles, since \KE is highly correlated with the total length of the particle's trajectory. In this paper, at least 15 hits are required. This number corresponds to approximately 40 \MeV of kinetic energy threshold for charged pions inside the ProtoDUNE-SP TPC. This requirement can be optimized depending on the targeted \KE range of the particles under study.

The second parameter is related to poor \dedx measurements in the first and last several hits of a track. These hits are affected by proximity to other ionizing particle activity, electric field distortions in the detector near the field cage, and the fact that the starting and the stopping points within the argon volume viewed by the last hit wire is unknown. The fact that charges are induced on nearby sense wires means that hits close to the track ends suffer from end effects, not just the last hit. In this paper, the first three hits and the last three hits are not included in the fitting. These are also tunable parameters of the algorithm. In addition, for a hit, \dedx value measured by the collection plane is used.

The third parameter comes from the nature of the energy loss of particles. Figure~\ref{fig:Figure_002} shows that \dedx follows the Vavilov distribution where particles have \KE of hundreds of $\MeV$. Therefore, hits with measured \dedx values in either the low tail or the high tail of the Vavilov distribution are assigned low probabilities. This can cause the fit to converge to a poor result. To improve the reliability of the fit, hits with extreme measured values of \dedx are truncated. In this paper, only hits with \dedx of $0.5 \MeV/\rm{cm} - 5 \MeV/\rm{cm}$ are used for charged pions.

The last parameters are the adjustable step size (${L}_{\rm{step}}$) and the maximum additional track length (${L}_{\rm{max}}$).
The value of ${L}_{\rm{max}}$ determines the maximum measurable \KE for a given track. The \KE given by CSDA using the original track length plus ${L}_{\rm{max}}$ becomes the maximum measurable \KE for that track. The ${L}_{\rm{step}}$ can affect the energy measurement resolution. Approximately $2.1 \MeV/\rm{cm} \times {L}_{\rm{step}}$ is the \KE step of the fitting in the \KE region up to several hundred MeV for charged pions. A longer ${L}_{\rm{max}}$ and a shorter ${L}_{\rm{step}}$ could improve the performance of the fits.
But it also leads to fits that require more CPU time because fits evaluate Eq.~\ref{eq:Gaussain_chi2} and Eq.~\ref{eq:tot_likelihood} multiple times, namely ${L}_{\rm{max}}$ divided by ${L}_{\rm{step}}$, to look for the best additional track length. In addition, having shorter ${L}_{\rm{step}}$ does not always provide better energy resolution. If the fitting itself has poorer energy resolution compared with $2.1 \MeV/\rm{cm} \times {L}_{\rm{step}}$, a smaller ${L}_{\rm{step}}$ does not improve the resolution and instead only takes more CPU time. In this paper, ${L}_{\rm{max}} = 450~\rm{cm}$ is used, since the 1~GeV/$c$ primary beam charged pion has about 870 \MeV of \KE where the CSDA gives about 1000 \MeV of \KE for charged pions with $450~\rm{cm}$ of track length.
For ${L}_{\rm{step}}$, $1~\rm{cm}$ is used targeting about 2 \MeV of \KE measurement resolution.

The tunable parameters described above are summarized in Table~\ref{table:tunable_parameters}.

\begin{table}[htbp]
\footnotesize
\caption{Summary of tunable parameters of the TLEFit algorithm.}
\centering
\begin{tabular}{ccc}
\hline\hline
Parameters & Used values & Impacts\\
\hline\hline
Minimum number of hits & 15 & \KE acceptance\\
& & \& resolution\\
\hline
Skip first and last hits & 3 & Fitting performance\\
\hline
Truncate outlying \dedx & Use $0.5 \MeV/\rm{cm} - 5 \MeV/\rm{cm}$ for ${\pi}^{\pm}$ & Fitting performance\\
\hline
Maximum additional length & $450~\rm{cm}$ for ${\pi}^{\pm}$ & Maximum measurable \KE\\
& & \& CPU time\\
\hline
Fitting step size & $1.0~\rm{cm}$ for ${\pi}^{\pm}$ & \KE resolution\\
& & \& CPU time\\
\hline\hline
\end{tabular}
\label{table:tunable_parameters}
\end{table}

\subsection{Performance in MC simulation sample using true \boldmath\KE as the reference}
\label{subsubsec:performance_MC_truth}
In this section, the energy measurement performance is presented in terms of resolution and fractional bias as functions of true \KE and the number of hits.

For example, if we have 50 hits for a charged pion, multiple fits are performed after vetoing several hits from the end of the track for fitting, using 49 hits, 48 hits, and so on. Then, resolution is measured for each fit. Resolution histograms are drawn as a function of \KEtrue and number of hits, and their peaks are fit with a Gaussian function to extract the resolution and the fractional bias.

\begin{figure*}[htbp]
\begin{center}
  \includegraphics[width=0.48\textwidth]{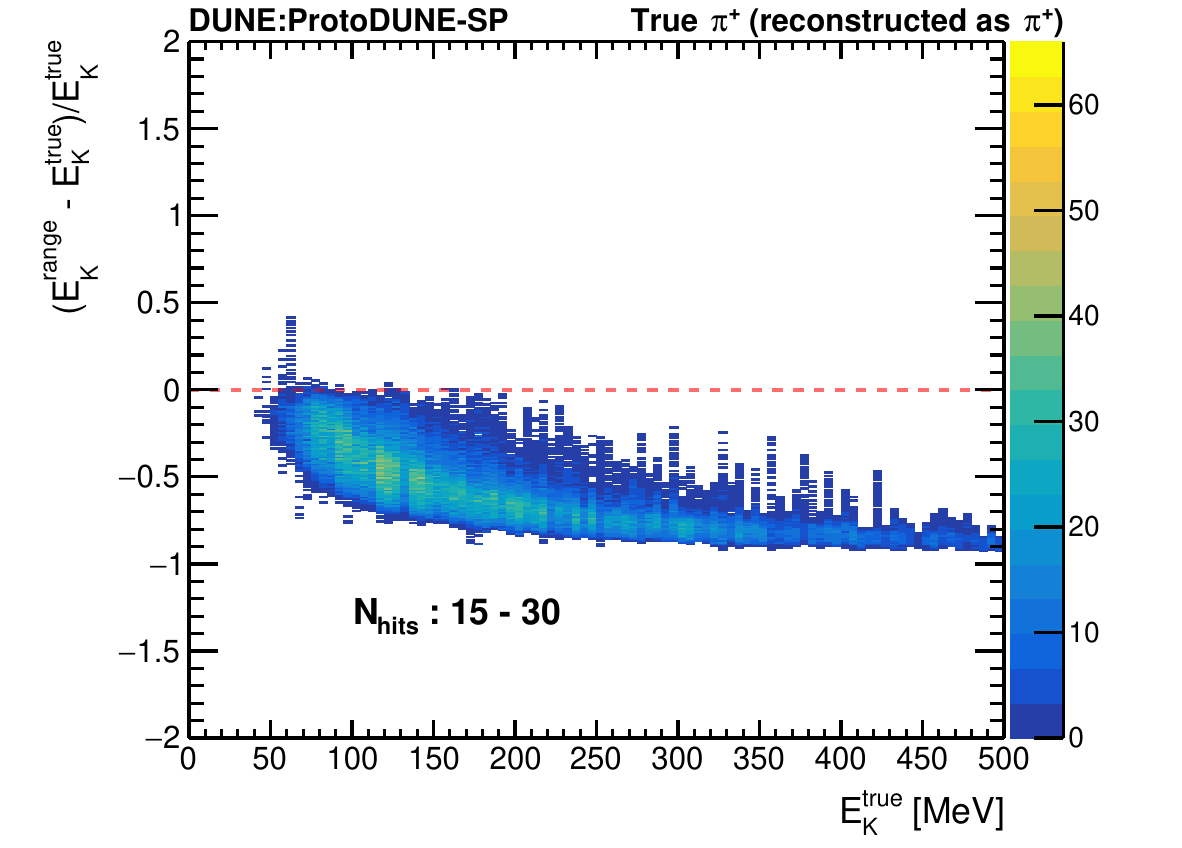}
  \includegraphics[width=0.48\textwidth]{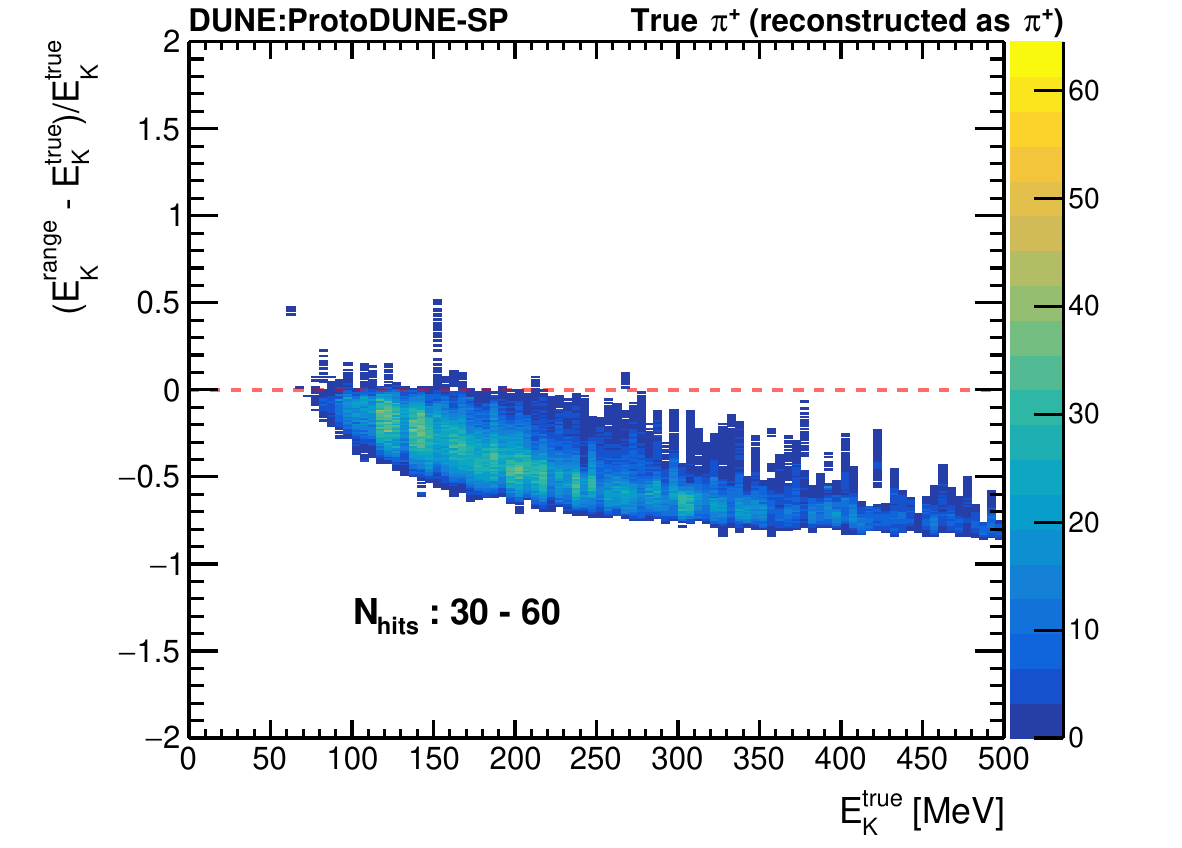}
  \caption{Two-dimensional distributions of the fractional energy residual from the CSDA with incomplete tracks, using 15 to 30 (left) and 30 to 60 (right) hits.
  }
  \label{fig:Figure_007}
\end{center}
\end{figure*}

\begin{figure*}[htbp]
\begin{center}
  \begin{subfigure}[b]{0.48\textwidth}
    \includegraphics[width=\textwidth]{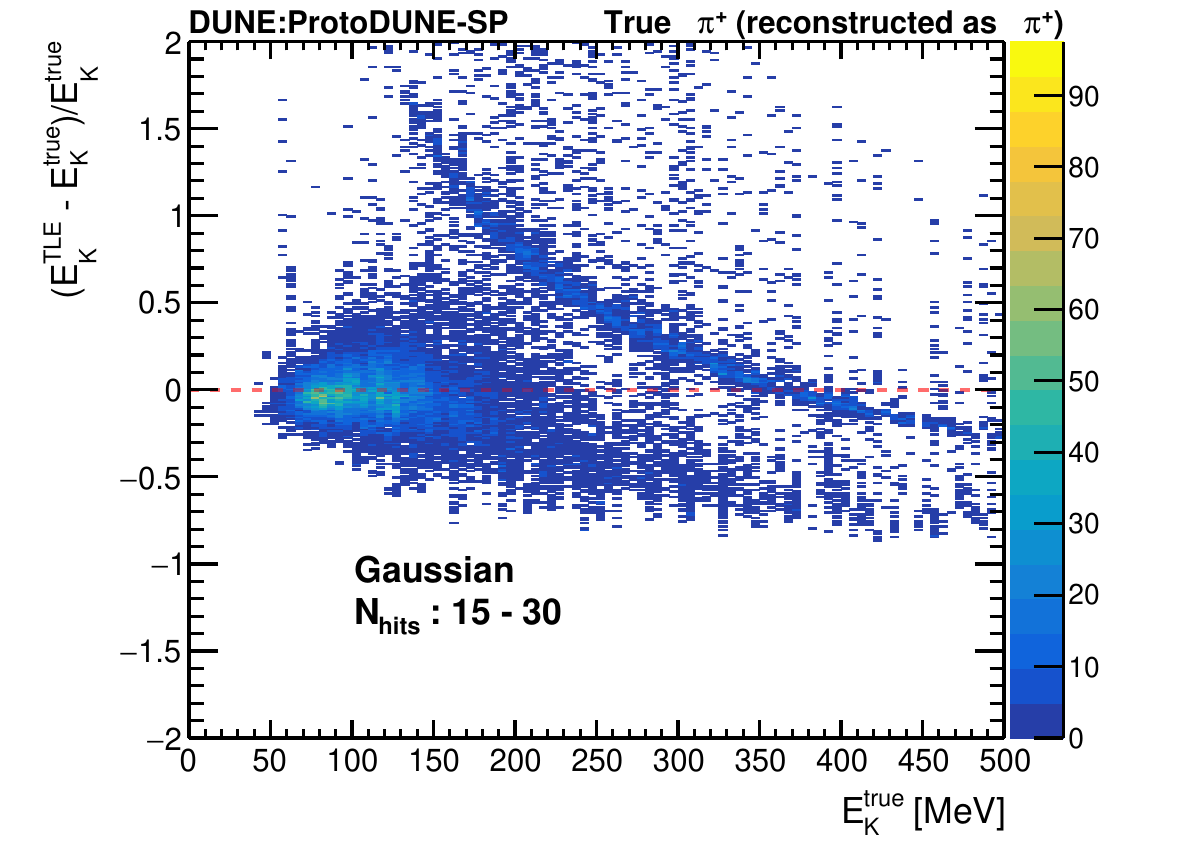}
    \caption{}
    \label{fig:Figure_008_a}
  \end{subfigure}
  \begin{subfigure}[b]{0.48\textwidth}
    \includegraphics[width=\textwidth]{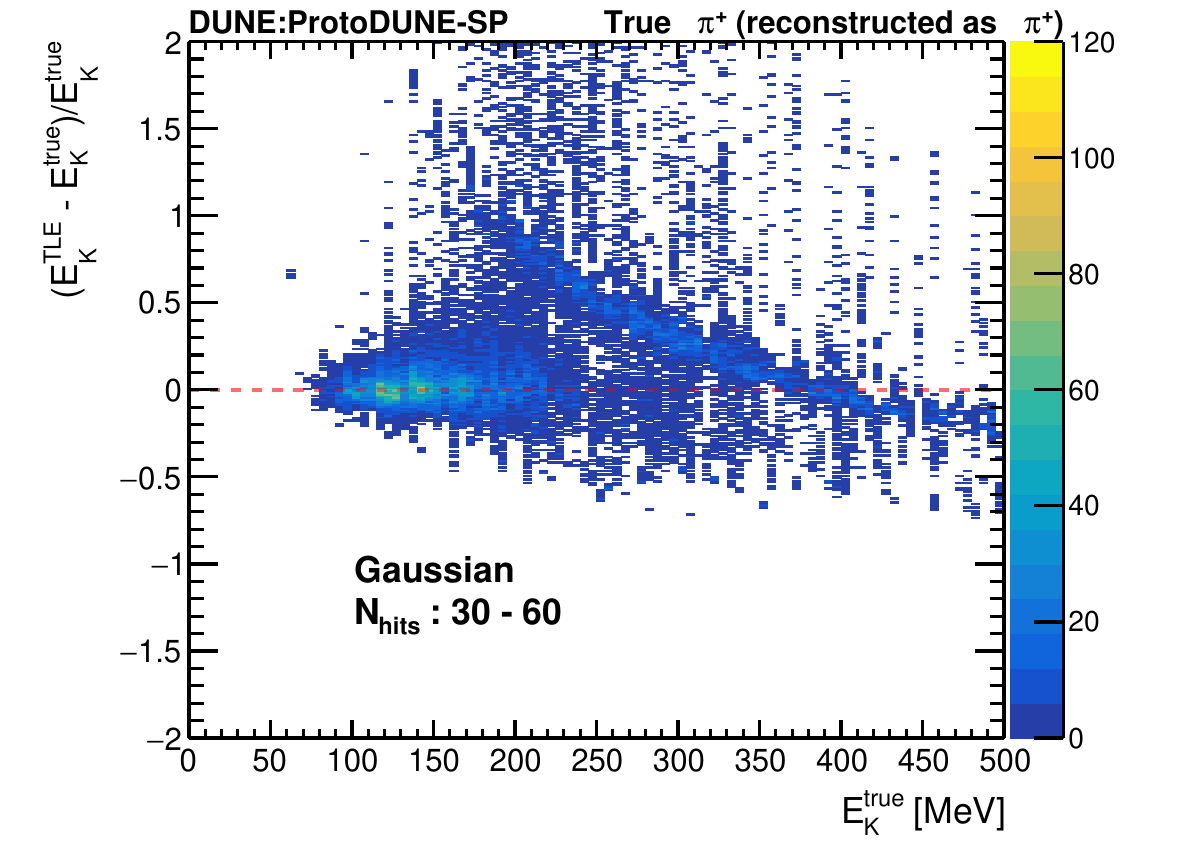}
    \caption{}
    \label{fig:Figure_008_b}
  \end{subfigure}
  \begin{subfigure}[b]{0.48\textwidth}
    \includegraphics[width=\textwidth]{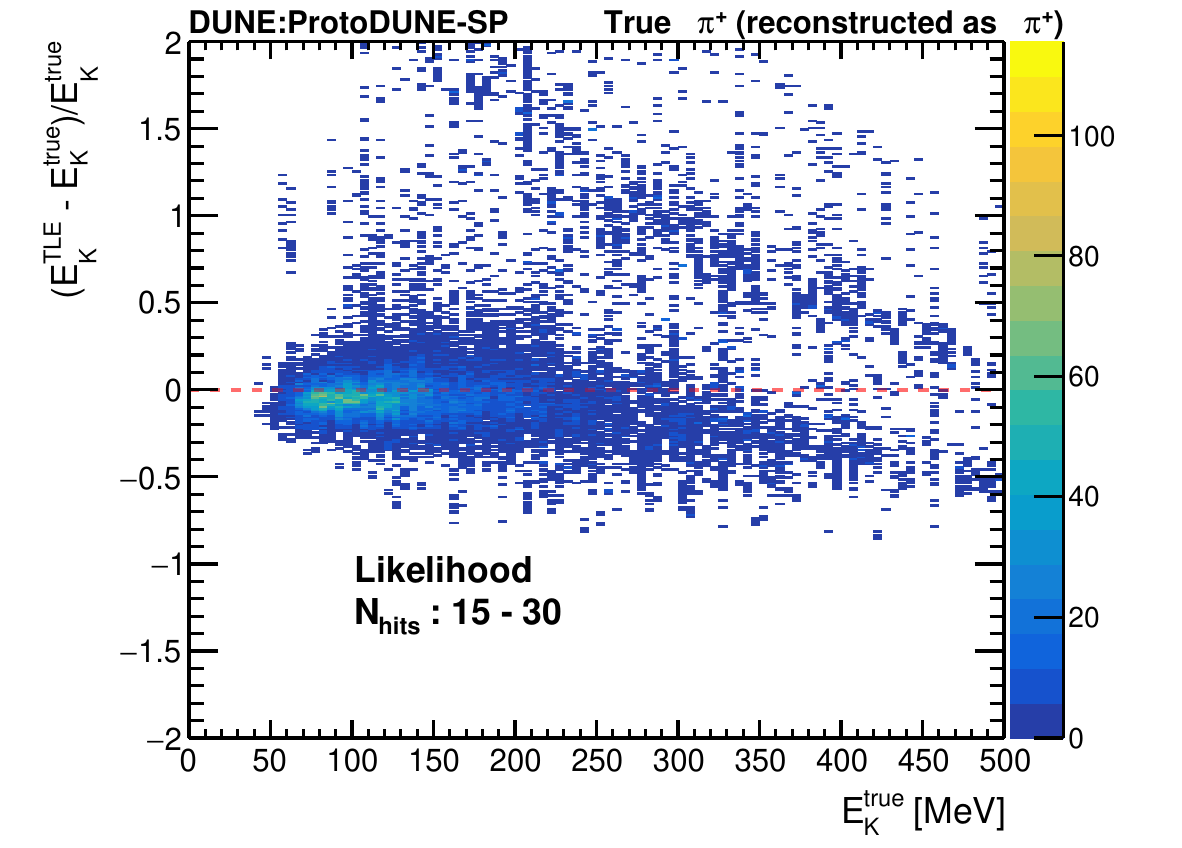}
    \caption{}
    \label{fig:Figure_008_c}
  \end{subfigure}
  \begin{subfigure}[b]{0.48\textwidth}
    \includegraphics[width=\textwidth]{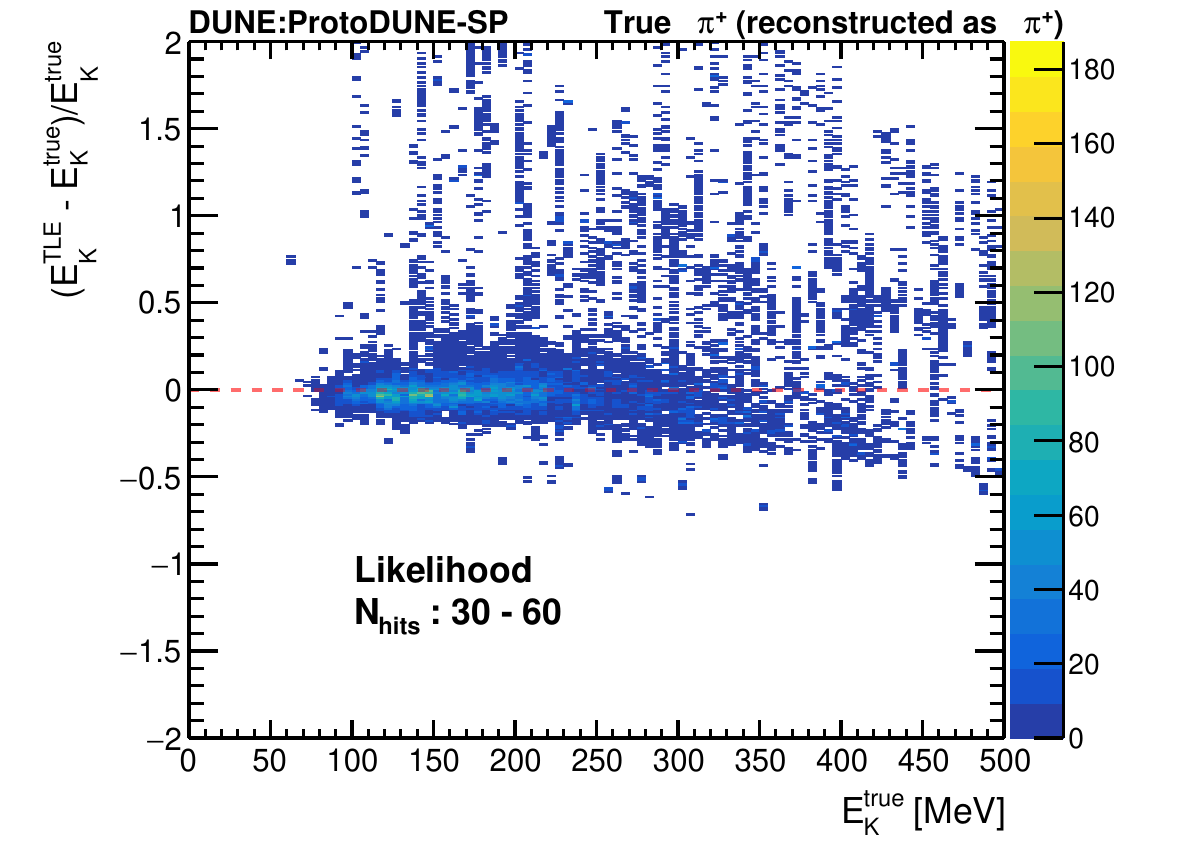}
    \caption{}
    \label{fig:Figure_008_d}
  \end{subfigure}
  \caption{Example two-dimensional distributions of the fractional energy residual from the Gaussian approximation method (top) and the maximum-likelihood method (bottom), using 15 to 30 (left) and 30 to 60 (right) hits.
  }
  \label{fig:Figure_008}
\end{center}
\end{figure*}

The CSDA using an incomplete track (\KErange) underestimates the charged pion's energy as shown in figure~\ref{fig:Figure_007}. The TLEFit algorithm can provide better energy measurement (\KEtle) by introducing the best additional track length. Figure~\ref{fig:Figure_008} shows two-dimensional distributions of the fractional energy residual from the Gaussian approximation and the maximum-likelihood methods. The Gaussian approximation shows clear bands corresponding to biased energy measurements. These measurements occur when the best additional track length is found to be in the minimum ionizing \KE. The maximum-likelihood method plots show populations with biased energy measurements, but these are not concentrated into a band, as seen in the Gaussian approximation plots.

Equation~\ref{eq:Gaussain_chi2} tends to have a minimum ${\chi}^{2}$ value where $\frac{dE}{dx}({\textrm{Bethe--Bloch}};~{\rm{range}}_{i} + {L}^{\prime})$ is near its minimum point regardless of the true charged pion \KE. The band corresponds to the constant fitted pion energy that gives the minimum \dedx in the Bethe-Bloch formula. This tendency becomes stronger with more hits in the MIP region. In the other words, this bias becomes more frequent with higher charged pion kinetic energies and with shorter reconstructed track lengths (smaller numbers of hits), which is well shown in figure~\ref{fig:Figure_008}.

\begin{figure*}[htbp]
\begin{center}
  \includegraphics[width=0.48\textwidth]{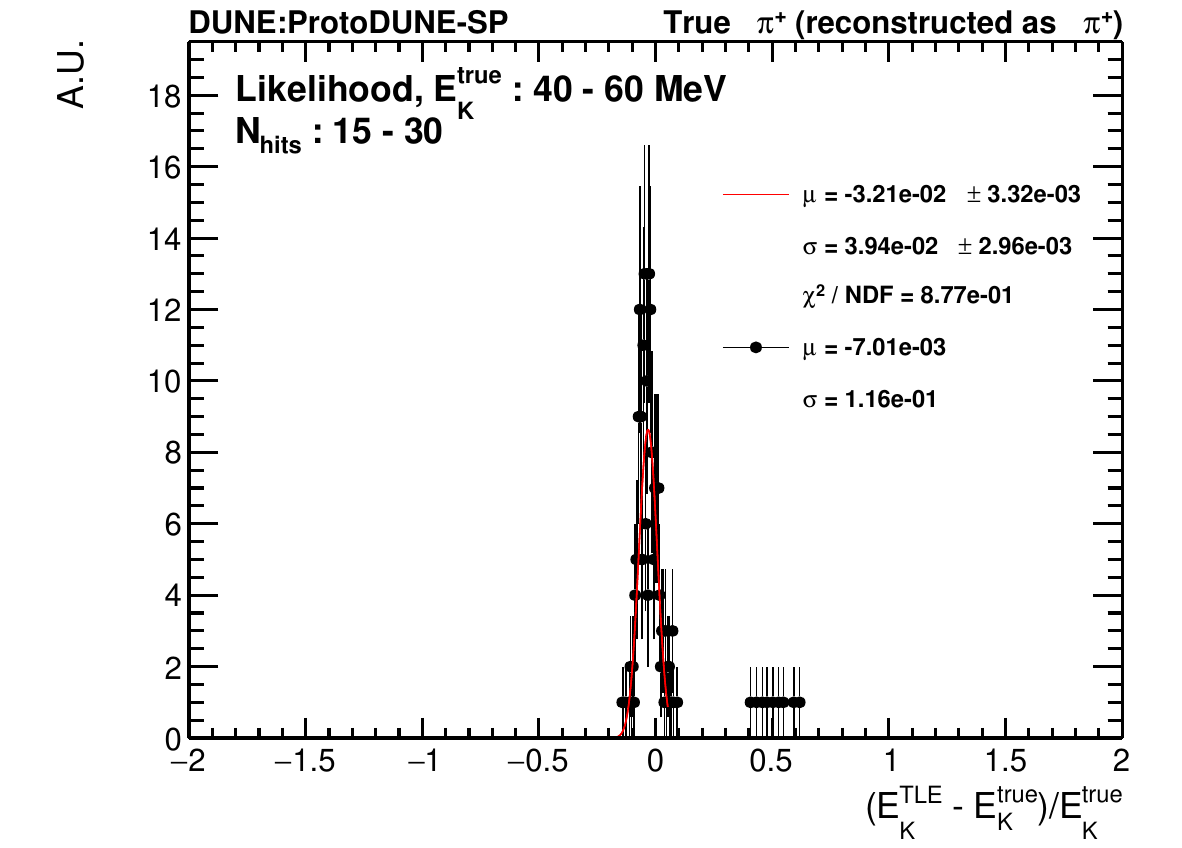}
  \includegraphics[width=0.48\textwidth]{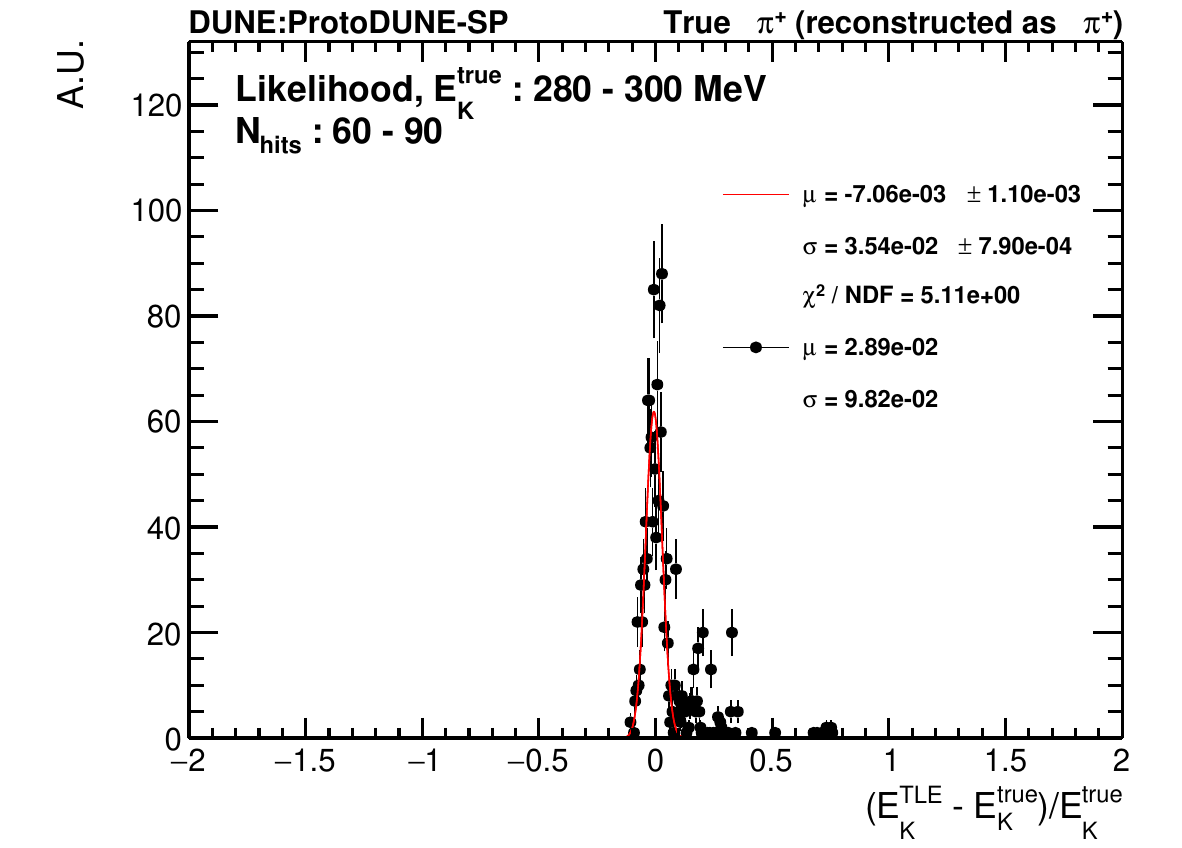}
  \caption{Example plots of energy measurement resolutions based on truth-level \KE. Distributions and Gaussian fit results with truth-level \KE from 40 to 60 \MeV and 280 to 300 \MeV with number of hits from 15 to 30 and 60 to 90 are shown in the top and bottom panels, respectively.
  }
  \label{fig:Figure_009}
\end{center}
\end{figure*}

This bias implies that the maximum-likelihood method is the most appropriate one for charged pions. Therefore, we fix the fitting method to maximum-likelihood. Figure~\ref{fig:Figure_009} shows examples of histograms which are used to measure resolutions and fractional biases. The energy measurement resolution and fractional bias for each distribution are calculated using data points within $\pm 2\sigma$ of the fitted Gaussian function.

\begin{figure*}[htbp]
\begin{center}
  \includegraphics[width=0.48\textwidth]{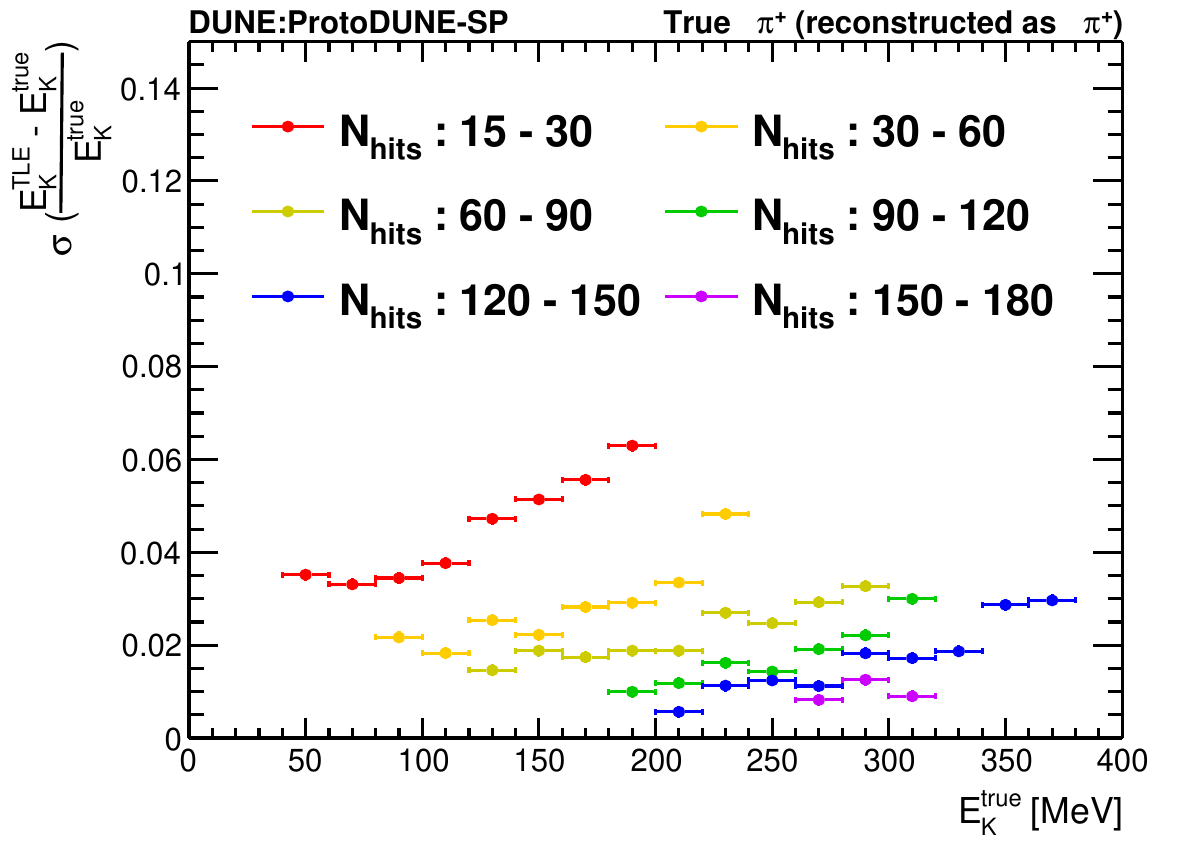}
  \includegraphics[width=0.48\textwidth]{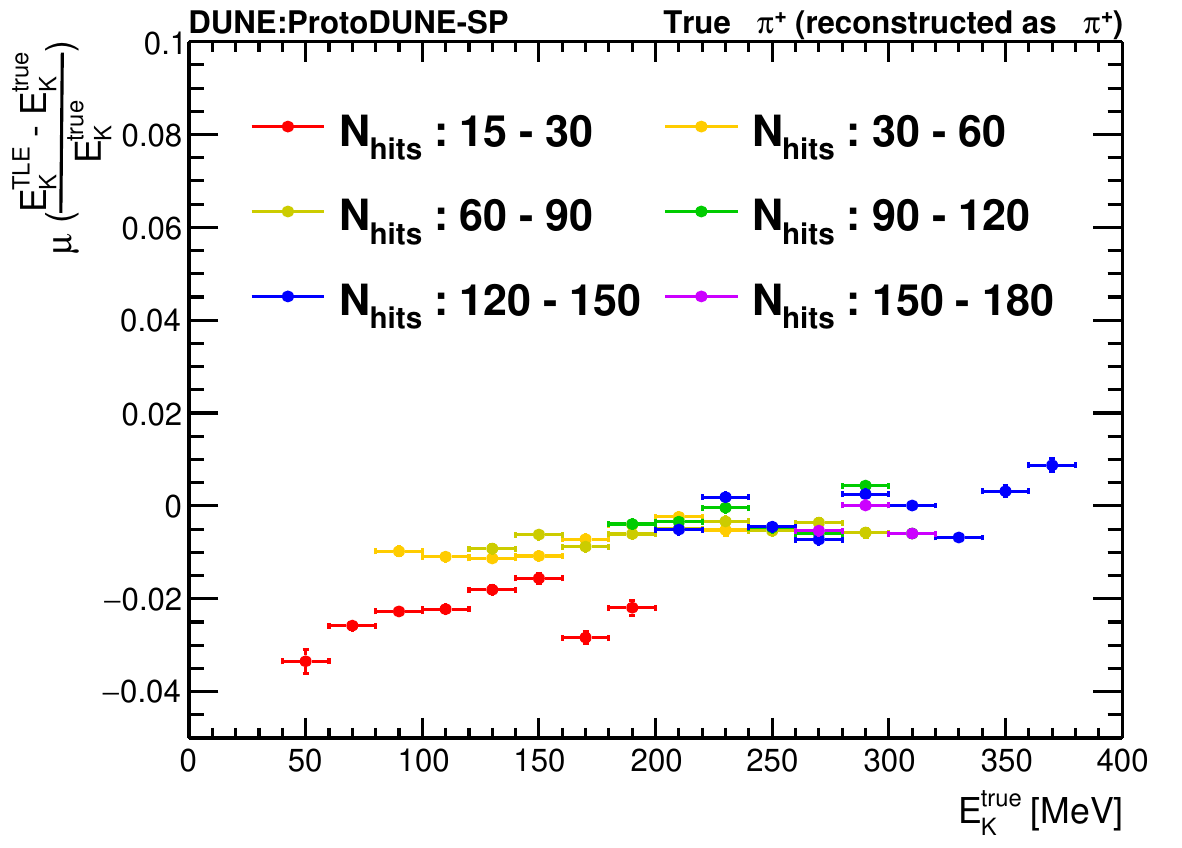}
  \caption{Summarized plots of energy measurement performance of the TLEFit method based on maximum-likelihood method using a MC sample. Resolutions (left) and fractional biases (right) are shown as functions of charged pions' true \KE and number of hits.
  }
  \label{fig:Figure_010}
\end{center}
\end{figure*}

Figure~\ref{fig:Figure_010} summarizes the extracted resolutions and fractional biases. The resolutions are better than 6.5\%, and the fractional biases are smaller than 4\%. Better resolutions are observed at lower \KE and with a greater number of hits. Lower \KE benefits from the steeper slope of \dedx as a function of \KE, while a greater number of hits provides stronger constraints. Sizes of fractional biases are smaller than 1.5\%, except for charged pions with 15 to 30 hits, which exhibit biases down to -3.5\% at lower \KE.

From a study using MC truth information, we find that tracks with fewer hits tend to have shorter reconstructed total track lengths compared to the true track lengths. For a given \KE, a smaller number of hits implies that the track’s angle relative to the anode plane is steeper, reducing the number of sense wires that can collect signals from the track. In such cases, the 3D spatial points of hits are reconstructed with larger uncertainties. Consequently, the residual ranges and their sum, the reconstructed total track length, are biased to be shorter than the true values. This effect is most pronounced in the smallest-hit region. However, the magnitude of the fractional bias, up to 3.5\%, remains relatively small compared to the biases observed when using CSDA for inelastically interacting charged pions, as shown in Figure 7.

\clearpage
\subsection{Performance in data and MC simulation samples using \boldmath\KEfull as the reference}
\label{subsubsec:performance_2}

\begin{figure*}[htbp]
\begin{center}
  \begin{subfigure}[b]{0.48\textwidth}
    \includegraphics[width=\textwidth]{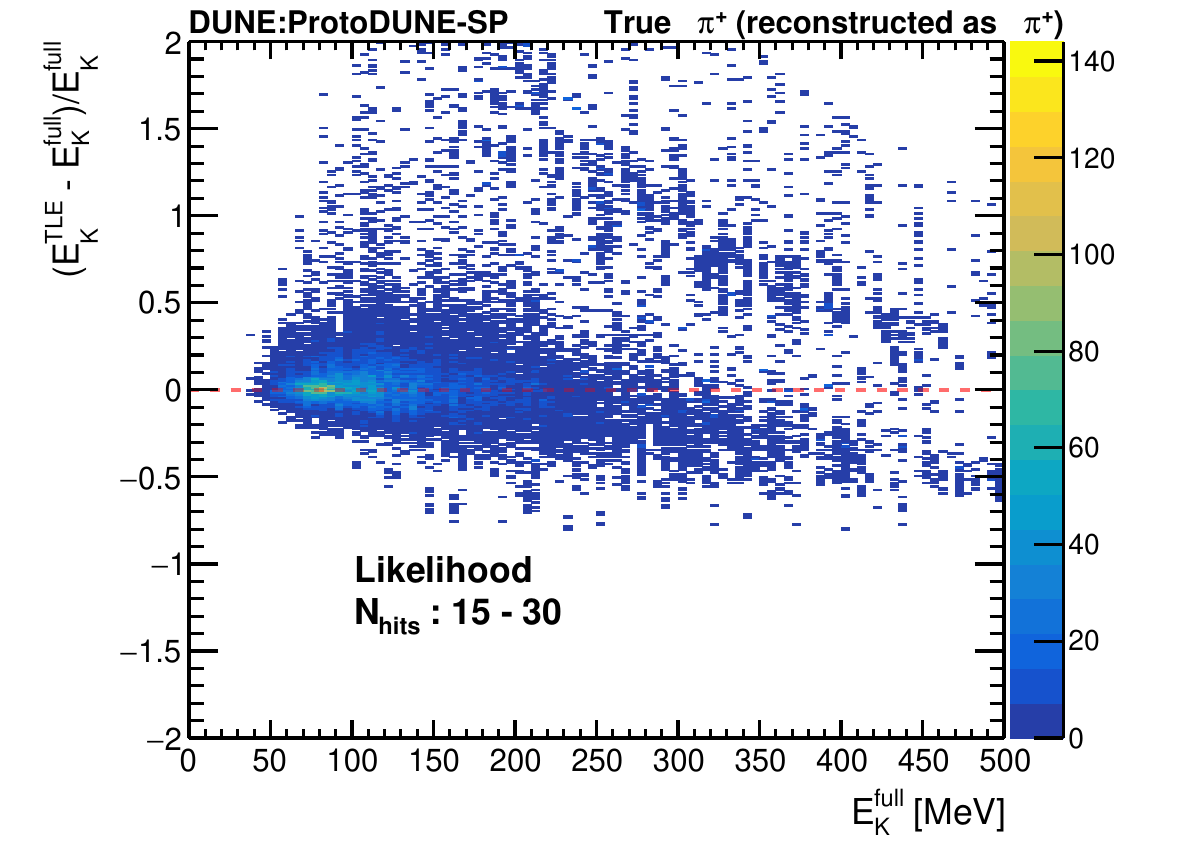}
    \caption{}
    \label{fig:Figure_011_a}
  \end{subfigure}
  \begin{subfigure}[b]{0.48\textwidth}
    \includegraphics[width=\textwidth]{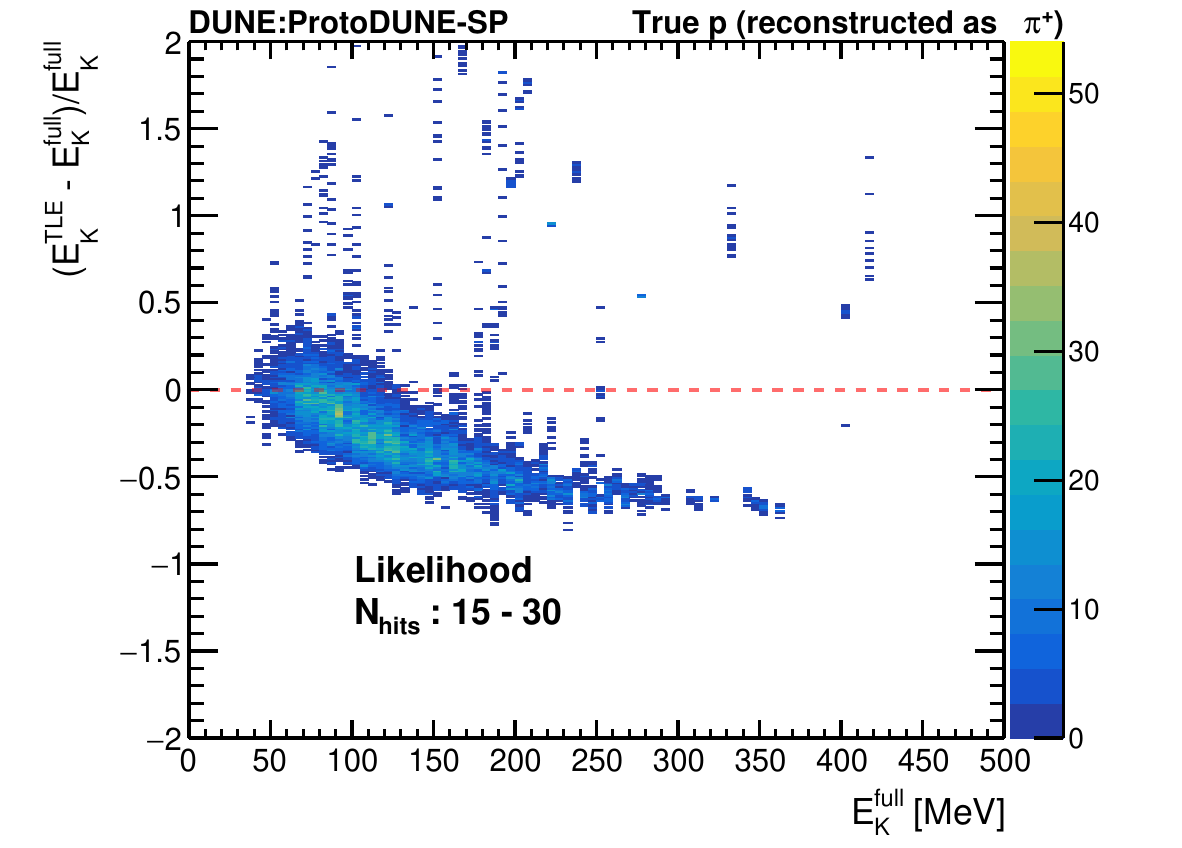}
    \caption{}
    \label{fig:Figure_011_b}
  \end{subfigure}
  \begin{subfigure}[b]{0.48\textwidth}
    \includegraphics[width=\textwidth]{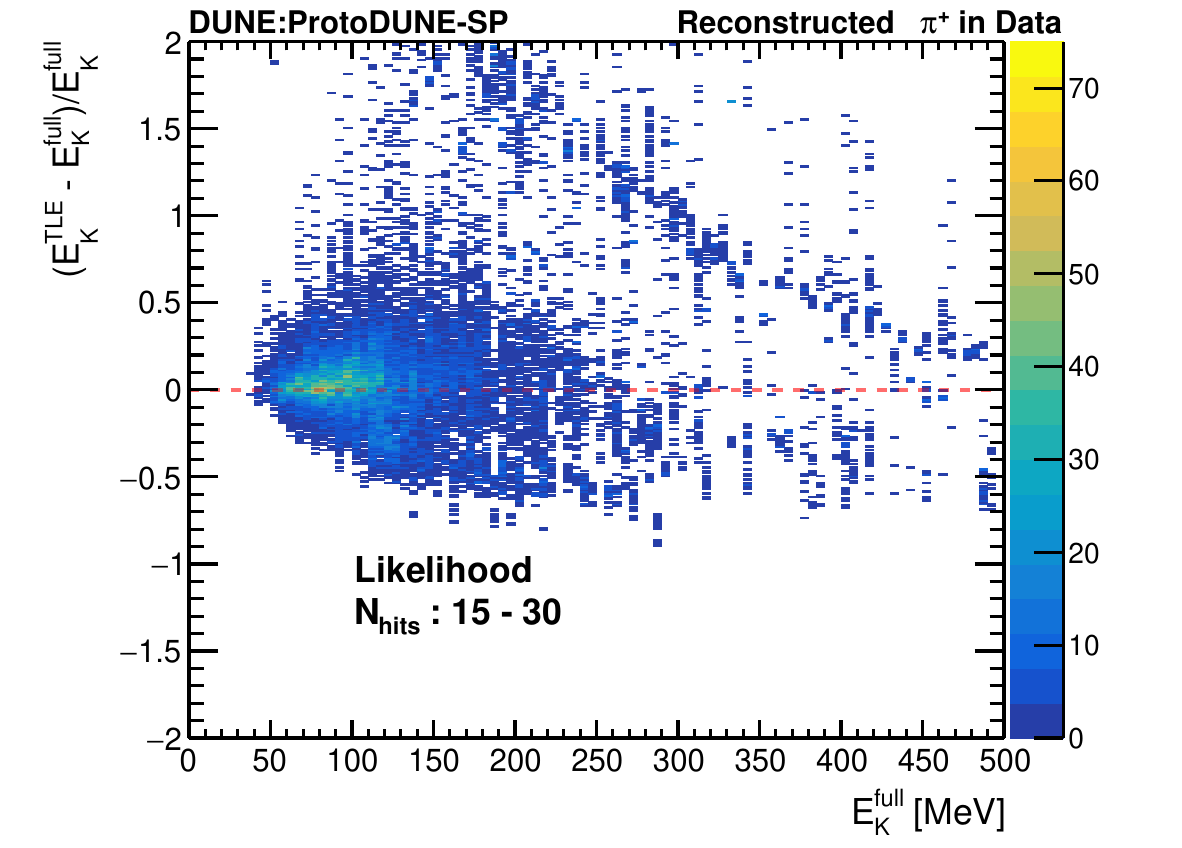}
    \caption{}
    \label{fig:Figure_011_c}
  \end{subfigure}
  \begin{subfigure}[b]{0.48\textwidth}
    \includegraphics[width=\textwidth]{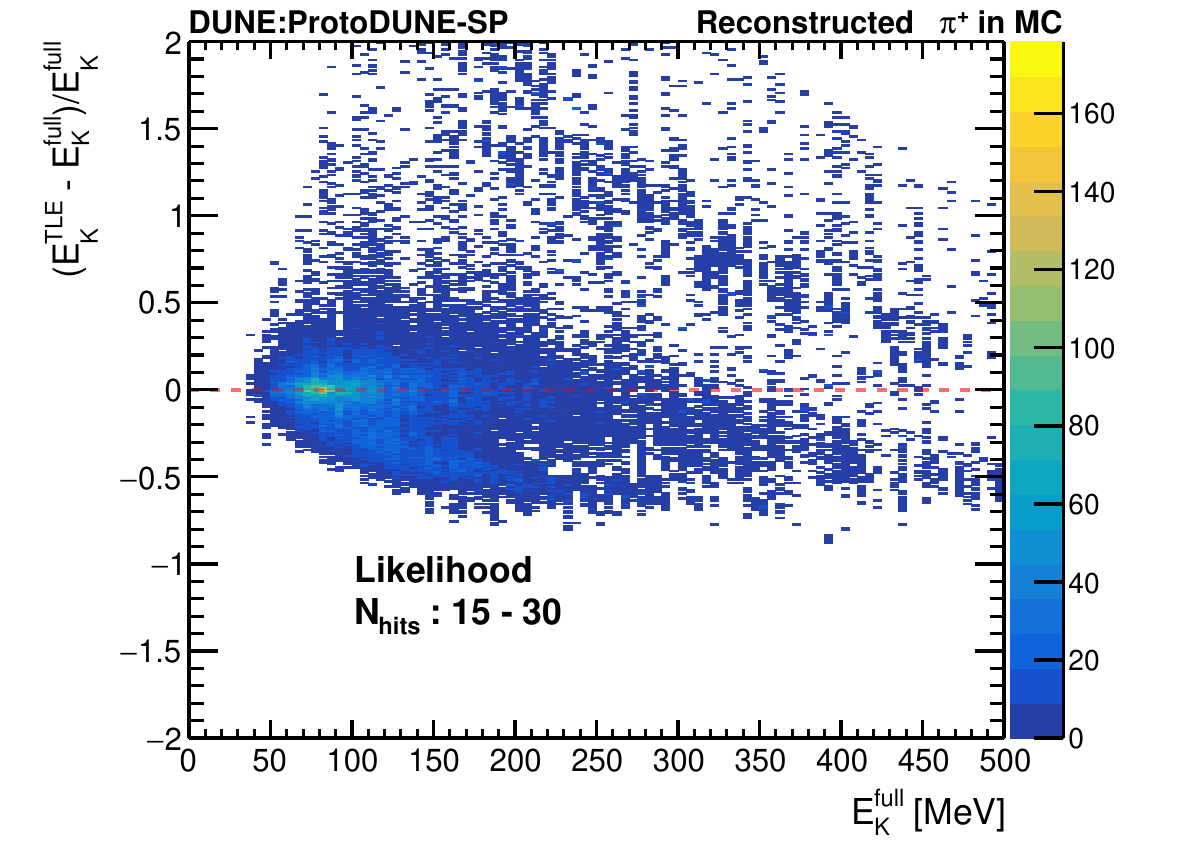}
    \caption{}
    \label{fig:Figure_011_d}
  \end{subfigure}
  \caption{Example two-dimensional distributions of the fractional energy residual using \KEfull as reference. Results are coming from reconstructed charged pions with 15 to 30 hits. Top plots show results for MC true pions (left) and true protons (right). Bottom plots show results using all reconstructed charged pions for data (left) and MC sample (right).
  }
  \label{fig:Figure_011}
\end{center}
\end{figure*}

The same study is performed using the \KE from the CSDA with \KEfull as the reference instead of the true energy to investigate performance for real data and to directly compare results from MC simulation and data. One notable point is that we cannot select pure secondary charged pion sample. There is non-negligible contribution coming from protons. Figure~\ref{fig:Figure_011} shows that both data and MC have proton contributions which overlap with charged pion distributions at low \KE regions.

\begin{figure*}[htbp]
\begin{center}
  \begin{subfigure}[b]{0.48\textwidth}
    \includegraphics[width=\textwidth]{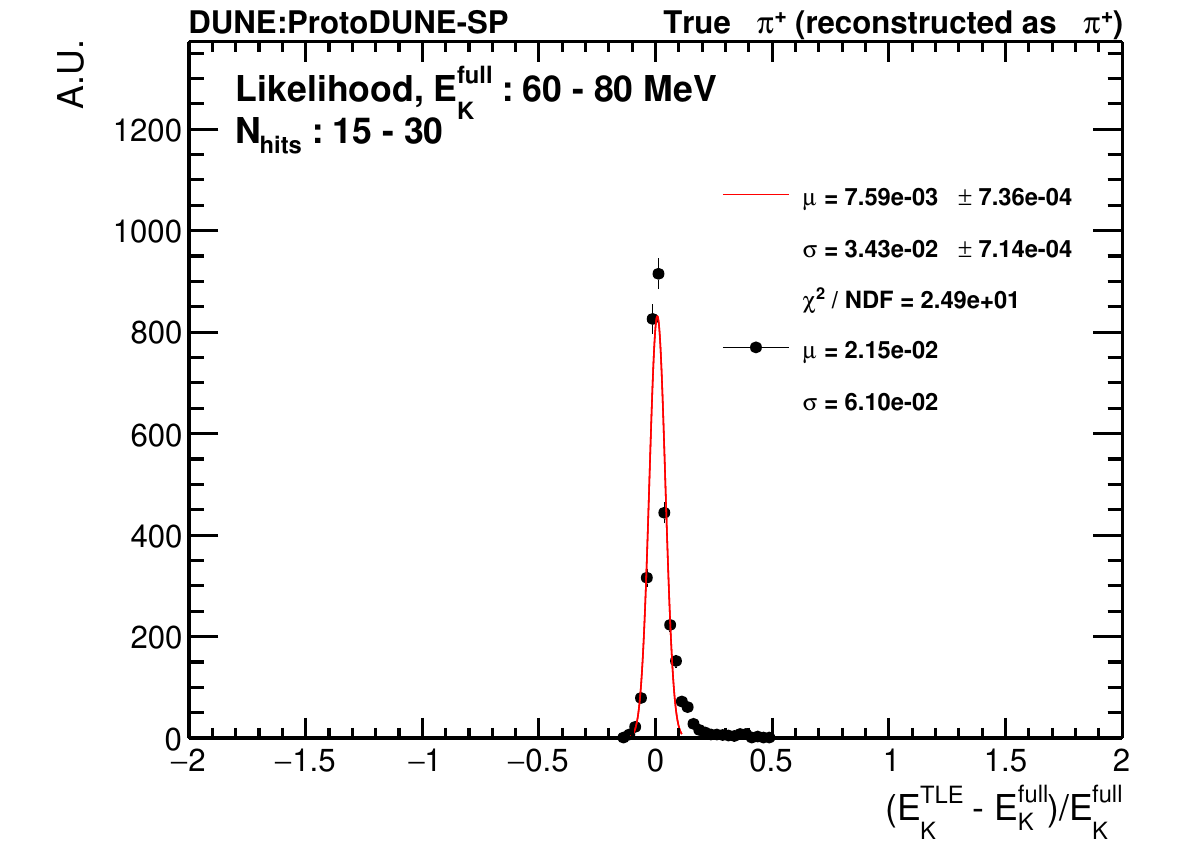}
    \caption{}
    \label{fig:Figure_012_a}
  \end{subfigure}
  \begin{subfigure}[b]{0.48\textwidth}
    \includegraphics[width=\textwidth]{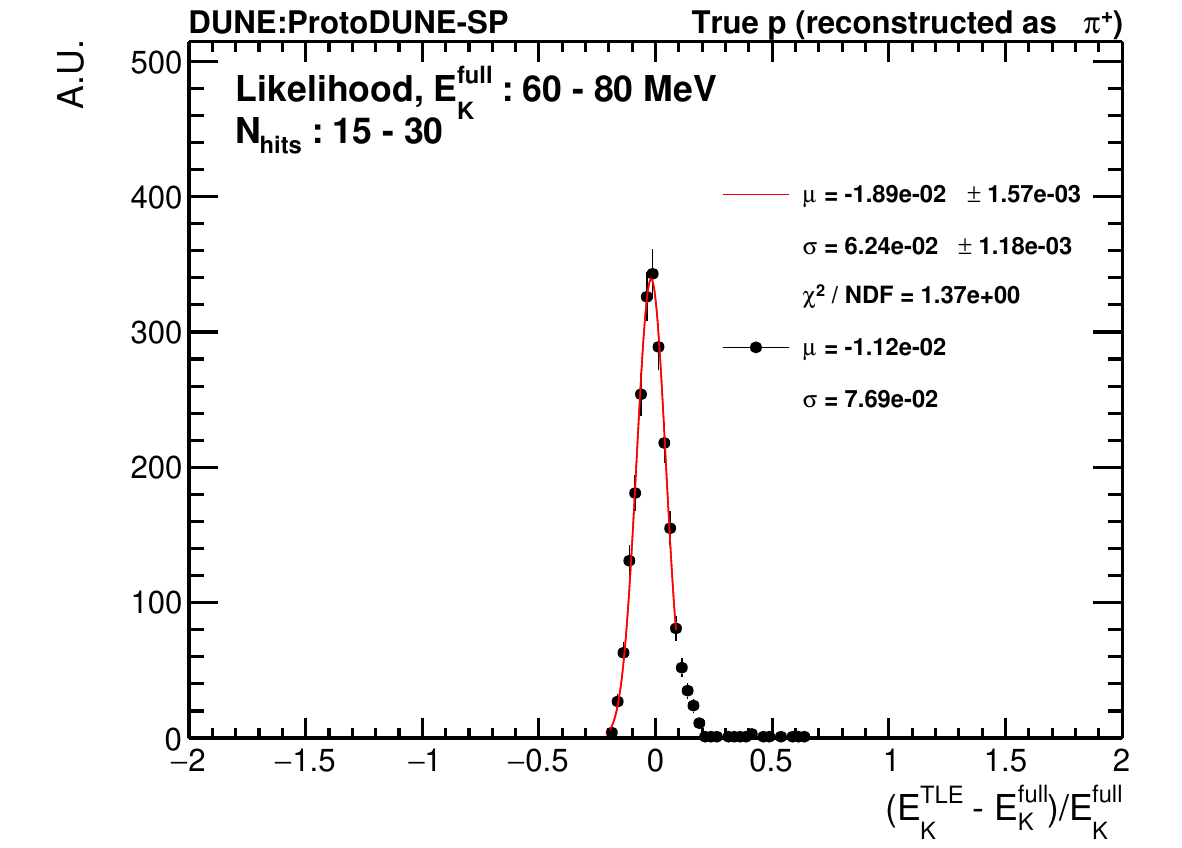}
    \caption{}
    \label{fig:Figure_012_b}
  \end{subfigure}
  \begin{subfigure}[b]{0.48\textwidth}
    \includegraphics[width=\textwidth]{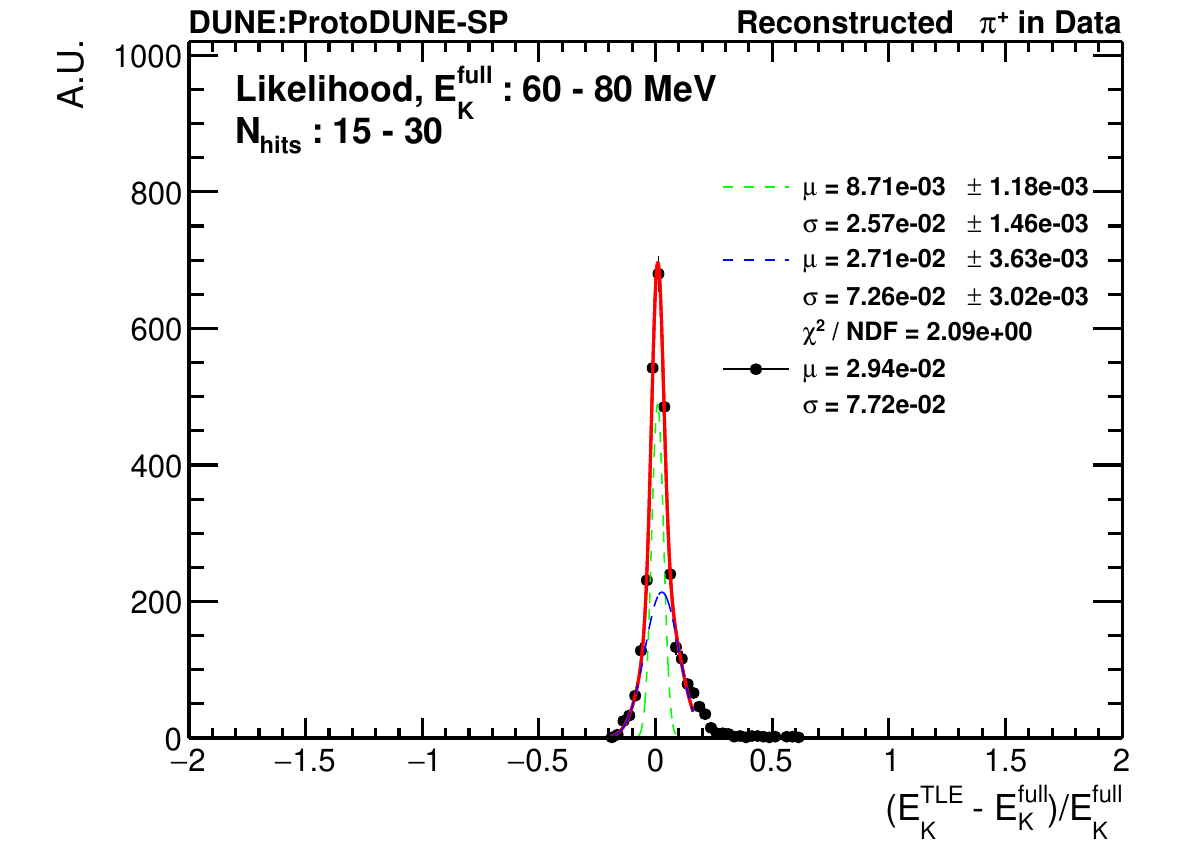}
    \caption{}
    \label{fig:Figure_012_c}
  \end{subfigure}
  \begin{subfigure}[b]{0.48\textwidth}
    \includegraphics[width=\textwidth]{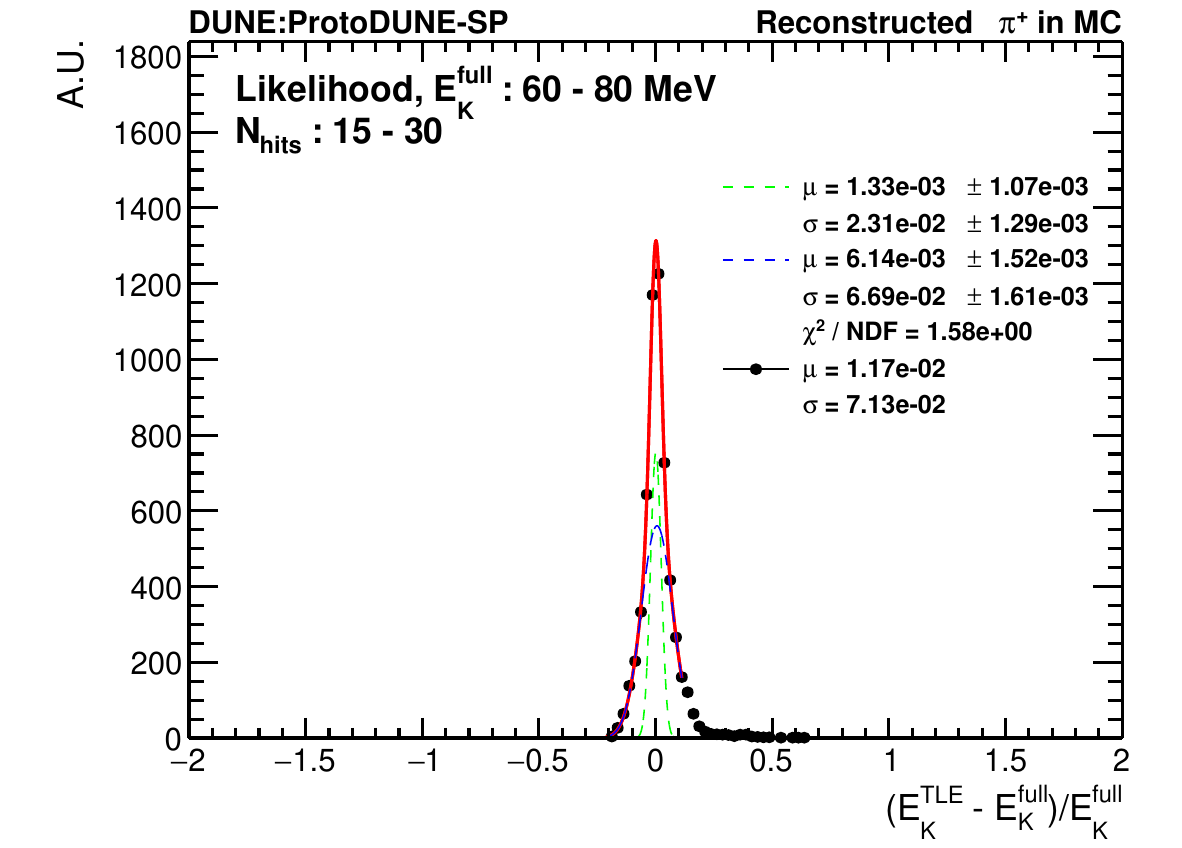}
    \caption{}
    \label{fig:Figure_012_d}
  \end{subfigure}
  \caption{Example distributions of the fractional energy residual using range-based \KE as reference. Results are coming from reconstructed charged pions with 15 to 30 hits and \KEfull from 60 MeV to 80 MeV. Top plots show results for MC true pions (a) and true protons (b). Bottom plots show results using all reconstructed charged pions for data (c) and MC sample (d), where green and blue Gaussian functions represent charged pion and proton contributions, respectively.
  }
  \label{fig:Figure_012}
\end{center}
\end{figure*}

\begin{figure*}[htbp]
\begin{center}
  \begin{subfigure}[b]{0.48\textwidth}
    \includegraphics[width=\textwidth]{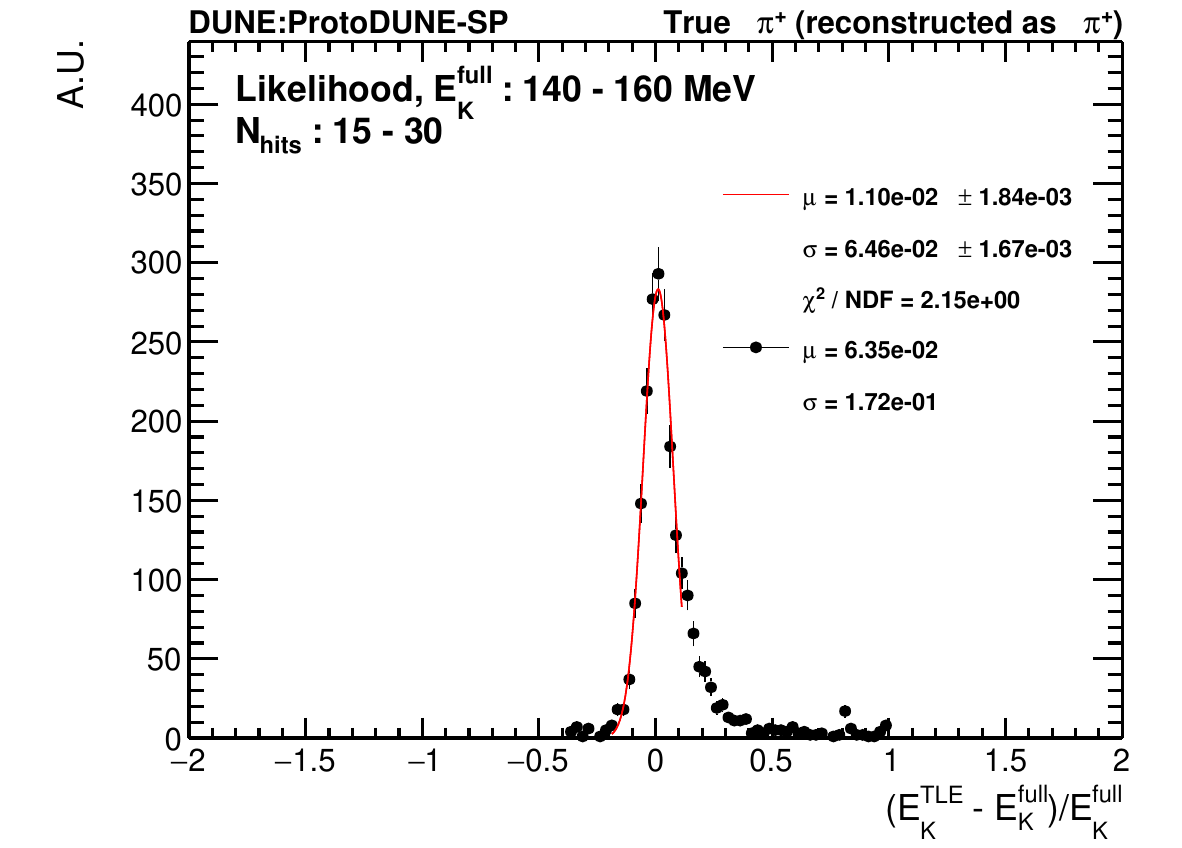}
    \caption{}
    \label{fig:Figure_013_a}
  \end{subfigure}
  \begin{subfigure}[b]{0.48\textwidth}
    \includegraphics[width=\textwidth]{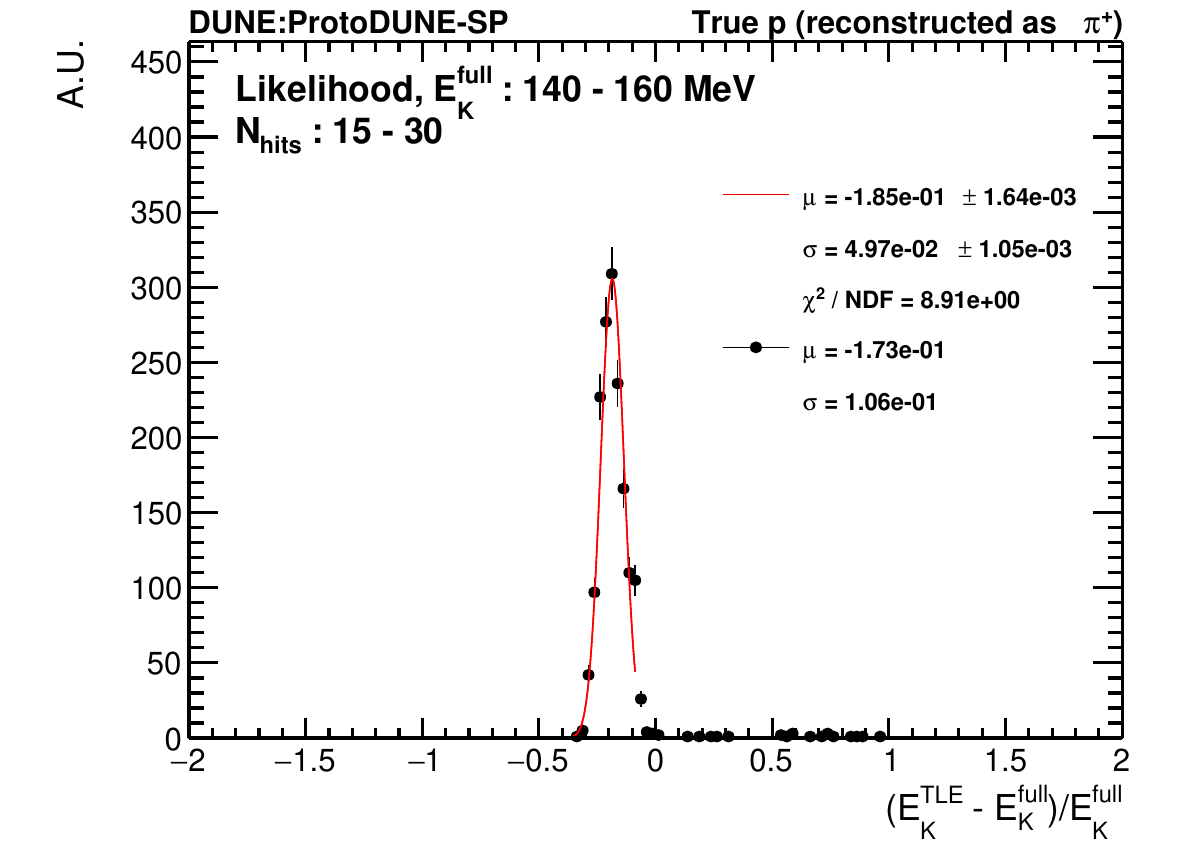}
    \caption{}
    \label{fig:Figure_013_b}
  \end{subfigure}
  \begin{subfigure}[b]{0.48\textwidth}
    \includegraphics[width=\textwidth]{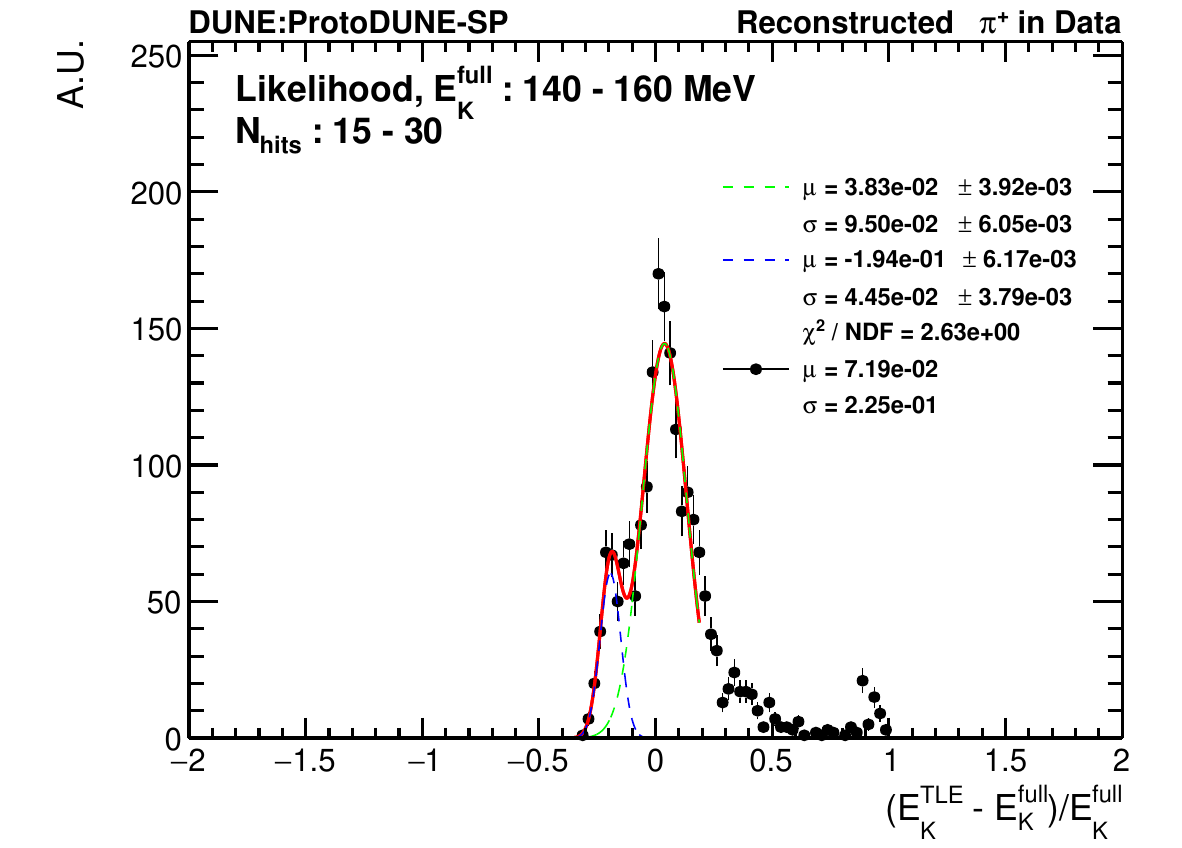}
    \caption{}
    \label{fig:Figure_013_c}
  \end{subfigure}
  \begin{subfigure}[b]{0.48\textwidth}
    \includegraphics[width=\textwidth]{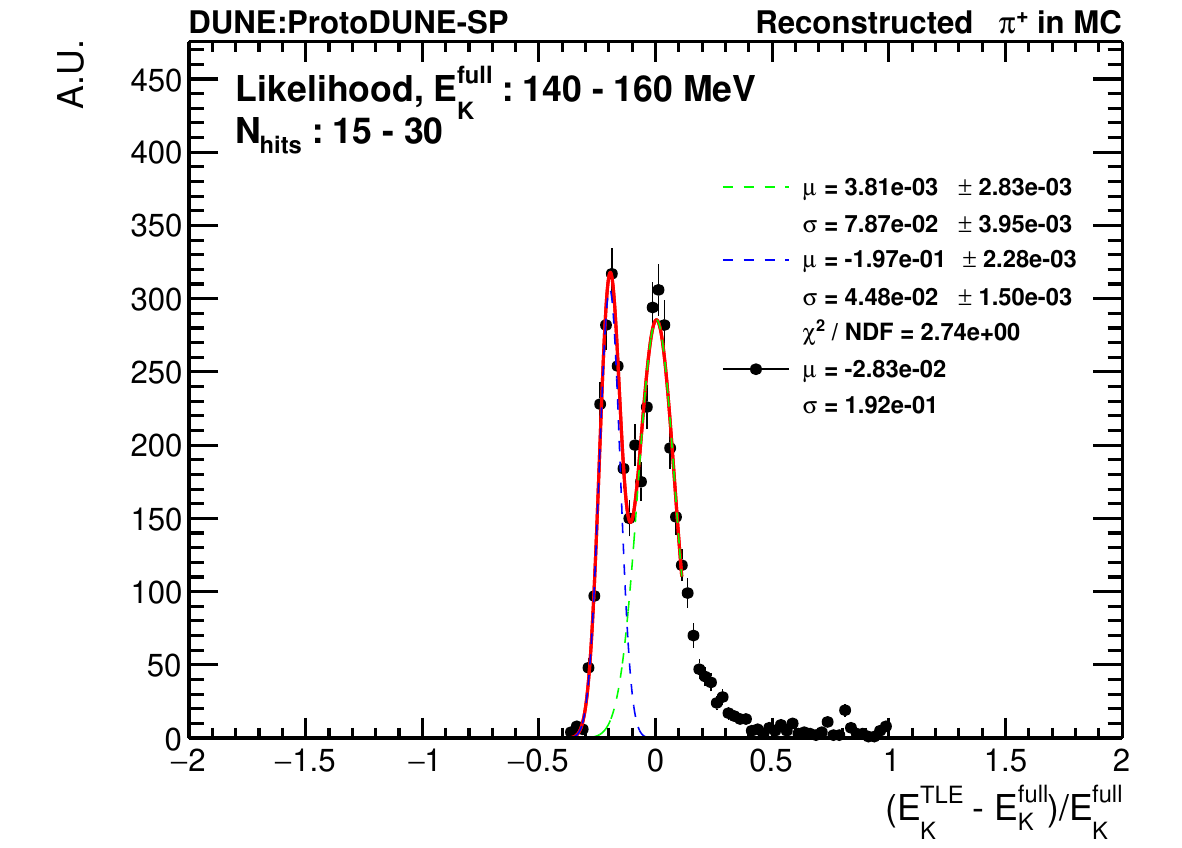}
    \caption{}
    \label{fig:Figure_013_d}
  \end{subfigure}
  \caption{Example distributions of the fractional energy residual using range-based \KE as reference. Results come from reconstructed charged pions with 15 to 30 hits and \KEfull from 140 MeV to 160 MeV. Top plots show results for MC true pions (a) and true protons (b). Bottom plots show results using all reconstructed charged pions for data (c) and the MC sample (d), where green and blue Gaussian functions represent charged pion and proton contributions, respectively.
  }
  \label{fig:Figure_013}
\end{center}
\end{figure*}

The same overlap is observed in one-dimensional distributions of the fractional energy residual as shown in figures~\ref{fig:Figure_012} and \ref{fig:Figure_013}.
MC simulation and data samples show different proton contributions. It implies that a selection efficiency correction should be made, which is beyond the scope of this study. To deal with the overlap between charged pions and protons, fractional bias distributions are fit using a double Gaussian function. Initial values of the two Gaussian function parameters are set using single Gaussian fit parameters coming from pure charged pion and pure proton distributions of the MC simulation sample. The energy measurement resolution and the fractional bias are calculated using data points within two standard deviations of the charged pion part Gaussian after subtracting the proton part Gaussian function's contribution. For the case where mean values of two Gaussian functions have difference smaller than 0.1, the proton contribution is not subtracted since the charged pion and the proton contributions are not distinguishable.

\begin{figure*}[htbp]
\begin{center}
  \includegraphics[width=0.48\textwidth]{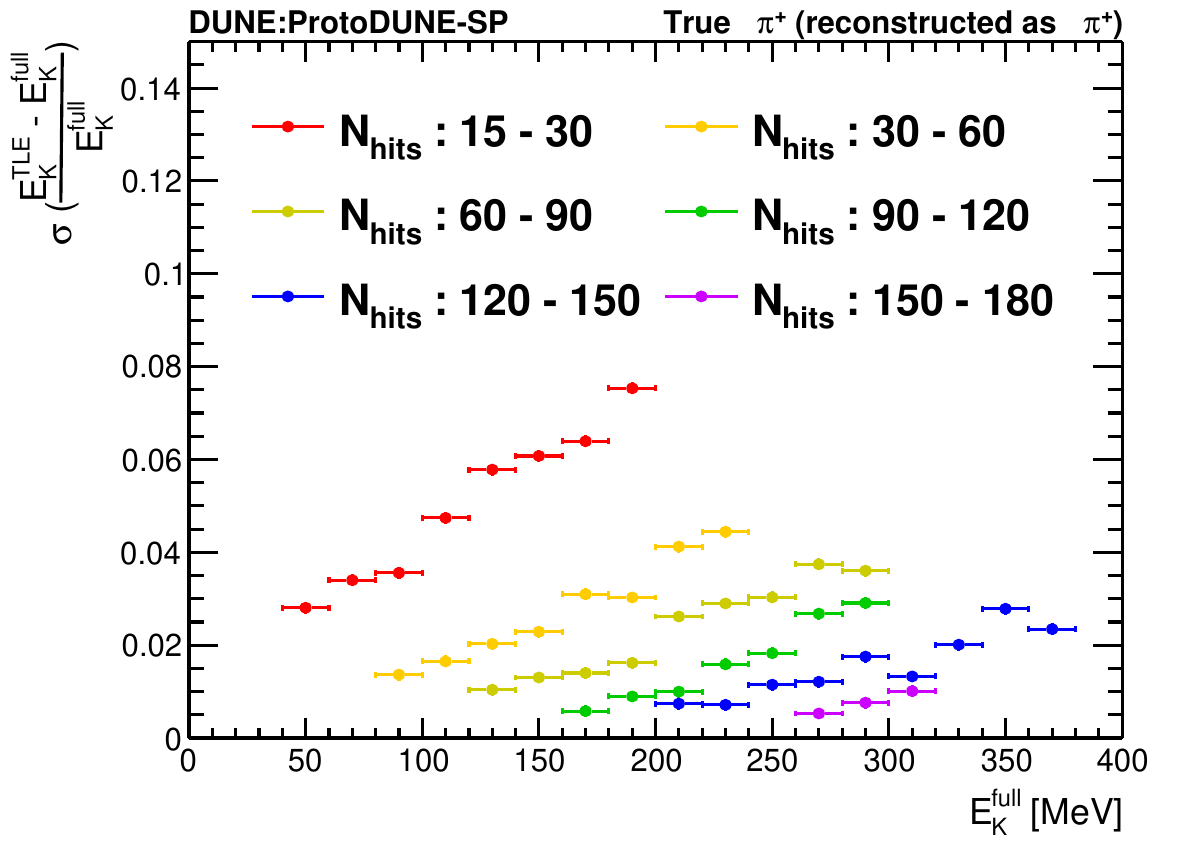}
  \includegraphics[width=0.48\textwidth]{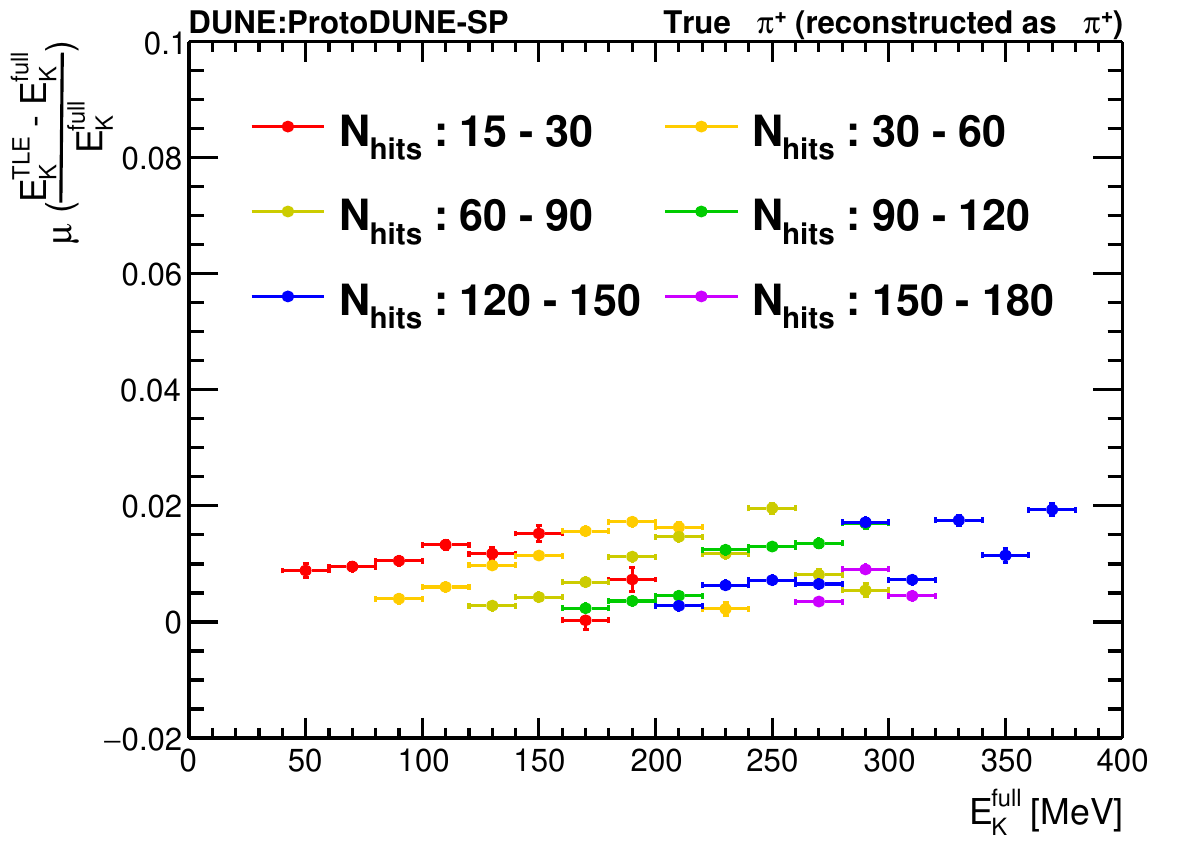}
  \caption{Summarized plots of energy measurement performance of the TLEFit method based on the maximum-likelihood method. Resolutions (left) and fractional biases (right) are shown as functions of charged pions' range-based \KE and number of hits. Pure secondary charged pions are selected from the MC sample.
  }
  \label{fig:Figure_014}
\end{center}
\end{figure*}

\begin{figure*}[htbp]
\begin{center}
  \includegraphics[width=0.90\textwidth]{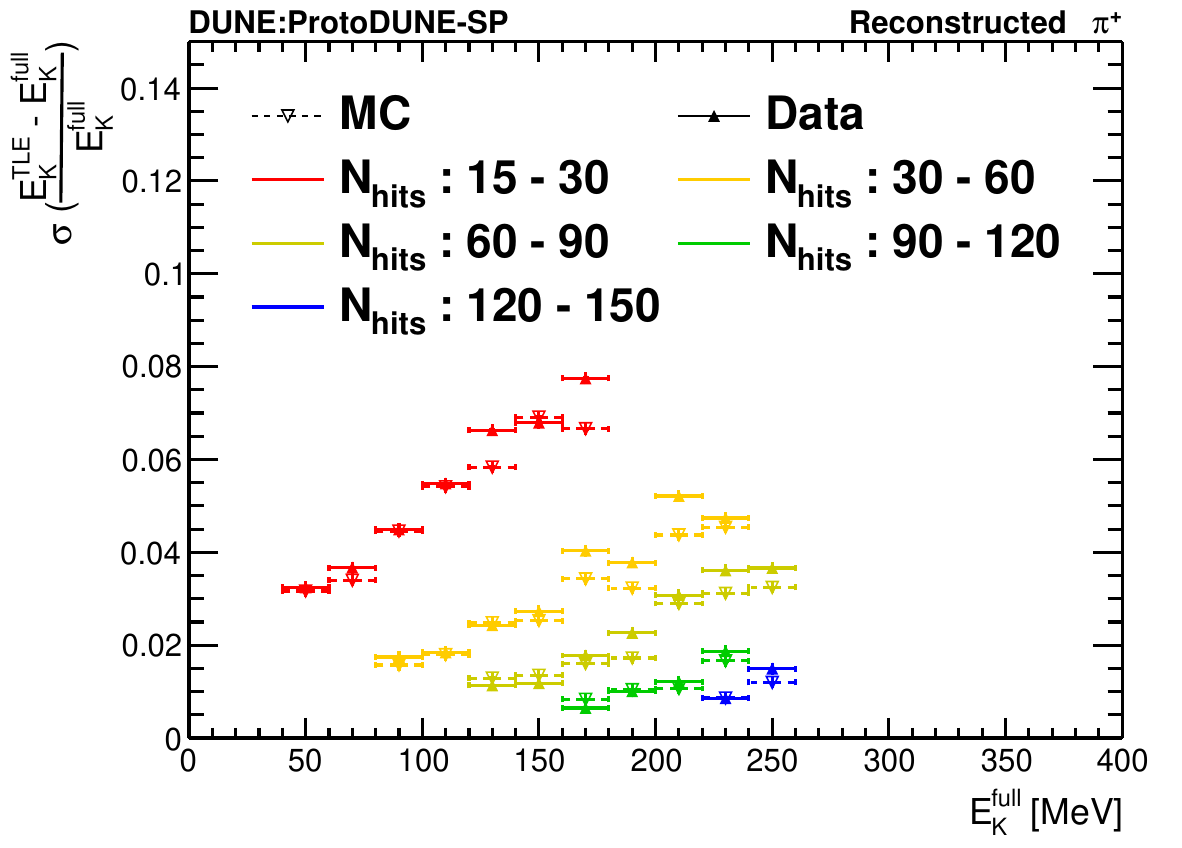}
  \includegraphics[width=0.90\textwidth]{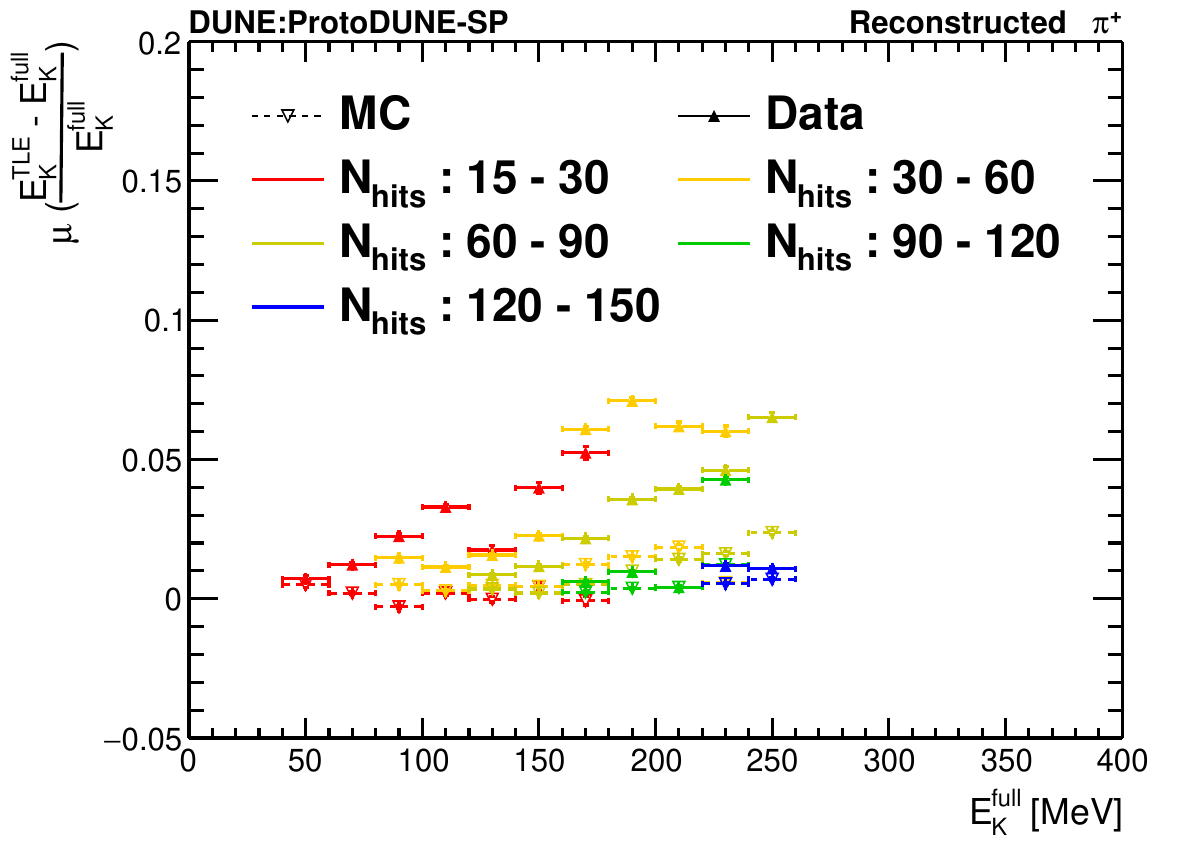}
  \caption{Summarized plots of energy measurement performance of the TLEFit method based on maximum-likelihood method. Resolutions (left) and fractional biases (right) are shown as functions of charged pions' range-based \KE and number of hits. All reconstructed secondary charged pions passing the stopping charged pion cut are used. Monte Carlo points are shown with dashed horizontal bars and the data points are shown with solid horizontal bars.
  }
  \label{fig:Figure_015}
\end{center}
\end{figure*}

Figure~\ref{fig:Figure_014} summarizes the energy measurement performance for a pure sample of secondary charged pions from the MC simulation. Figure~\ref{fig:Figure_015} shows a direct performance comparison between data and MC simulation samples. The upper \KE cutoff for each number of hits range is lower than that in figure~\ref{fig:Figure_014} to ensure good modeling of the proton contribution. The top plot of figure~\ref{fig:Figure_015} shows good agreement between MC and data in resolution with values smaller than 8\%. The bottom plot of figure~\ref{fig:Figure_015} shows a fractional bias result that is smaller than 10\% with discrepancy between MC simulation and data. It means that there should be studies of the energy scale correction and the corresponding systematic uncertainty in order to utilize the TLEFit method. The energy scale correction can be measured at any LArTPC using stopping secondary charged pions from neutrino interactions as introduced in this section, but as a function of \KEtle rather than the \KEfull. Section~\ref{subsubsec:understanding_dEdx} describes the corresponding systematic uncertainty estimates, separately for each source.

An additional comment is that the TLEFit method does not work well for a track for which the \dedx values have yet to sample the Bragg peak. It is a major reason why there is an upper \KE cutoff for each number of hits range. It is an energy measurement method mainly for charged pions with \KE less than 400 \MeV. In this energy region, the most dominant inelastic scattering topology between charged pions and argon nuclei is absorption.

\clearpage

\subsection{Impact of {\boldmath{\dedx}} modeling}
\label{subsubsec:understanding_dEdx}
Small discrepancies are seen between data and MC simulation, and these motivate ionization scale and resolution corrections with corresponding systematic uncertainties.
Estimations of these corrections are needed in order to use the TLEFit method for physics analyses.
In this section, we present the impact of \dedx modeling on results in figure~\ref{fig:Figure_015} to understand the discrepancy between MC and data in the fractional bias.
Motivation of testing \dedx modeling is that data shows up to about 7\% fractional bias with respect to the range-based energy using full track length while MC sample has less than 2.5\% fractional bias.
We observe that \dedx modeling could explain the discrepancy.

Reconstruction of \dedx for a hit is performed with the following steps. First, ADC values are integrated over time ticks to measure total collected electric charge for the hit. Subsequently, a correction is applied to remove the effect of electron attachment during the drift.
Then, a calibration factor ($C_{\rm{cal}}$) is multiplied to convert the total electric charge into number of electrons. Finally, total number of electrons is converted into deposited energy based on an assumption that the particle deposits energy only through ionizing argon atoms. The methods for determining $C_{\rm{cal}}$ and the electron attachment rate from the experimental data are described in Ref.~\cite{Abi_2020}.

Recombination between argon ions and electrons has a significant impact on conversion between collected electric charge per unit length \dqdx [electrons/cm] and \dedx [MeV/cm]. Thus, it should be modeled.
The ProtoDUNE-SP experiment uses the modified box model~\cite{ArgoNeuT:2013kpa} to consider electron recombination. In this model, the relation between \dqdx and \dedx is
\begin{equation}
{\frac{dQ}{dx} = \frac{1}{\rho \epsilon \beta' {W}_{\rm{Ion}}}\log \left(\rho \epsilon \beta'\frac{dE}{dx} + \alpha \right)},
  \label{eq:recom_mod_box}
\end{equation}
where $\epsilon$ is the electric field strength, ${W}_{\rm{Ion}}$ is mean ionization energy of an argon atom in \MeV, and $\alpha$ and $\beta'$ are the model parameters. For both data production and simulation, the modified box model parameters presented in~\cite{ArgoNeuT:2013kpa}, $\alpha = 0.93 \pm 0.02$ and $\mathrm{\beta'=0.212 \pm 0.002 ~(kV/cm)(g/{cm}^{2})/MeV}$, are used.

The calibration constant is measured using stopping cosmic muons and their minimum ionizing hits.
The minimum ionizing hits are selected using kinetic energies measured by the CSDA in the range from 250 \MeV to 450 \MeV. For a fixed $C_{\rm{cal}}$ value, \dedx distributions are drawn as a function of muon \KE bin. Each \dedx distribution is fitted using a convolution of a Landau function and a Gaussian to measure the most probable value (MPV). Then, a $\mathrm{{\chi}_{cal}^{2}}$ variable is defined as
\begin{equation}
\mathrm{{\chi}_{cal}^{2} = {\Sigma}_{i} {({MPV}_{fitted, i} - MPV_{Vavilov, i})}^{2} / {\sigma}_{{MPV}_{fitted}, i}^{2}},
  \label{eq:calib_chi2}
\end{equation}
where the index $\mathrm{i}$ runs for different \KE bins, $\mathrm{{MPV}_{Vavilov,i}}$ is expected MPV given by the theory~\cite{RevModPhys.60.663}, and $\mathrm{{\sigma}_{{MPV}_{fitted}, i}}$ is the error of the fitted MPV. The $C_{\mathrm{cal}}$ value that gives the minimum ${\chi}_{\rm{cal}}^{2}$ is selected as the calibration constant. In this way, we can achieve a good agreement between data and MC simulation for the MPV of the \dedx distribution of MIPs.

\begin{figure*}[htbp]
\begin{center}
  \begin{subfigure}[b]{0.48\textwidth}
    \includegraphics[width=\textwidth]{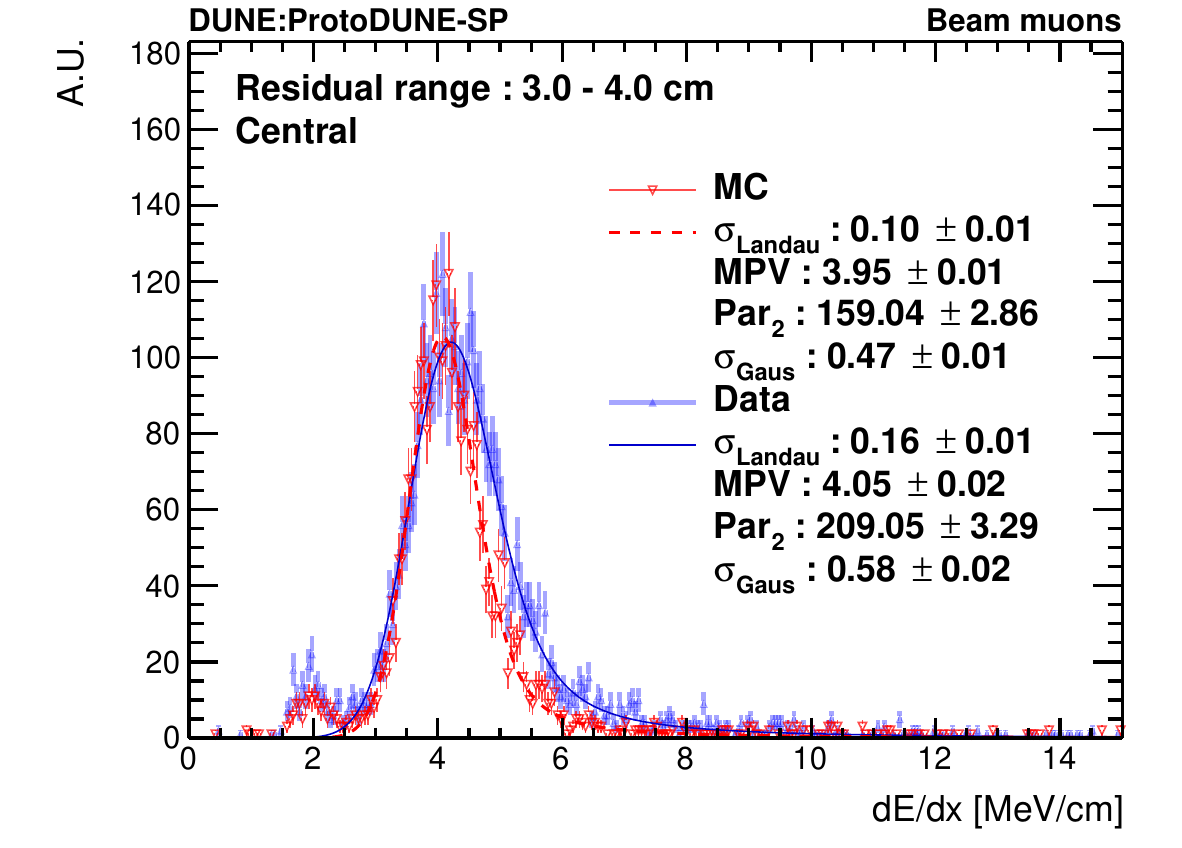}
    \caption{}
    \label{fig:Figure_016_a}
  \end{subfigure}
  \begin{subfigure}[b]{0.48\textwidth}
    \includegraphics[width=\textwidth]{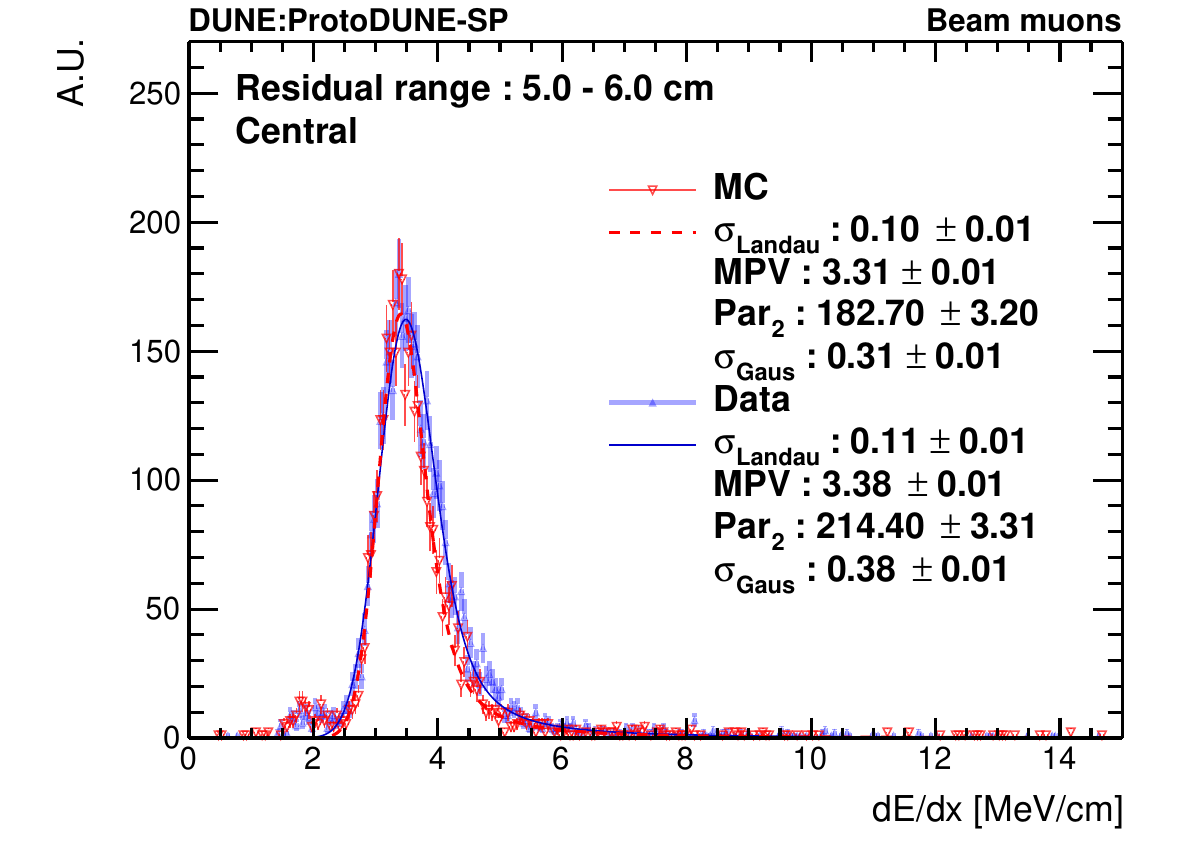}
    \caption{}
    \label{fig:Figure_016_b}
  \end{subfigure}
  \begin{subfigure}[b]{0.48\textwidth}
    \includegraphics[width=\textwidth]{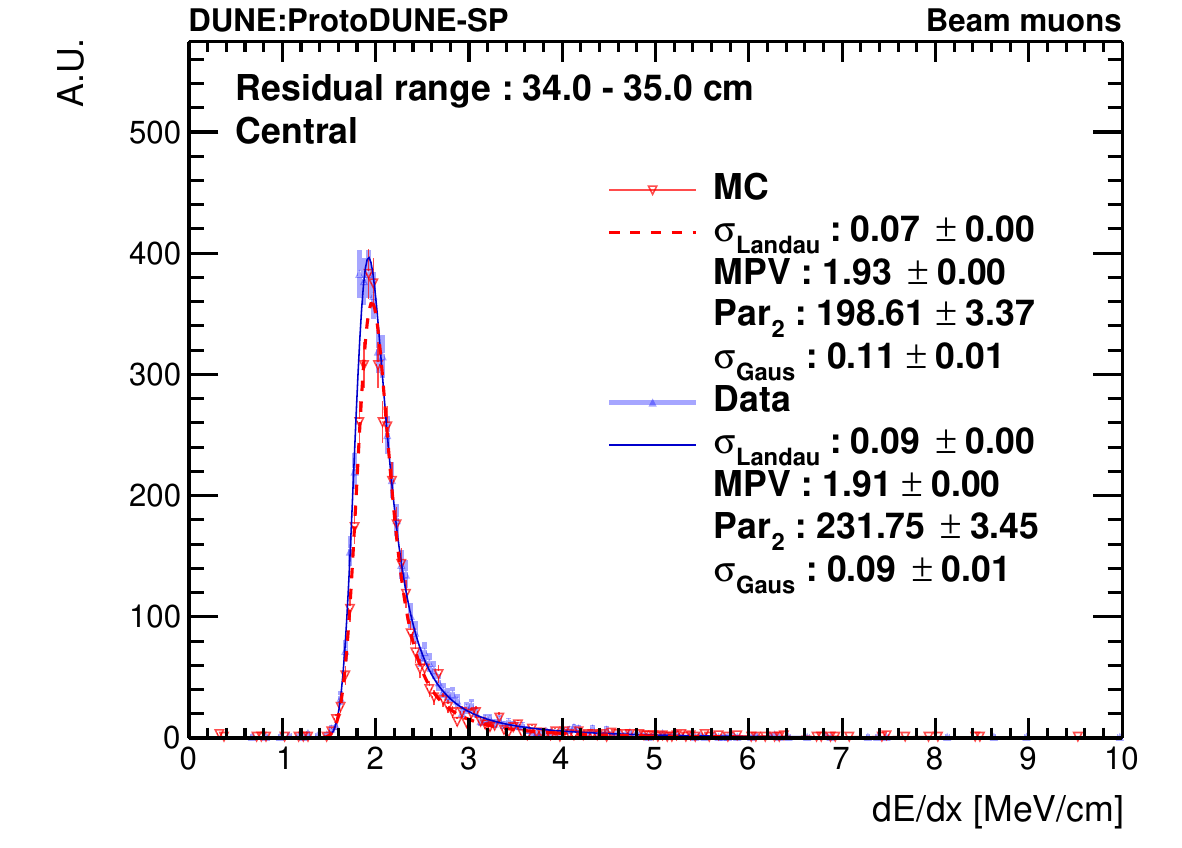}
    \caption{}
    \label{fig:Figure_016_c}
  \end{subfigure}
  \begin{subfigure}[b]{0.48\textwidth}
    \includegraphics[width=\textwidth]{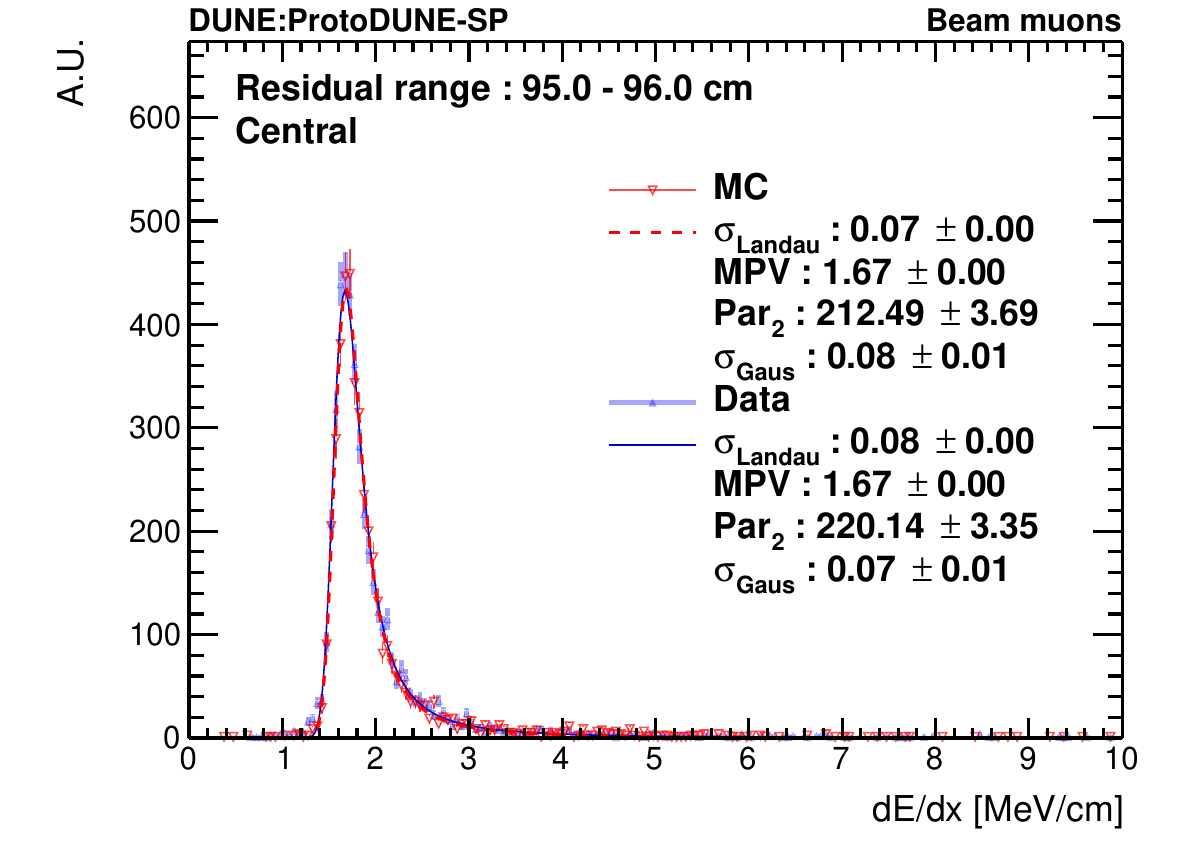}
    \caption{}
    \label{fig:Figure_016_d}
  \end{subfigure}
  \caption{The \dedx of beam muons is shown as a function of residual range for both data (blue) and MC (red). The results of fits to a Landau function convoluted with a Gaussian are also shown in the legends. The $\mathrm{\sigma_{Landau}}$ is intrinsic width of the Landau function, MPV is fitted most probable value, $\mathrm{{Par}_{2}}$ is normalization factor, and $\mathrm{\sigma_{Gaus}}$ is the width of the Gaussian.
  }
  \label{fig:Figure_016}
\end{center}
\end{figure*}

\begin{figure*}[htbp]
\begin{center}
  \begin{subfigure}[b]{0.48\textwidth}
    \includegraphics[width=\textwidth]{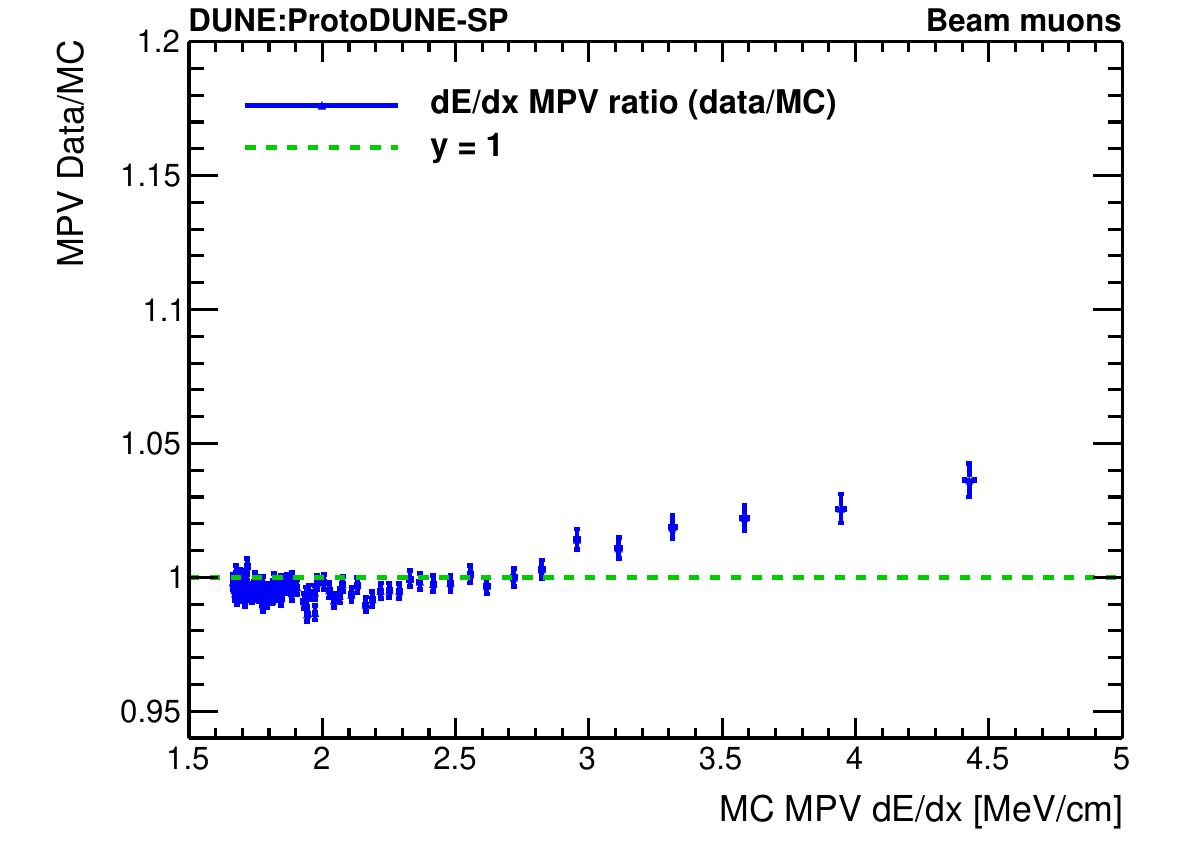}
    \caption{}
    \label{fig:Figure_017_a}
  \end{subfigure}
  \begin{subfigure}[b]{0.48\textwidth}
    \includegraphics[width=\textwidth]{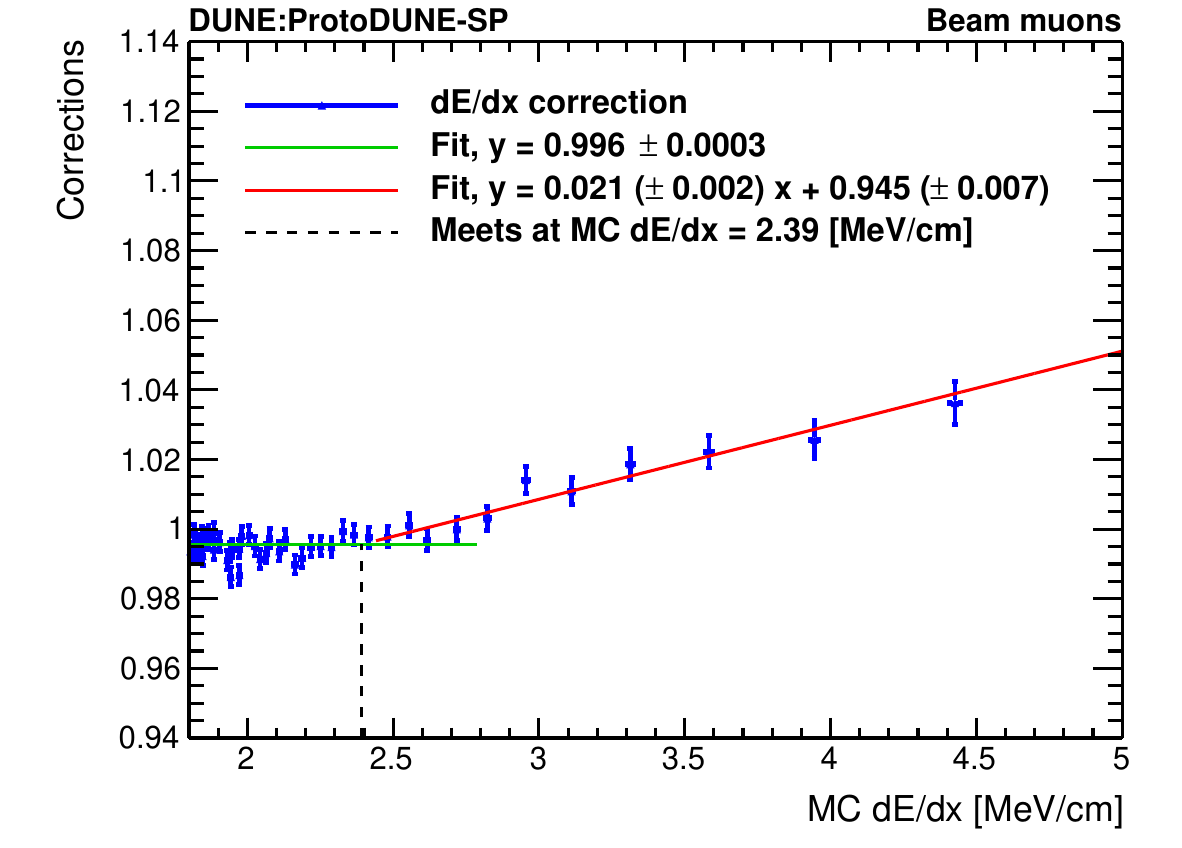}
    \caption{}
    \label{fig:Figure_017_b}
  \end{subfigure}
  \caption{Ratio of fitted MPV values between MC and data as a function of MC MPV. Each MPV value is fitted using beam muon's hits with a 1 $\mathrm{cm}$ interval in residual range from 2 to 100 $\mathrm{cm}$. Ratio is calculated for each residual range interval. Vertical error bars show statistical uncertainties only. Figure (a) shows the ratio distribution with respect to the unity. Figure (b) shows how the \dedx correction for the Bragg peak region is derived using a linear fit.
  }
  \label{fig:Figure_017}
\end{center}
\end{figure*}

As a result, figure~\ref{fig:Figure_016}-(d) shows good agreement between data and MC simulation for \dedx distributions of beam muon hits in a residual range region from 95 to 96~cm, which corresponds to the MIP region. But, for hits near the Bragg peak region, the agreement between data and MC simulation deteriorates as shown in the plots with shorter residual ranges of figure~\ref{fig:Figure_016}. It means that the modified box model with parameters in~\cite{ArgoNeuT:2013kpa} cannot describe the recombination effect in ProtoDUNE-SP in the MIP region and the Bragg peak region at the same time. It is well illustrated in figure~\ref{fig:Figure_017_a}. The abscissa shows fitted MPVs of MC simulation sample for beam muon hits in residual range region from 2 to 100 $\mathrm{cm}$ with 1 $\mathrm{cm}$ step. The ordinate shows the ratio of fitted MPV between MC simulation and data in each residual range region. MPV ratio is close to unity where MC simulation MPV is smaller than 1.7 $\mathrm{MeV/cm}$. With increasing MC simulation MPV values, the ratio goes down to about 0.985 and goes up to about 1.04. To understand the impact of such differences in the \dedx probability density function between data and MC simulation on fractional bias results shown in figure~\ref{fig:Figure_015}, we performed studies described below.

\subsubsection{Reproduction of data {\boldmath{\dedx}} distributions}
\begin{figure*}[htbp]
\begin{center}
  \begin{subfigure}[b]{0.48\textwidth}
    \includegraphics[width=\textwidth]{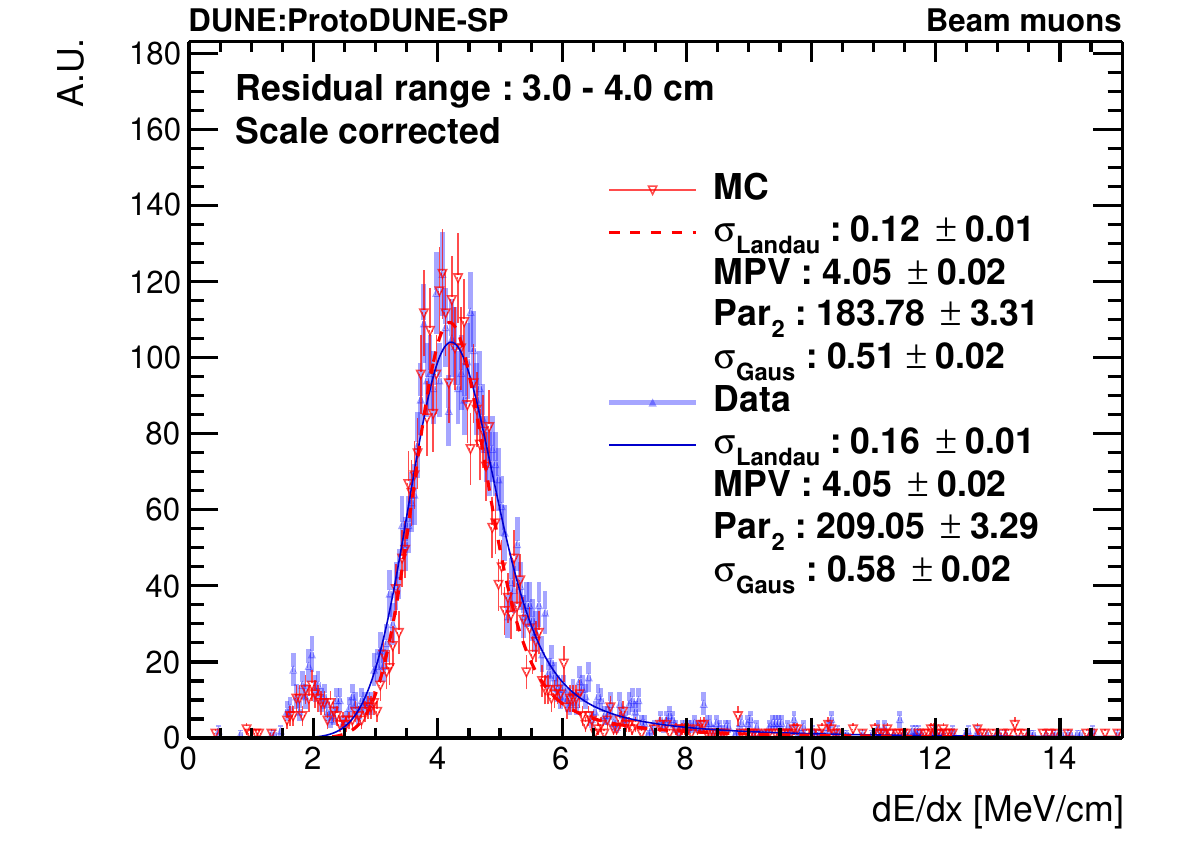}
    \caption{}
    \label{fig:Figure_018_a}
  \end{subfigure}
  \begin{subfigure}[b]{0.48\textwidth}
    \includegraphics[width=\textwidth]{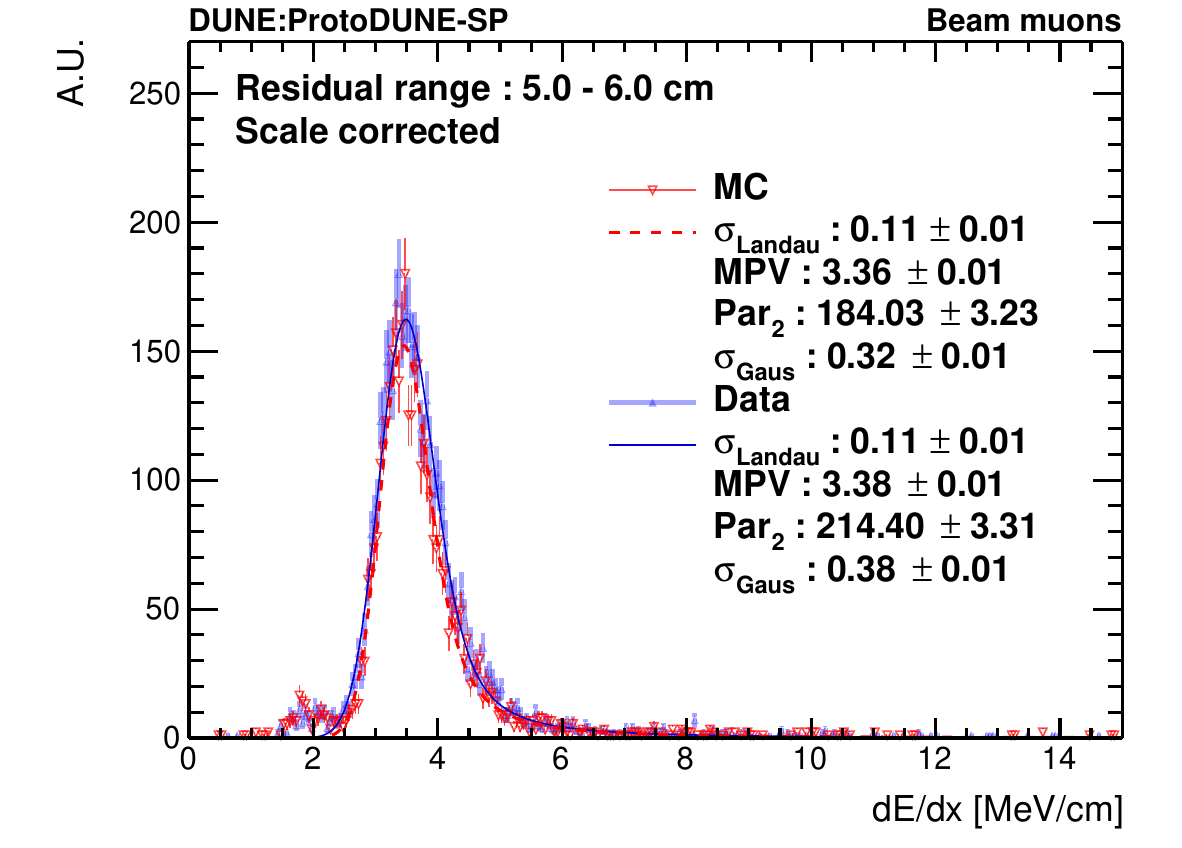}
    \caption{}
    \label{fig:Figure_018_b}
  \end{subfigure}
  \begin{subfigure}[b]{0.48\textwidth}
    \includegraphics[width=\textwidth]{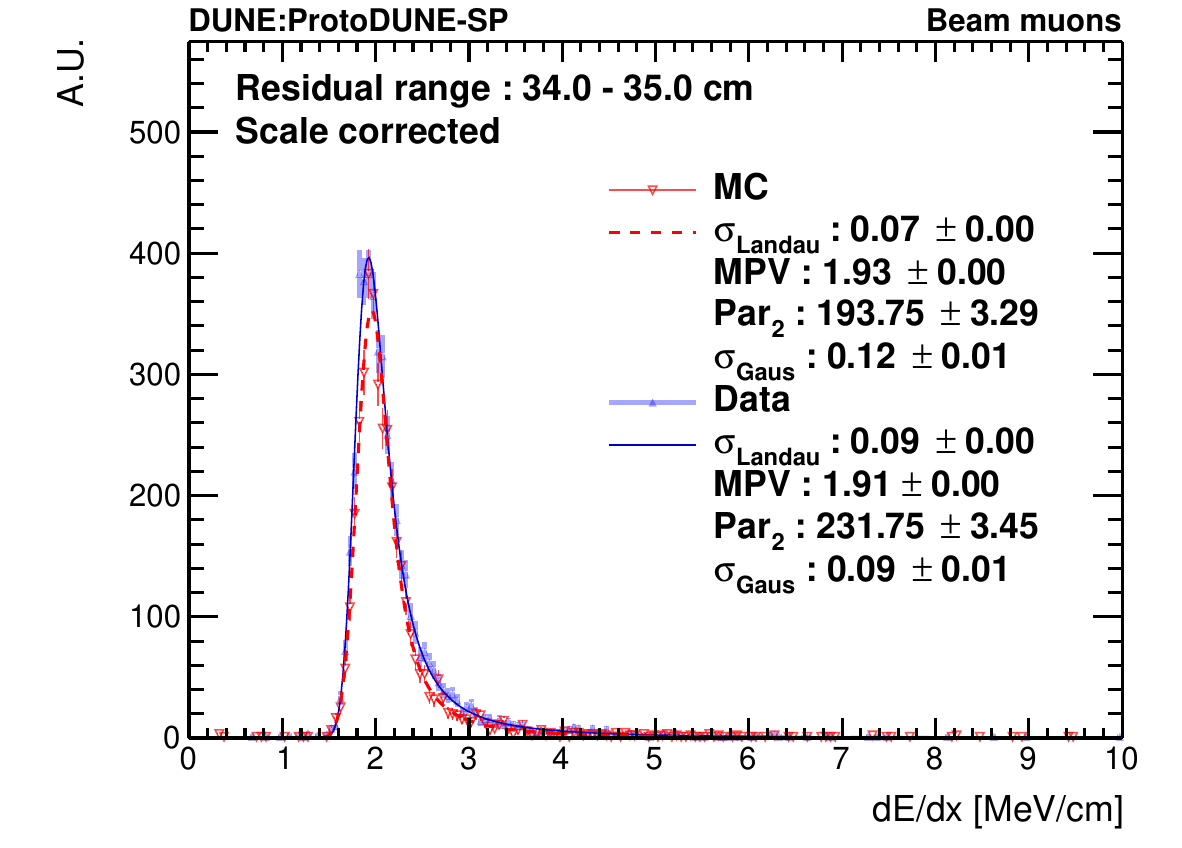}
    \caption{}
    \label{fig:Figure_018_c}
  \end{subfigure}
  \begin{subfigure}[b]{0.48\textwidth}
    \includegraphics[width=\textwidth]{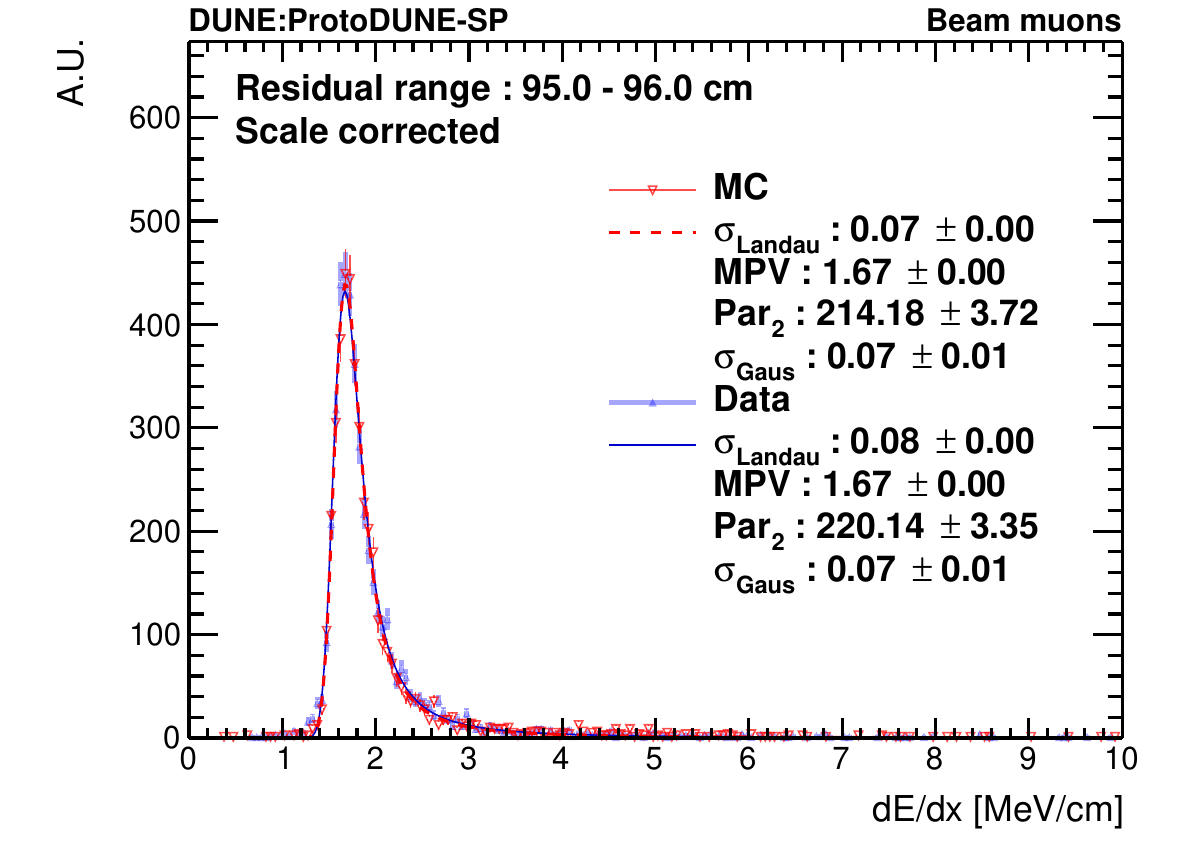}
    \caption{}
    \label{fig:Figure_018_d}
  \end{subfigure}
  \caption{The measured \dedx values for hits along beam muon tracks are shown as a function of residual range for both data (blue) and MC (red). Fitting results with the convoluted functions of the Gaussian and the Landau functions are also shown in the legends. The parameter $\mathrm{\sigma_{Landau}}$ is the intrinsic width of the Landau function, MPV is the fitted most probable value, $\mathrm{{Par}_{2}}$ is the normalization factor, and $\mathrm{\sigma_{Gaus}}$ is the width of the Gaussian part. The scale correction shown in figure~\ref{fig:Figure_017_b} for \dedx values is applied for MC sample.
  }
  \label{fig:Figure_018}
\end{center}
\end{figure*}

We attempt to reproduce the data's \dedx distributions by applying scale corrections to the MC simulation sample's \dedx values. The distribution of the MPV ratio between MC simulation and data, as shown in figure~\ref{fig:Figure_017_b}, is fitted using two linear functions: one flat function for the MIP region and the other with a slope for the Bragg peak region, to derive the scale corrections. It is shown that the two linear functions intersect at a \dedx value of 2.39 $\mathrm{MeV/cm}$. Consequently, the scale correction for the Bragg peak region is applied to hits with \dedx greater than 2.39 $\mathrm{MeV/cm}$. As a result, figure~\ref{fig:Figure_018} shows improved agreement in MPV values between data and MC simulation in the Bragg peak region compared to the state before applying the correction, as shown in figure~\ref{fig:Figure_016}. Note that widths of the Gaussian contribution also have better agreements in the Bragg peak region after applying the scale correction, but with non-negligible differences with respect to their statistical uncertainties. For the MIP region, we test two constant scale corrections with 0.985 and 0.975 to consider the discrepancy for \dedx from 1.8 $\mathrm{MeV/cm}$ to 2.3 $\mathrm{MeV/cm}$ shown in figure~\ref{fig:Figure_017_b}.

\subsubsection{Modified box model parameter uncertainties}
\begin{figure*}[htbp]
\begin{center}
  \begin{subfigure}[b]{0.48\textwidth}
    \includegraphics[width=\textwidth]{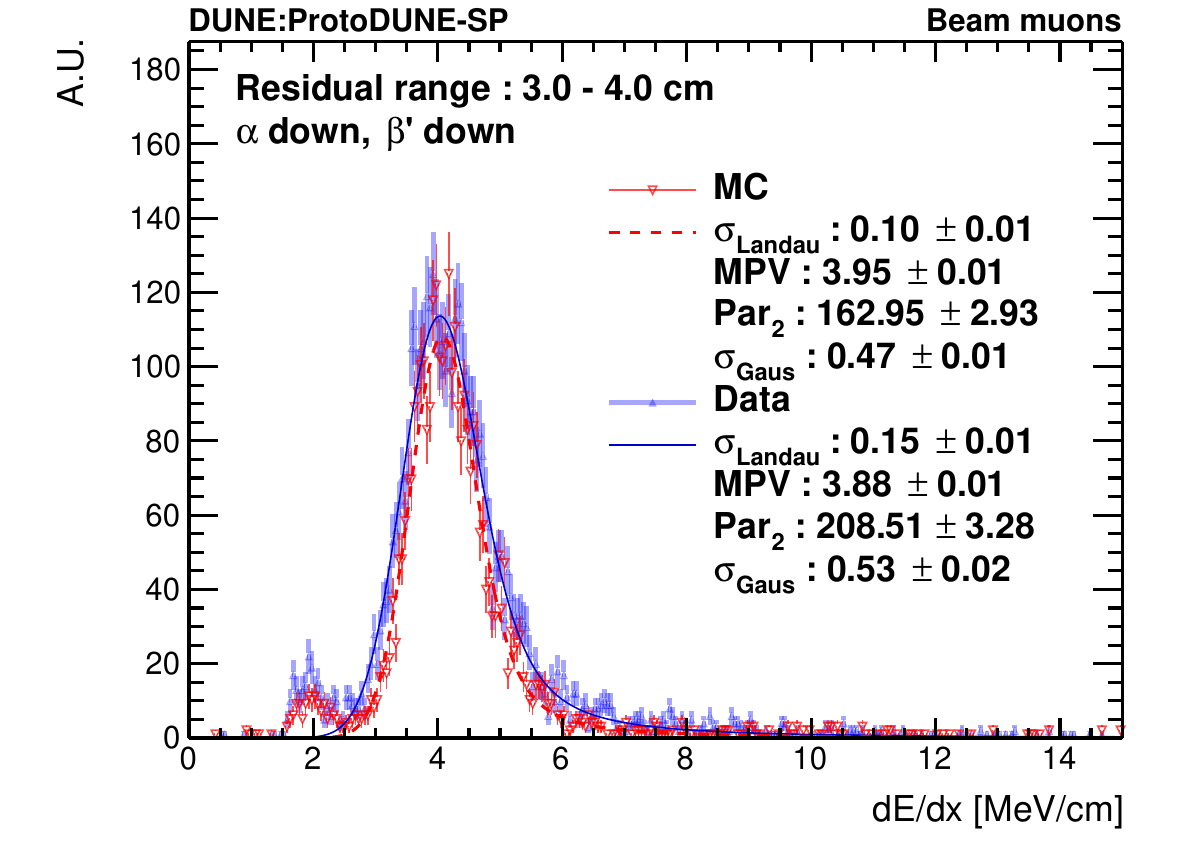}
    \caption{}
    \label{fig:Figure_019_a}
  \end{subfigure}
  \begin{subfigure}[b]{0.48\textwidth}
    \includegraphics[width=\textwidth]{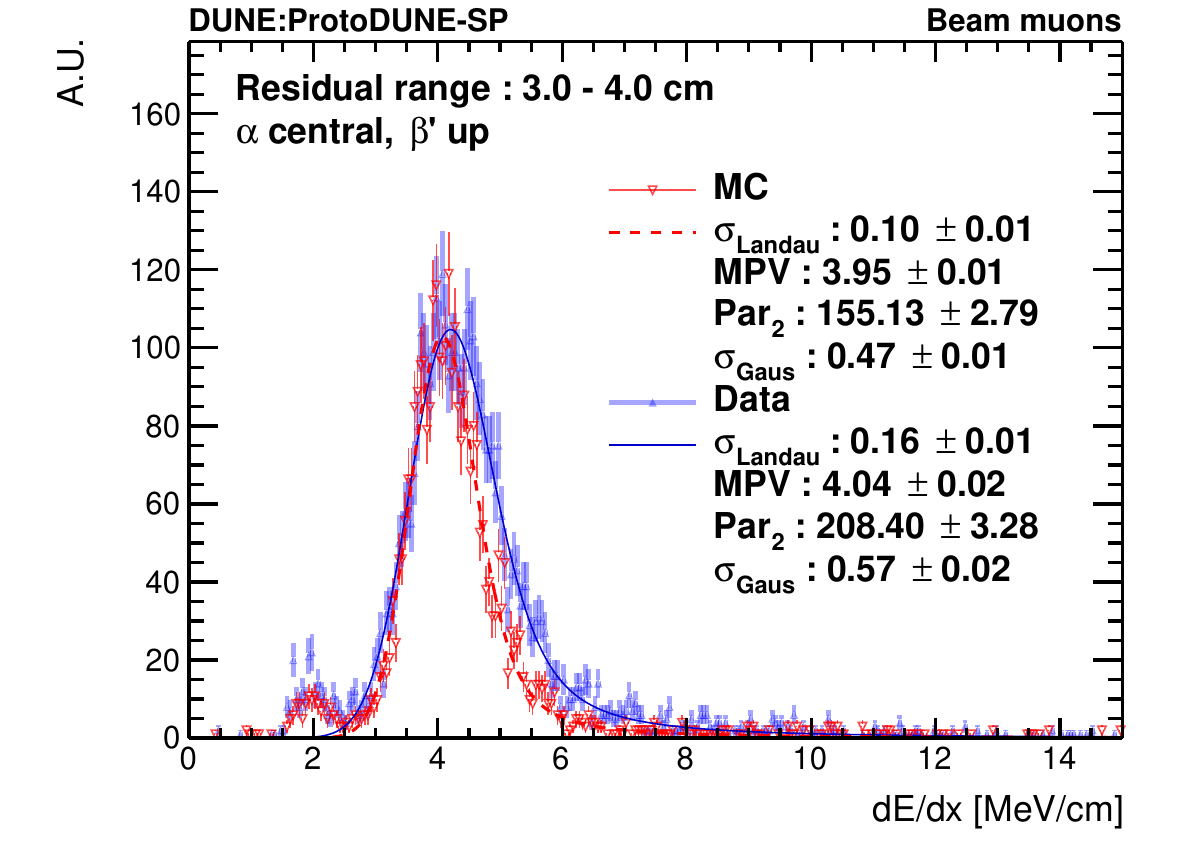}
    \caption{}
    \label{fig:Figure_019_b}
  \end{subfigure}
  \begin{subfigure}[b]{0.48\textwidth}
    \includegraphics[width=\textwidth]{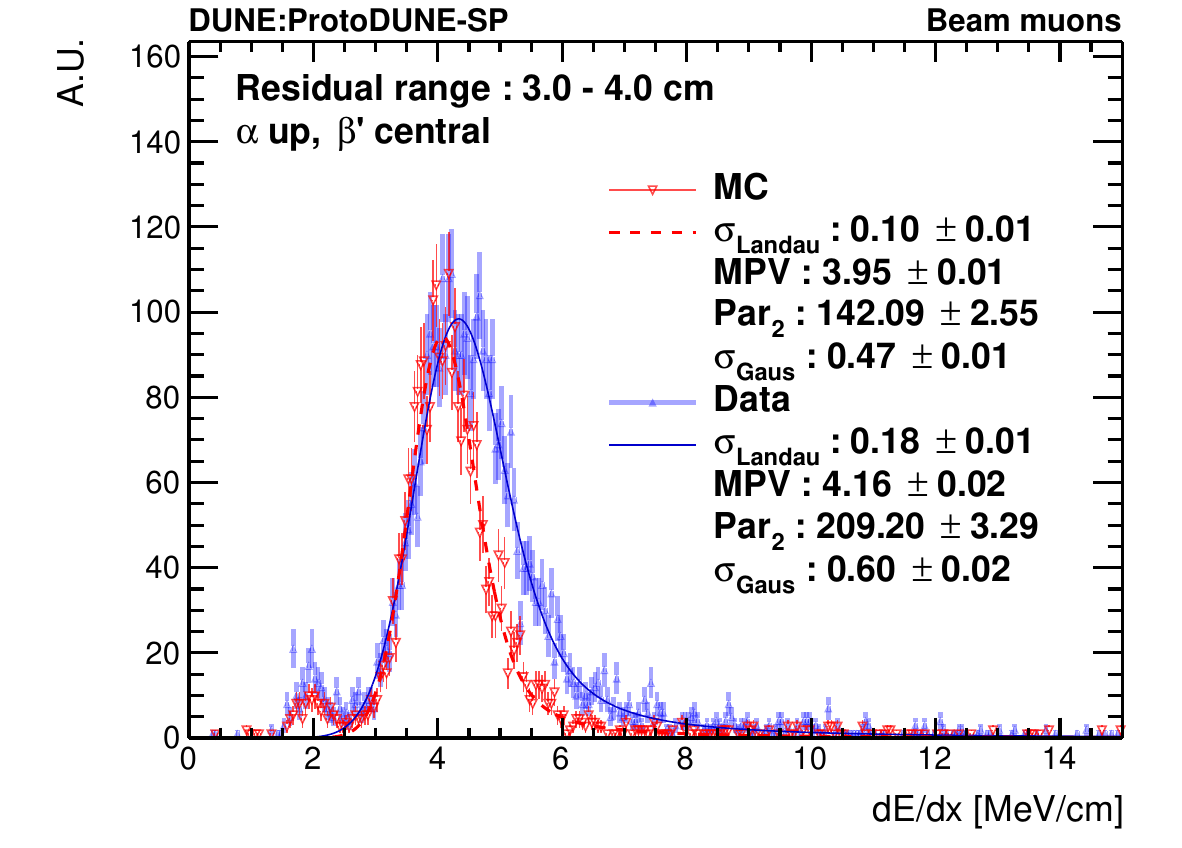}
    \caption{}
    \label{fig:Figure_019_c}
  \end{subfigure}
  \begin{subfigure}[b]{0.48\textwidth}
    \includegraphics[width=\textwidth]{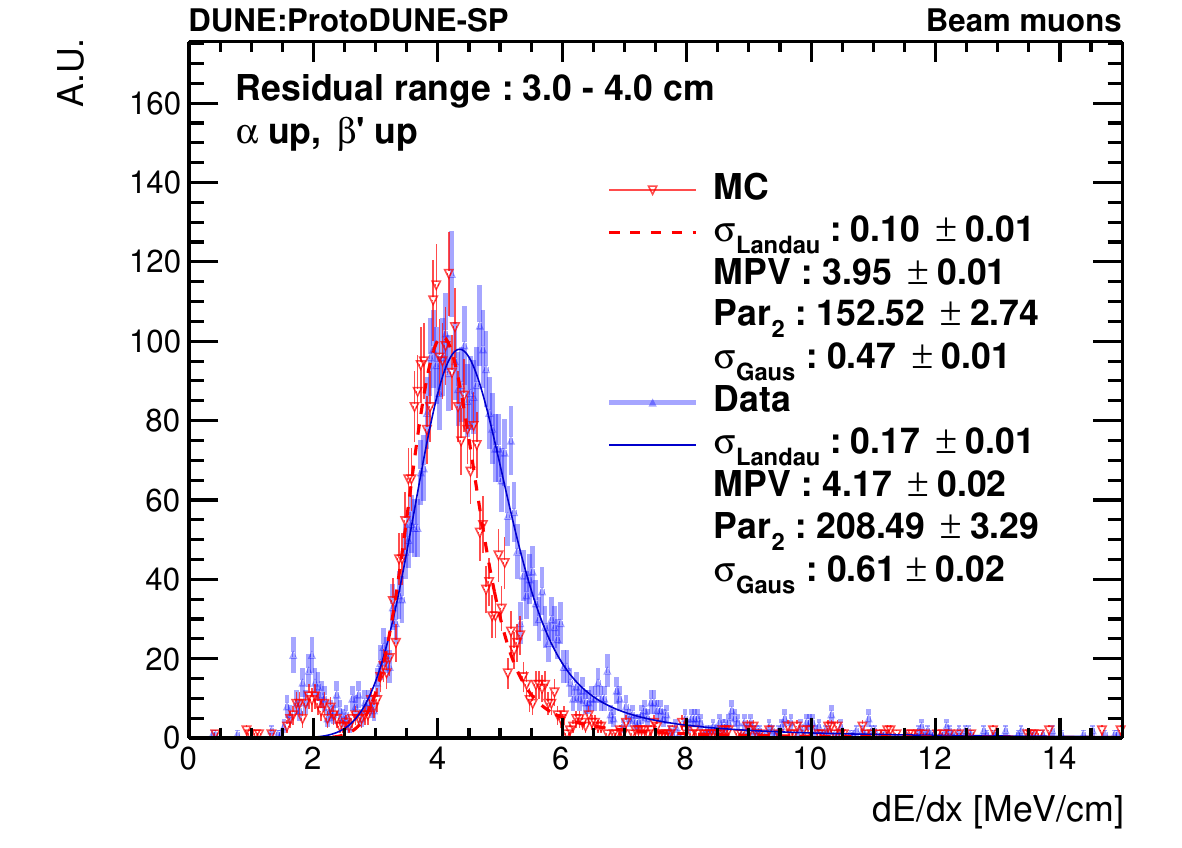}
    \caption{}
    \label{fig:Figure_019_d}
  \end{subfigure}
  \caption{The measured \dedx values of hits along beam muon tracks are shown for residual ranges between 3 and 4~cm for data (blue) and MC (red). Fitting results with the convoluted functions of the Gaussian and the Landau functions are also shown in the legends. The parameter $\mathrm{\sigma_{Landau}}$ is the intrinsic width of the Landau function, MPV is the fitted most probable value, $\mathrm{{Par}_{2}}$ is the normalization factor, and $\mathrm{\sigma_{Gaus}}$ is the width of the Gaussian component. Modified box model parameters are each shifted by one standard deviation~\cite{ArgoNeuT:2013kpa} for data. Distributions of MC are the same in each of the four plots.
  }
  \label{fig:Figure_019}
\end{center}
\end{figure*}

Another important point to note here is that shifting the modified box model parameters by one standard deviation can change \dedx distributions significantly in the Bragg peak region as shown in figure~\ref{fig:Figure_019}. Therefore, a corresponding systematic uncertainty should be considered for the TLEFit algorithm's performance results in figure~\ref{fig:Figure_015} for data. The study shown in Section~\ref{subsubsec:performance_2} is repeated for data with the modified box model parameters shifted up and down by one standard deviations which gives 8 sets of $\alpha$ and $\beta'$ besides the central set as shown in Table~\ref{table:recom_syst}. The envelope among the 8 systematic variations is taken as systematic uncertainty for each fractional bias and resolution data point.

\begin{table}[htbp]
\footnotesize
\caption{Recombination parameter shifts that are used for the study based on ArgoNeuT's measurement~\cite{ArgoNeuT:2013kpa}.}
\centering
\begin{tabular}{ccc}
\hline\hline
 & $\alpha$ & $\mathrm{\beta'~[(kV/cm)(g/{cm}^{2})/MeV}]$\\
\hline\hline
Central & 0.93 & 0.212\\
\hline
Systematic variations & 0.93 & 0.210\\
 & 0.93 & 0.214\\
 & 0.91 & 0.210\\
 & 0.91 & 0.212\\
 & 0.91 & 0.214\\
 & 0.95 & 0.210\\
 & 0.95 & 0.212\\
 & 0.95 & 0.214\\
\hline\hline
\end{tabular}
\label{table:recom_syst}
\end{table}

\subsubsection{Impact on $\mathbf{{\chi}^{2}_{{\pi}^{\pm}}}$}
Differences in the \dedx probability density functions between data and MC also affect the stopping charged pion selection.
It is required that reconstructed secondary charged pions should have ${\chi}^{2}_{{\pi}^{\pm}}$ smaller than 6.
The amount of contribution coming from interacting charged pions is different between data and MC simulation with the same ${\chi}^{2}_{{\pi}^{\pm}}$ cut because measured \dedx values are included in Eq.~\ref{eq:chi2_pion} and there are discrepancies between the data and MC simulation \dedx measurements. Therefore, the ${\chi}^{2}_{{\pi}^{\pm}}$ with \dedx values after corrections are 
used to select stopping charged pions in two studies above.

\subsubsection{Results}
\begin{figure*}[htbp]
\begin{center}
  \begin{subfigure}[b]{0.48\textwidth}
    \includegraphics[width=\textwidth]{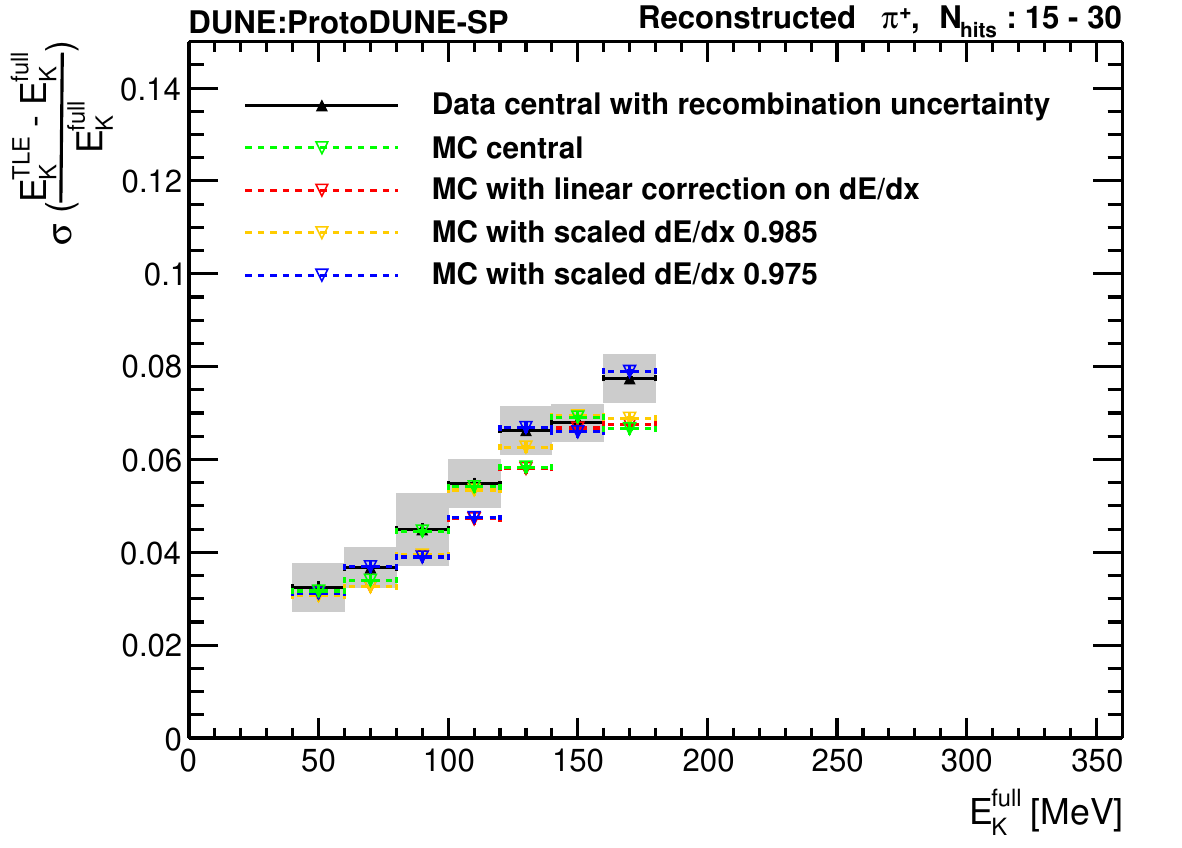}
    \caption{}
    \label{fig:Figure_020_a}
  \end{subfigure}
  \begin{subfigure}[b]{0.48\textwidth}
    \includegraphics[width=\textwidth]{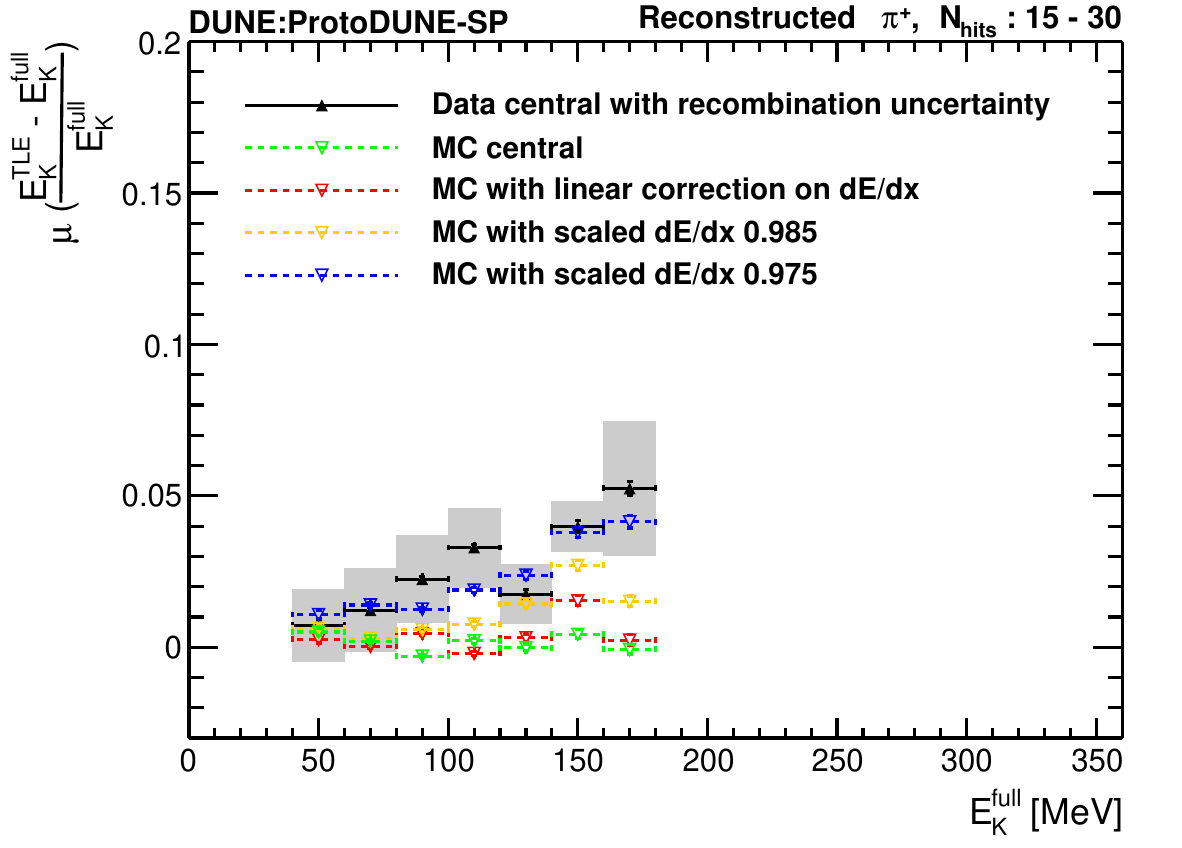}
    \caption{}
    \label{fig:Figure_020_b}
  \end{subfigure}
  \begin{subfigure}[b]{0.48\textwidth}
    \includegraphics[width=\textwidth]{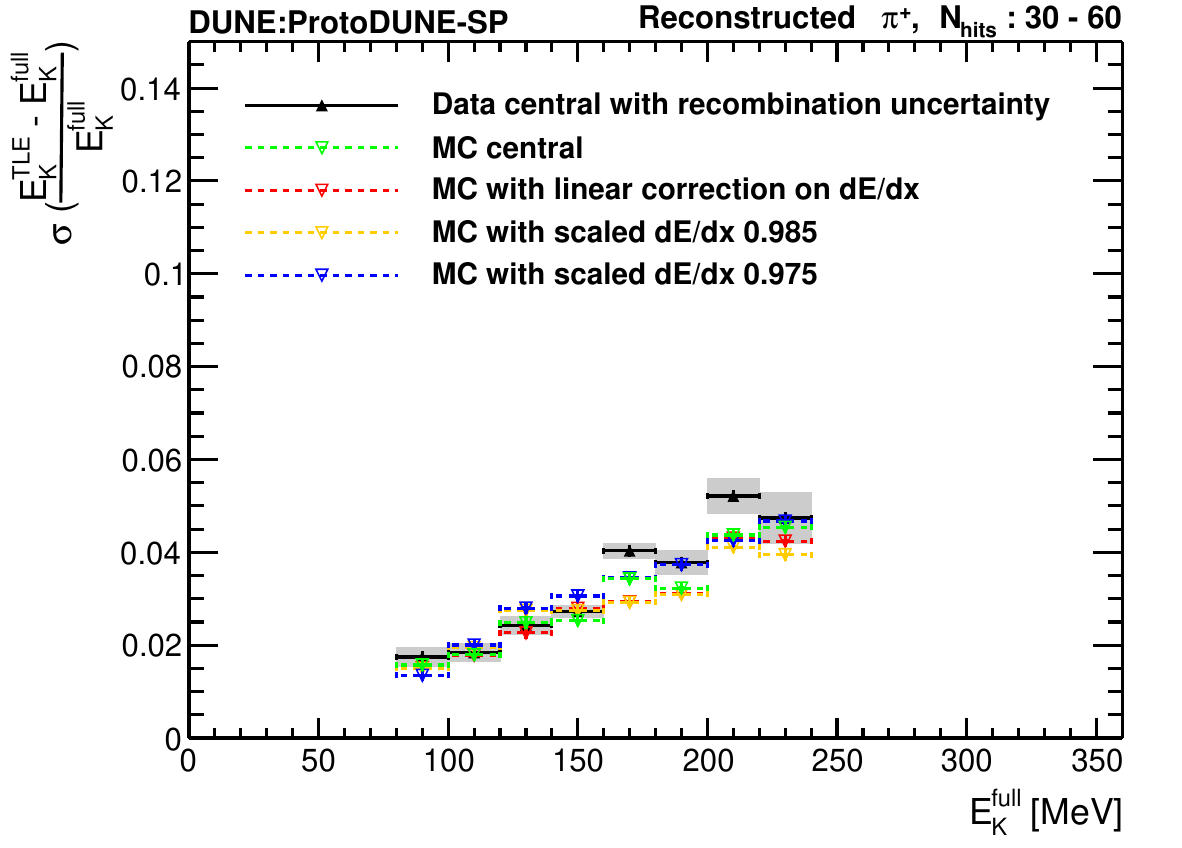}
    \caption{}
    \label{fig:Figure_020_c}
  \end{subfigure}
  \begin{subfigure}[b]{0.48\textwidth}
    \includegraphics[width=\textwidth]{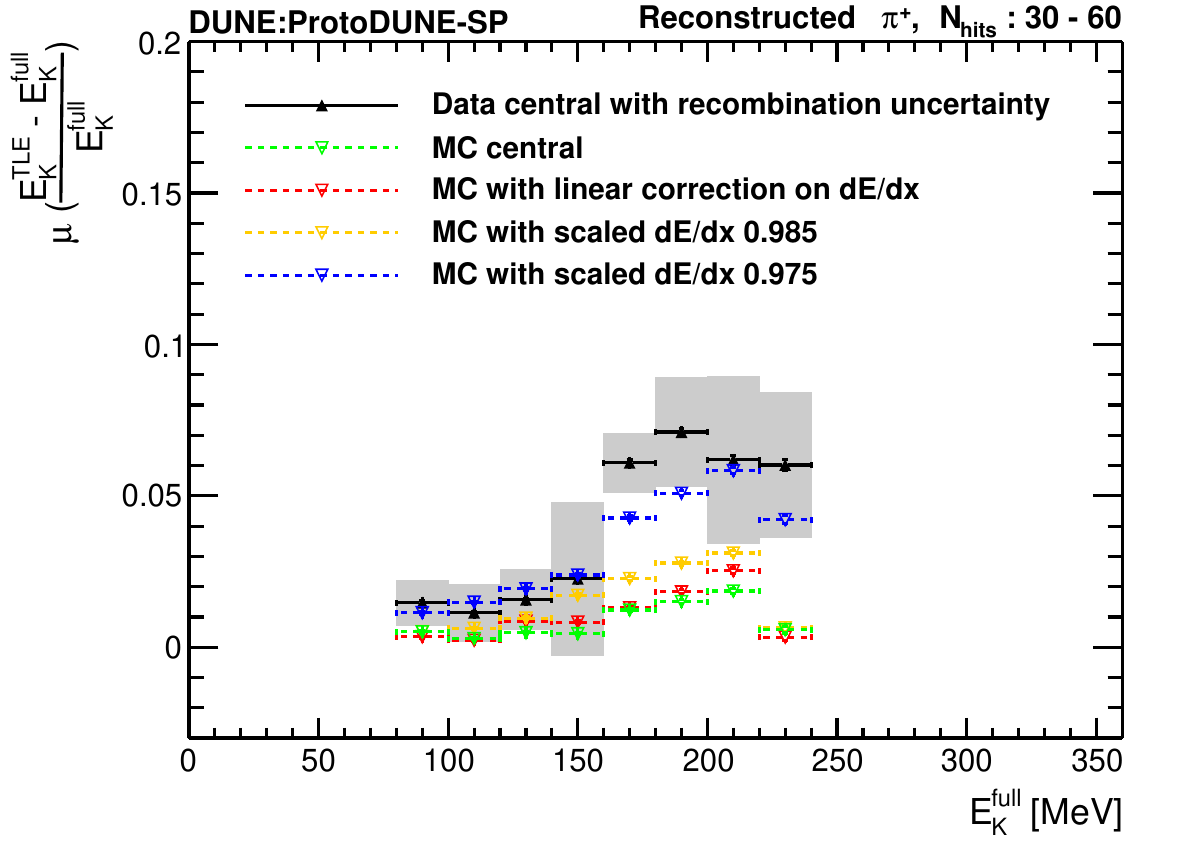}
    \caption{}
    \label{fig:Figure_020_d}
  \end{subfigure}
  \begin{subfigure}[b]{0.48\textwidth}
    \includegraphics[width=\textwidth]{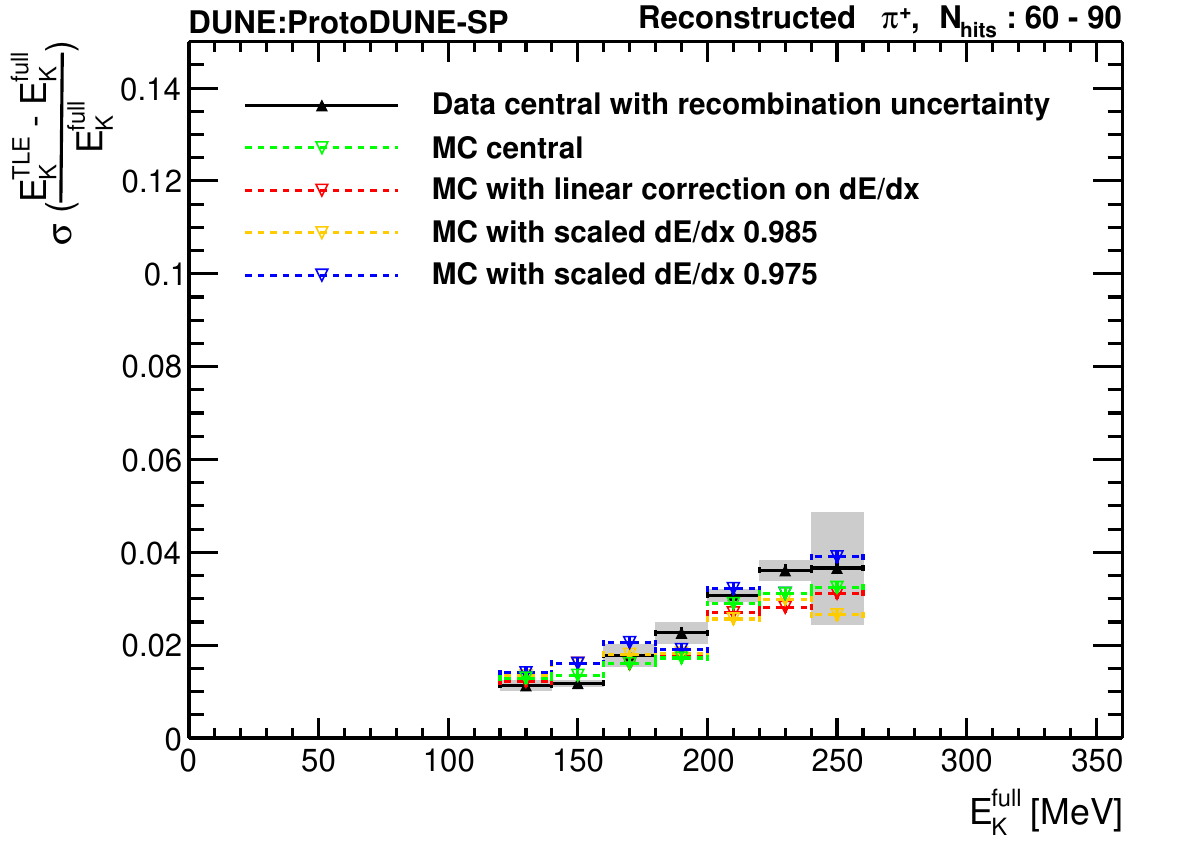}
    \caption{}
    \label{fig:Figure_020_e}
  \end{subfigure}
  \begin{subfigure}[b]{0.48\textwidth}
    \includegraphics[width=\textwidth]{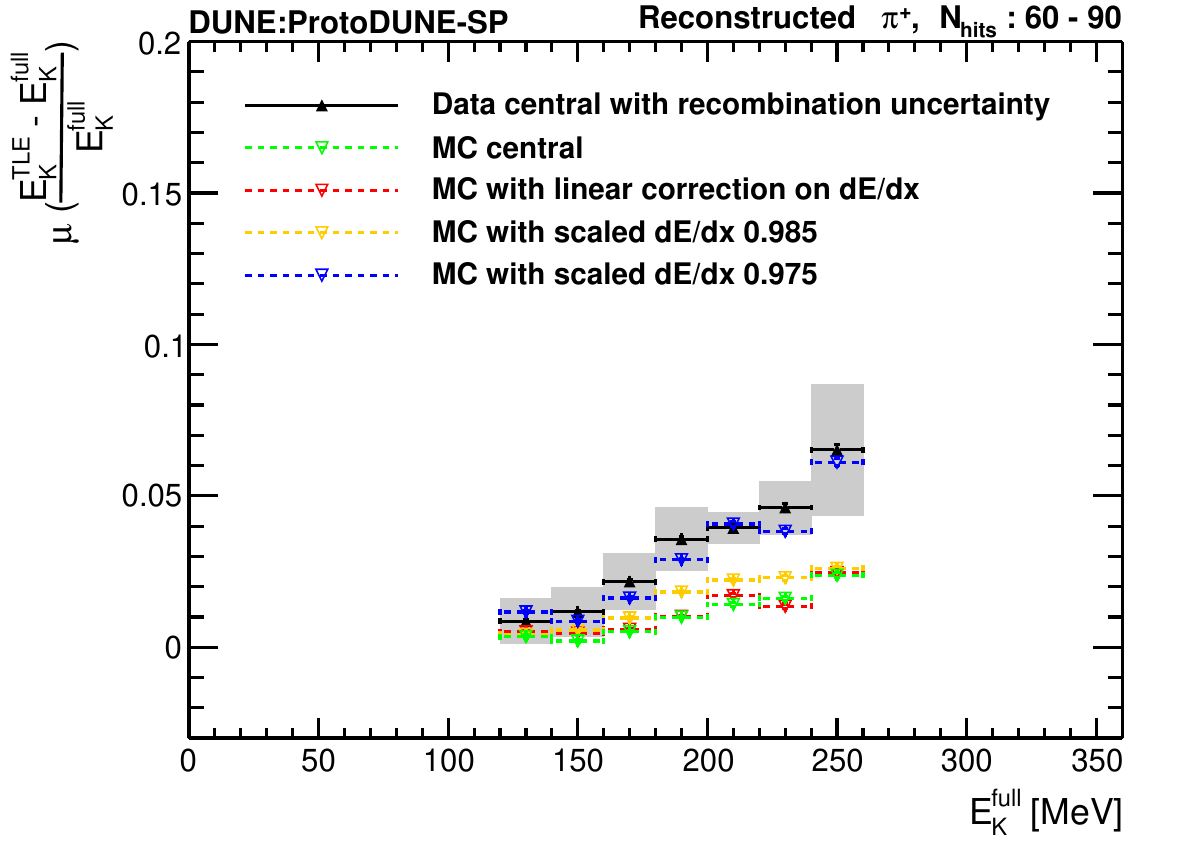}
    \caption{}
    \label{fig:Figure_020_f}
  \end{subfigure}
  \caption{Summary of studies considering the impact of \dedx modeling. Resolutions (left) and fractional biases (right) are shown as functions of charged pions' \KEfull and the number of hits (top: 15 to 30 hits, middle: 30 to 60 hits, and bottom: 60 to 90 hits). Black points show data results with gray error bars that are measured with biggest differences between data and 8 sets of shifted modified box model parameters. Green points show central MC results. Red points show results with linear scale correction on MC that is shown as a red solid line in figure~\ref{fig:Figure_017_b}. Orange and blue points show results with constant \dedx scale corrections on MC with 0.985 and 0.975, respectively.
  }
  \label{fig:Figure_020}
\end{center}
\end{figure*}

\begin{figure*}[htbp]
\begin{center}
  \begin{subfigure}[b]{0.48\textwidth}
    \includegraphics[width=\textwidth]{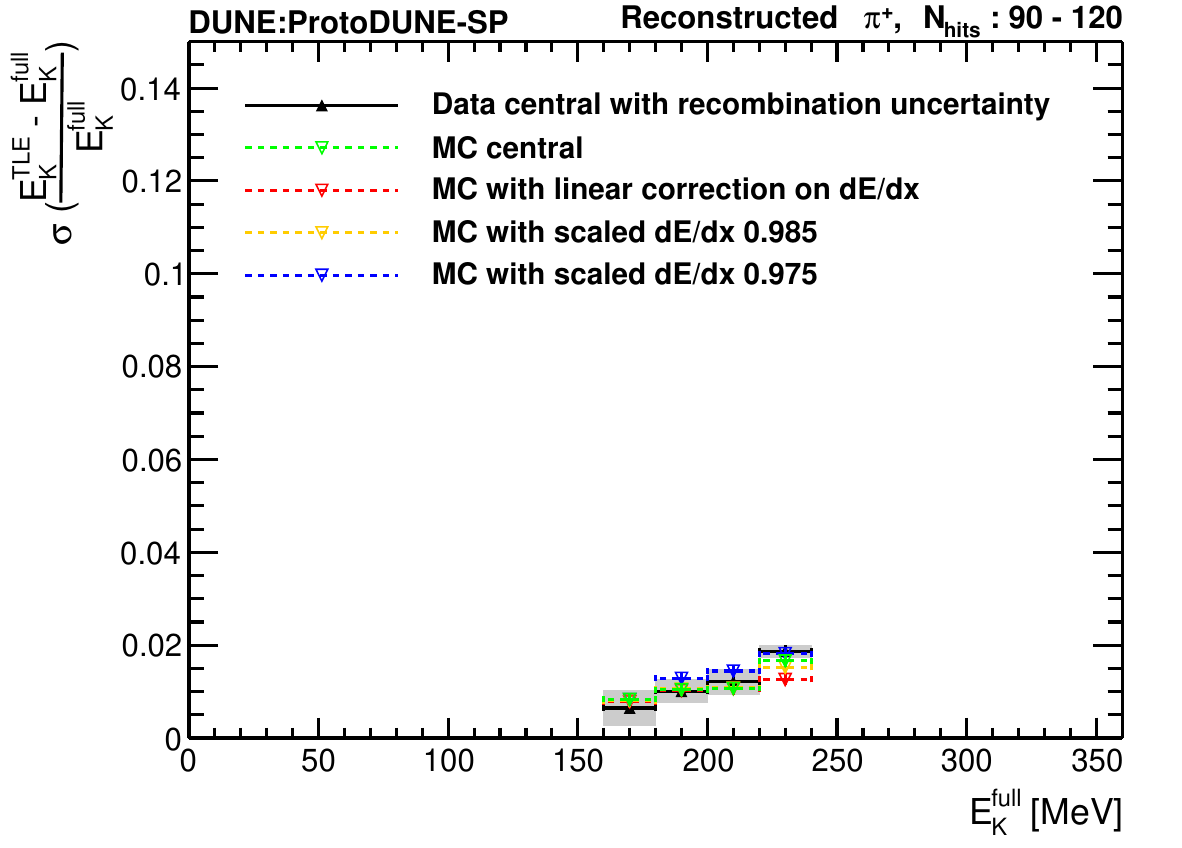}
    \caption{}
    \label{fig:Figure_021_a}
  \end{subfigure}
  \begin{subfigure}[b]{0.48\textwidth}
    \includegraphics[width=\textwidth]{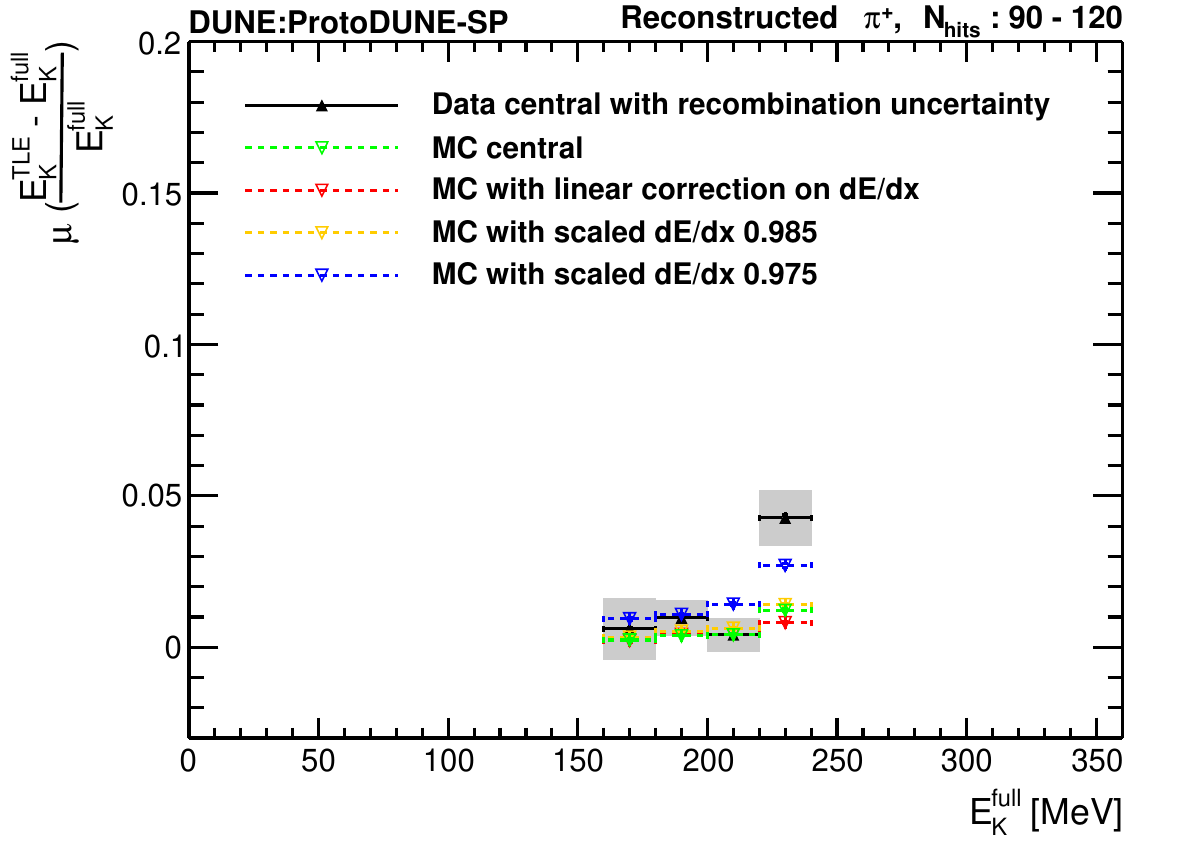}
    \caption{}
    \label{fig:Figure_021_b}
  \end{subfigure}
  \begin{subfigure}[b]{0.48\textwidth}
    \includegraphics[width=\textwidth]{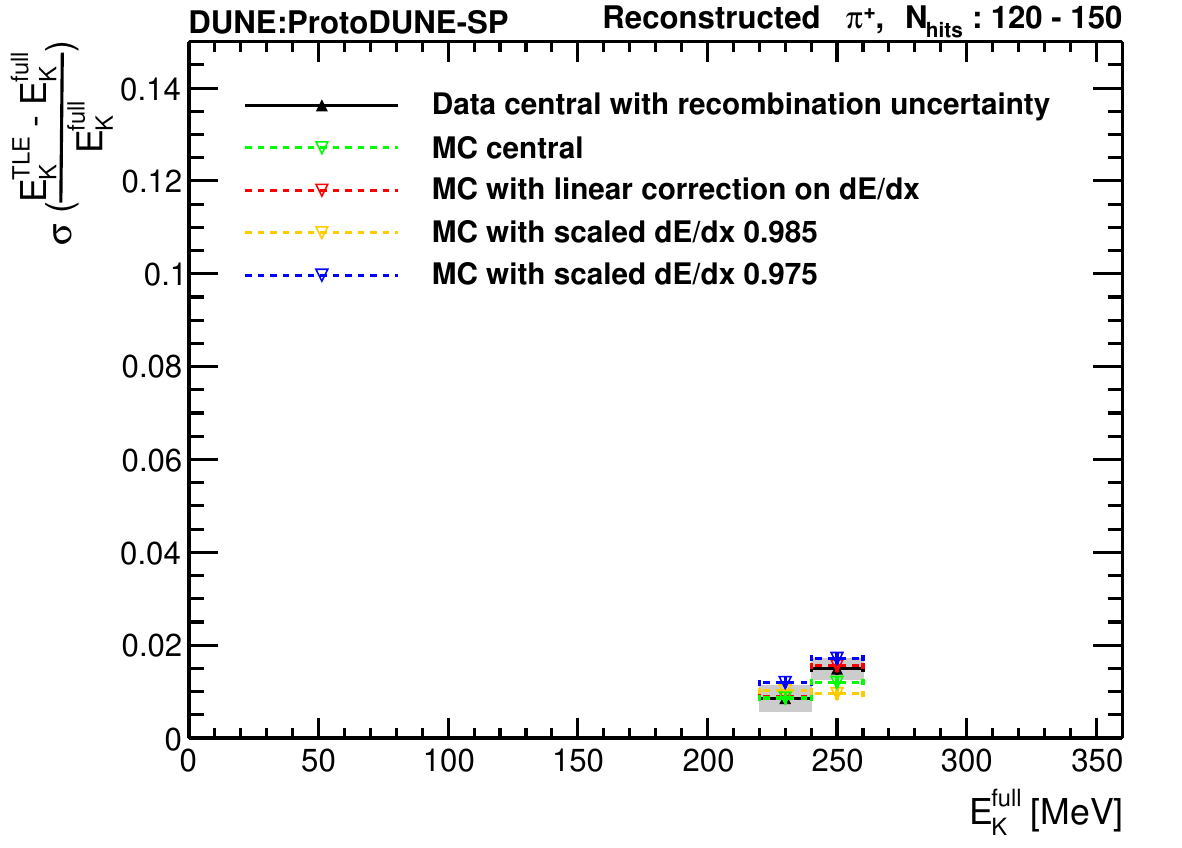}
    \caption{}
    \label{fig:Figure_021_c}
  \end{subfigure}
  \begin{subfigure}[b]{0.48\textwidth}
    \includegraphics[width=\textwidth]{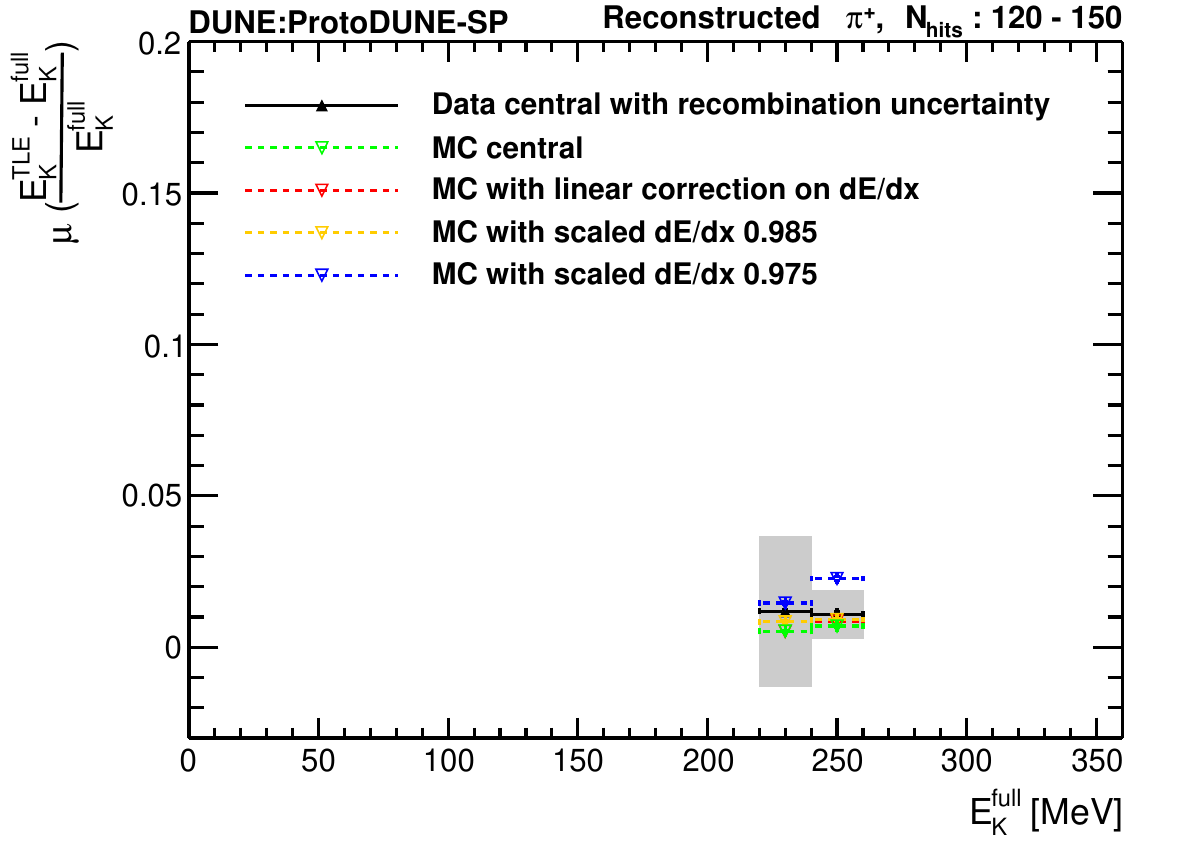}
    \caption{}
    \label{fig:Figure_021_d}
  \end{subfigure}
  \caption{Summary of studies considering the impact of \dedx modeling. Resolutions (left) and fractional biases (right) are shown as functions of charged pions' \KEfull and the number of hits (top: 90 to 120 hits and bottom: 120 to 150 hits). Black points show data results with gray error bars that are measured with biggest differences between data and 8 sets of shifted modified box model parameters. Green points show central MC results. Red points show results with linear scale correction on MC that is shown as a red solid line in figure~\ref{fig:Figure_017_b}. Orange and blue points show results with constant \dedx scale corrections on MC with 0.985 and 0.975, respectively.
  }
  \label{fig:Figure_021}
\end{center}
\end{figure*}

Figures~\ref{fig:Figure_020} and \ref{fig:Figure_021} summarize the track length-fitting method's performance results derived from the studies described above. Energy measurement resolution results are stable. For fractional biases, there are two points to note here. The first point is about \dedx scale corrections for the MC sample. While the correction factor for \dedx values exceeding 2.39 MeV/cm can reach up to 5\%, and the correction for \dedx values in the range of 1.8 to 2.3 MeV/cm can reach up to 2.5\%, the latter correction has a more significant impact on the fractional bias result. This can be understood from the slope of the Bethe-Bloch formula as a function of residual range shown in figure~\ref{fig:Figure_001}. A few-percent shift in \dedx could lead to a larger change in the residual range for small \dedx values compared to large \dedx values since the slope of the energy loss function is falling down as a function of residual range. It can also explain the tendency shown in central data that charged pions with higher \KE show larger fractional biases in their measured \KE. The second point is that shifts in modified box model parameters for data have significant impact on fractional bias results. These two points conclude that better understanding of \dedx measurements leads to smaller systematic uncertainty on the scale of measured \KE by the TLEFit method.

\clearpage

\section{Summary}
\label{sec:summary}
We introduce the track-length extension fitting (TLEFit) algorithm for measuring the kinetic energies of inelastically interacting particles in liquid argon time projection chambers.
The algorithm's performance for charged pion energy measurement is studied in detail using secondary charged pions from data collected by the ProtoDUNE-SP detector with a $1~{\rm{GeV}}/c$ charged pion beam and in a corresponding Monte Carlo sample. For charged pion tracks with kinetic energy (\KE) less than 400 \MeV and with the Bragg peak signature, the energy resolution is better than 6.5\%, and the fractional bias is less than 4\% for MC sample using true \KE as reference. For direct comparison of the performance between the data and the MC samples, range-based \KE is used as a reference after a validation that it can describe the true \KE well for charged pions passing a selection based on ${\chi}^{2}_{{\pi}^{\pm}}$. After subtracting the proton contribution from distributions of the fractional energy residual, the resolution is measured to be better than 8\% (7\%), and the fractional bias is smaller than 8\% (3\%) for the data (MC simulation) sample. Differences between the data and MC simulation samples are considered in order to calculate the energy scale correction and the systematic uncertainty on the scale and resolution. Additional studies find that the resolution results are stable and the fractional bias results are sensitive to the \dedx scale in the MIP region and also to the recombination model. Since the algorithm fits for the expected total track length until the particle would stop for an incomplete track, it can be used to measure the energies of charged pions absorbed by argon nuclei in the detector material. Absorption is the dominant inelastic scattering interaction between charged pions and argon nuclei for charged-pion kinetic energies less than 300~\MeV. In addition, the method of using stopping secondary charged pions to characterize the energy measurement performance can be used in any LArTPC by collecting neutrino interaction events with stopping charged pions in the final state.

\acknowledgments
The ProtoDUNE-SP detector was constructed and operated on the CERN Neutrino Platform.
We gratefully acknowledge the support of the CERN management, and the
CERN EP, BE, TE, EN and IT Departments for NP04/Proto\-DUNE-SP.

This document was prepared by the DUNE collaboration using the
resources of the Fermi National Accelerator Laboratory 
(Fermilab), a U.S. Department of Energy, Office of Science, 
HEP User Facility. Fermilab is managed by Fermi Research Alliance, 
LLC (FRA), acting under Contract No. DE-AC02-07CH11359.

This work was supported by
CNPq,
FAPERJ,
FAPEG and 
FAPESP,                         Brazil;
CFI, 
IPP and 
NSERC,                          Canada;
CERN;
M\v{S}MT,                       Czech Republic;
ERDF, 
Horizon Europe, 
MSCA and NextGenerationEU,      European Union;
CNRS/IN2P3 and
CEA,                            France;
INFN,                           Italy;
FCT,                            Portugal;
NRF,                            South Korea;
Generalitat Valenciana, 
Junta de Andalucıa-FEDER, 
MICINN, and 
Xunta de Galicia,               Spain;
SERI and 
SNSF,                           Switzerland;
T\"UB\.ITAK,                    Turkey;
The Royal Society and 
UKRI/STFC,                      United Kingdom;
DOE and 
NSF,                            United States of America.

\bibliographystyle{JHEP}
\bibliography{main}

\end{document}